\documentclass[12pt]{article} 
\pdfoutput=1
\usepackage{amsmath,amssymb,amsfonts}
\usepackage{psfrag}
\usepackage{color}
\definecolor{darkblue}{rgb}{0.1,0.1,.7}
\usepackage[colorlinks, linkcolor=darkblue, citecolor=darkblue, urlcolor=darkblue, linktocpage]{hyperref} 
\usepackage[square, comma, compress,numbers]{natbib}
\usepackage[]{amsmath}
\usepackage[]{graphicx}
\usepackage[]{latexsym}
\usepackage[utf8]{inputenc}
\usepackage{geometry}
\usepackage{amscd}
\usepackage[all,cmtip]{xy}
\usepackage{mathrsfs}
\usepackage{bbold}
\usepackage{subfigure}
\usepackage[margin=10pt,font=small,labelfont=bf]{caption}
\usepackage{dsdshorthand}
\usepackage{simplewick}
\usepackage{changepage}
\usepackage[multiple]{footmisc}
\numberwithin{equation}{section}

\renewcommand{\be}{\begin{eqnarray}}
\renewcommand{\ee}{\end{eqnarray}}
\newcommand{\bea}{\begin{eqnarray}}
\newcommand{\eea}{\end{eqnarray}}

\newcommand{\nn}{\nonumber}

\newcommand*\Bell{\ensuremath{\boldsymbol\ell}}

\begin{document}

\vspace*{-.6in}
\thispagestyle{empty}
\vspace{.2in}
\begin{flushright}
	CALT-TH 2017-043
\end{flushright}
{\Large
\begin{center}
{\bf The 3d Stress-Tensor Bootstrap}
\end{center}
}
\vspace{.2in}
\begin{center}
{\bf 
Anatoly~Dymarsky$^{a,b}$,
Filip~Kos$^{c,d}$, 
Petr~Kravchuk$^{e}$,\\
David~Poland$^{f}$,
David~Simmons-Duffin$^{e,g}$
}
\\
\vspace{.2in} 
$^a${\it  Department of Physics and Astronomy, University of Kentucky,\\ Lexington, KY 40506, USA}\\
$^b${\it Skolkovo Institute of Science and Technology, Skolkovo Innovation Center, \\ Moscow, Russia 143026}\\
$^c${\it Berkeley Center for Theoretical Physics, Department of Physics, \\ University of California, Berkeley, CA 94720, USA}\\
$^d${\it Theoretical Physics Group, Lawrence Berkeley National Laboratory, CA 94720, USA}\\
$^e${\it Walter Burke Institute for Theoretical Physics, Caltech, Pasadena, CA 91125, USA}\\
$^f${\it Department of Physics, Yale University, New Haven, CT 06520, USA}\\
$^g${\it School of Natural Sciences, Institute for Advanced Study, Princeton, NJ 08540, USA}\\
\end{center}

\vspace{.2in}

\begin{abstract}
We study the conformal bootstrap for 4-point functions of stress tensors in parity-preserving 3d CFTs. To set up the bootstrap equations, we analyze the constraints of conformal symmetry, permutation symmetry, and conservation on the stress-tensor 4-point function and identify a non-redundant set of crossing equations. Studying these equations numerically using semidefinite optimization, we compute bounds on the central charge as a function of the independent coefficient in the stress-tensor 3-point function. With no additional assumptions, these bounds numerically reproduce the conformal collider bounds and give a general lower bound on the central charge. We also study the effect of gaps in the scalar, spin-2, and spin-4 spectra on the central charge bound. We find general upper bounds on these gaps as well as tighter restrictions on the stress-tensor 3-point function coefficients for theories with moderate gaps. When the gap for the leading scalar or spin-2 operator is sufficiently large to exclude large $N$ theories, we also obtain upper bounds on the central charge, thus finding compact allowed regions. Finally, assuming the known low-lying spectrum and central charge of the critical 3d Ising model, we determine its stress-tensor 3-point function and derive a bound on its leading parity-odd scalar.
\end{abstract}

\newpage

\renewcommand{\baselinestretch}{0.8}\normalsize
\tableofcontents
\renewcommand{\baselinestretch}{1.0}\normalsize

\newpage

\section{Introduction}
\label{sec:introduction}
The conformal bootstrap~\cite{Ferrara:1973yt,Polyakov:1974gs,Mack:1975jr,Rattazzi:2008pe} (see \cite{Rychkov:2016iqz,Simmons-Duffin:2016gjk,Poland:2016chs} for reviews) uses basic consistency conditions to bound the space of conformal field theories. By making fewer assumptions about the theories being studied, one can derive more universal bounds.\footnote{By contrast, one can study a specific theory by inputting characteristic features that distinguish the theory in question. In this sense, the conformal bootstrap was successfully applied to extract precise properties of the 3d Ising model \cite{ElShowk:2012ht,El-Showk:2014dwa,Gliozzi:2014jsa,Kos:2014bka,Kos:2016ysd,Simmons-Duffin:2016wlq}. Families of critical $O(N)$ models \cite{Kos:2016ysd,Rattazzi:2010yc,Kos:2013tga,Chester:2014gqa,Kos:2015mba}, Gross-Neveu-Yukawa models \cite{Iliesiu:2015qra,Iliesiu:2017nrv}, and various supersymmetric theories \cite{Beem:2013qxa,Chester:2014fya,Beem:2014zpa,Bobev:2015vsa,Bobev:2015jxa,Chester:2015qca,Beem:2015aoa,Poland:2015mta,Lemos:2015awa,Chester:2015lej,Lin:2015wcg,Beem:2016wfs,Lemos:2016xke,Li:2017ddj} have also been studied in this way.} The original bounds \cite{Rattazzi:2008pe,Rychkov:2009ij,Caracciolo:2009bx,Poland:2010wg,Rattazzi:2010gj,Vichi:2011ux,Poland:2011ey,Rychkov:2011et} apply to theories with scalar operators of various dimensions. Bounds from fermionic correlators \cite{Iliesiu:2015qra,Iliesiu:2015akf,Iliesiu:2017nrv} apply to theories with fermions, and the recent bounds in \cite{Dymarsky:2017xzb} apply to any 3d CFT with a continuous global symmetry.

Perhaps the minimal possible assumption about a CFT is the existence of a stress tensor. Indeed, a stress tensor (i.e.\ a conserved spin-2 operator whose integrals are the conformal charges) is necessarily present in any local CFT.\footnote{Examples of theories without a stress tensor include boundary/defect theories \cite{Liendo:2012hy,Gaiotto:2013nva,Gliozzi:2015qsa} and nonlocal theories like the Long-Range Ising model \cite{Paulos:2015jfa,Behan:2017dwr,Behan:2017emf}.} In this work, we study the constraints of conformal symmetry and unitarity on a four-point function of stress tensors in 3d CFTs. For simplicity, we also assume a parity symmetry, so our bounds apply universally to any unitary parity-preserving local 3d CFT. This birds-eye view of local CFTs with spacetime symmetry $O(3,2)$ is similar in spirit to the views of superconformal theories achieved in \cite{Beem:2013qxa,Chester:2014fya,Beem:2015aoa,Beem:2016wfs}.

An advantage of a numerical approach is that we can make contact with analytic results, but we also have the flexibility to perform more sophisticated studies that are currently not analytically tractable. For instance, we numerically recover the conformal collider bounds \cite{Hofman:2008ar,Buchel:2009sk,Hofman:2016awc,Hartman:2016lgu}, but we can additionally study how these bounds are modified under various assumptions about the spectrum of the CFT.  As we discuss below, we also find a host of new universal bounds constraining e.g.~the spectrum of low-dimension scalar operators. 

The bootstrap equations are consistency conditions on the conformal block decomposition of 4-point functions. Written in terms of CFT data, they are quadratic constraints on OPE coefficients. Self-consistency or ``feasibility'' of these constraints can be efficiently analyzed using semidefinite programming \cite{Kos:2014bka,Poland:2011ey,Simmons-Duffin:2015qma,Simmons-Duffin:2016gjk}.  Formulating the bootstrap constraints for stress tensors in a way suitable for semidefinite programming involves several steps, which we briefly describe below. First is the task of writing 3- and 4-point functions of stress tensors in an explicitly conformally-invariant way. We do this using a combination of the embedding formalism of \cite{Costa:2011mg} and the conformal frame formalism of \cite{Kravchuk:2016qvl}.  The second step is to get rid of the degeneracies associated with permutation symmetry and conservation. This is done by identifying a minimal set of linearly-independent crossing equations, slightly refining the approach of~\cite{Dymarsky:2013wla}. These steps are explained in detail in section \ref{sec:conformalstructures}. Finally, the third step is the calculation of conformal blocks which is done in section \ref{sec:conformalblocks} by translating the approach of \cite{Costa:2011dw} to the conformal frame formalism. In this way we obtain a set of bootstrap equations suitable for numerical analysis.

In the rest of the paper we analyze the bootstrap constraints supplemented by various additional assumptions about the spectrum. In section \ref{sec:generaltheories}  we numerically reproduce, in full generality, the conformal collider bounds on the ``central charges'' of unitary theories \cite{Hofman:2008ar,Buchel:2009sk}, previously discussed in the context of the analytic bootstrap in~\cite{Li:2015itl,Hofman:2016awc}. Our main result here is a lower bound on the central charge $C_T$ as a function of the independent parameter in the stress-tensor three-point function, characterized by the angle $\theta$ defined in \eqref{thetadef}. In section \ref{sec:scalargaps} we study constraints on the spectrum of the lightest parity-even and parity-odd scalars in general unitary 3d CFTs. Some of the results are shown in figure~\ref{fig:scalarGapsExclusionPlot}. In particular, we find that any unitary CFT must necessarily have both light parity-even and light parity-odd singlet scalars in its spectrum. This is similar to a recent finding that unitary 3d CFTs with global symmetries must have low-dimension scalars in the OPE of two conserved currents \cite{Dymarsky:2017xzb}.

Quite generally, we find that when the gaps in the spectrum of scalar operators are sufficiently large to exclude large $N$ theories (by excluding some double-trace operators), the allowed region for OPE coefficients $C_T$ and $\theta$ is compact -- in particular, there exists an upper bound on the central charge. This suggests that theories with large $C_T$ must necessarily have double-trace operators in $T\times T$ OPE. Furthermore, this may potentially point to the existence of new strongly-coupled theories residing inside these compact regions. We observe the same phenomenon when imposing a gap on the dimension of the second lightest spin-2 operator in section \ref{Spin-2}.

In section \ref{Spin-4} we discuss  theories with a gap $\Delta_4$ in the spectrum of spin-4 parity-even operators. In full consistency with the Nachtmann theorem, we observe that when $\Delta_4$ approaches $6$, the lower bound on $C_T$ grows indefinitely for all $\theta$, in accord with the expectation that the corresponding theory is dual to weakly coupled gravity in AdS${}_4$. Finally, section \ref{sec:Ising} is devoted to studies of the 3d Ising model. Under the assumption of no relevant parity-odd scalars, and by imposing the known values of the central charge and the dimensions of certain light operators, we obtain a window $0.01 < \theta < 0.05$. Under stronger but still plausible assumptions we obtain a tighter bound $0.010 < \theta < 0.019$. We also find an upper bound on the parity-odd scalar gap $\Delta_{\text{odd}} < 11.2$. We conclude with a discussion in section \ref{Discussion}.
 
\section{Conformal structures}
\label{sec:conformalstructures}

\subsection{3-point structures}
\label{sec:threeptstructures}

To set up the bootstrap equations for the 4-point function $\<TTTT\>$ in 3d CFTs preserving parity, we first need to understand the possible 3-point functions $\<TT\cO\>$ between the stress tensor $T^{\mu\nu}$ and various operators $\cO$ in the CFT. The purpose of this section is to classify such 3-point functions, and thus the operators which can be exchanged in the OPE decomposition of $\<TTTT\>$.

First of all, only bosonic operators $\cO$ can appear in $T \times T$ OPE, and so without loss of generality we can assume that $\cO$ is a traceless symmetric tensor primary of spin $\ell$. Furthermore, since $T$ is a singlet under all global symmetries, $\cO$ must be a singlet as well. However, $\cO$ may be even or odd under space parity.

The 3-point functions $\<TT\cO\>$ should be conformally-invariant, symmetric with respect to permutation of the two $T$ insertions, and satisfy the conservation equation for the stress tensor,
\be
	\ptl_\mu T^{\mu\nu}=0+\text{contact terms}.
\ee
Such 3-point functions have the form
\be
	\<TT\cO\>=\sum_{a=1}^{N_{TT\cO}}\lambda^{(a)}_{TT\cO}\<TT\cO\>_{(a)},
\ee
where $\<TT\cO\>_{(a)}$ are 3-point tensor structures which form a basis of solutions to the above constraints, and $\lambda^{(a)}_{TT\cO}$ are OPE coefficients. We can always choose a basis such that $\lambda_{TT\cO}^{(a)}$ are real.

The 3-point tensor structures $\<TT\cO\>_{(a)}$ can be classified using e.g.\ the conformal frame formalism of~\cite{Kravchuk:2016qvl}. We will also need to perform manipulations with explicit expressions, which we can obtain by constructing the tensor structures using the 5d embedding space formalism of~\cite{Costa:2011mg,Costa:2011dw}.

In this latter formalism, the parity-even 3-point tensor structures are constructed from basic invariants denoted by $H_{ij}$ and $V_i$, where $i$ and $j$ index the operators in the 3-point function. The structure $H_{ij}$ increases the spin by one unit for operators $i$ and $j$, while $V_i$ does so only for the operator $i$. For example, a general 3-point structure for $\<TT\phi\>$ with a scalar $\phi$ of dimension $\Delta$ is given by\footnote{We assume that the stress tensors are at positions $1$ and $2$, while the intermediate operator is at position $3$.}
\be
\<TT\phi\>=\frac{\alpha H_{12}^2+\beta H_{12}V_1 V_2 +\gamma V_1^2V_2^2}{
	(-2X_1\cdot X_2)^{\frac{10-\Delta}{2}}
	(-2X_2\cdot X_3)^{\frac{\Delta}{2}}
	(-2X_3\cdot X_1)^{\frac{\Delta}{2}}
},
\ee
where the constants $\alpha,\beta,\gamma$ are subject to linear constraints coming from conservation of $T$ and permutation symmetry, while $X_i$ are the embedding space coordinates of the operators~\cite{Costa:2011mg}. For sufficiently large $\ell$ there are 14 different combinations of $H_{ij}$ and $V_{i}$ which give the correct spins for the three operators in $\<TT\cO\>$. Not all of them are independent, since there exist non-linear relations between the invariants $H$ and $V$, which were classified in~\cite{Costa:2011mg}. In our case there is a single redundant structure
\be\label{eq:redundant_tensor_structure}
	H_{12}H_{23}H_{31}V_3^{\ell-2},
\ee
which can be expressed in terms of other structures.

Using the results of~\cite{Costa:2011mg}, it is straightforward to impose permutation and conservation constraints on these tensor structures. An analogous construction works for parity-odd tensor structures~\cite{Costa:2011mg}. We will not need the explicit expressions for the tensor structures in this ``algebraic'' basis, but rather in the so called differential basis, which we describe in section~\ref{sec:conformalblocks}.\footnote{We will still use input from the algebraic basis to perform calculations in the differential basis.} The explicit expressions in the differential basis are provided in appendix~\ref{app:tensor}.

Here, let us summarize the counting of 3-point tensor structures. Let $\cO_\ell$ denote a primary operator of spin $\ell$ and a scaling dimension $\De$ strictly above the unitarity bound. This restriction is important since the number of solutions to conservation equations can increase at special values of $\De$.\footnote{Note that the conservations constraints are linear with coefficients dependent on $\Delta$. The rank of a parameter-dependent linear system is always constant at generic values of the parameters and can only decrease at special values.} In fact, this is what happens for $\De=3$ and $\ell=2$, i.e.\ when $\cO_{\ell=2}=T$ is the stress tensor itself. With these conventions, the counting of 3-point tensor structures is given by the table:
\begin{center}
	\begin{tabular}{c|c}
		$\cO$ & $N_{TT\cO}$ \\
		\hline
		$\cO_0$ & $1^++1^-$\\
		\hline
		$\cO_2$ & $1^++1^-$\\
		\hline
		$T$ & $2^+$+$1^-$\\
		\hline
		$\cO_{2n}, n\geq 2$ & $2^++1^-$\\
		\hline
		$\cO_{2n+1}, n\geq 2$ & $1^-$\\
	\end{tabular}
\end{center}
where we have separated parity-even and parity-odd tensor structures (indicated by the $\pm$ superscripts). For $\cO=T$, the tensor structures are invariant under permutations of all three operators. Note that the parity-odd tensor structure for $\<TTT\>$ does not appear in a parity-preserving theory, since $T$ is necessarily parity-even, as can be seen from the Ward identity discussed below.

\subsubsection{Ward identities}

As mentioned above, the 3-point function $\<TTT\>$ has two allowed parity-even tensor structures, which can be realized in the theories of a free real scalar and a free Majorana fermion,
\be
	\<TTT\>=n_B\<TTT\>_B+n_F\<TTT\>_F.
\ee
There exists a non-trivial Ward identity for this correlator. Indeed, one can construct the dilatation current $J_D^\mu=x_\nu T^{\mu\nu}$ from one of the three stress-tensor operators, and integrate it over a surface surrounding another stress-tensor operator put at $x=0$ to obtain, schematically,
\be
	\int x\<TTT\> dS = \Delta_T\<TT\>.
\ee
This Ward identity implies a linear relation between the coefficients $n_B,n_F$ and the 2-point function $\<TT\>$. The latter can be parametrized as 
\be
	\<TT\>=C_T \<TT\>_B,
\ee
where $\<TT\>_B$ is the 2-point function $\<TT\>$ in the theory of a free real scalar and $C_T$ is the ``central charge." The Ward identity then must be of the form
\be\label{eq:wardidentity}
	C_B n_B+C_F n_F = C_T.
\ee
The constants $C_B, C_F$ are simply the central charges of the free real scalar and free Majorana fermion respectively, where our normalization for $C_T$ implies $C_B = C_F = 1$. However, in the sections below we will often write results in terms of the ratio $C_T/C_B$ so that they also hold for other normalizations of $C_T$.

\subsection{4-point structures}

The 4-point function $\<TTTT\>$ should satisfy the following properties, which interact with each other in nontrivial ways:
\begin{itemize}
\item conformal invariance,
\item permutation symmetry,
\item conservation,
\item regularity (analyticity).
\end{itemize}
We will address each property in turn, culminating in a minimal set of crossing symmetry equations suitable for applying numerical bootstrap techniques.

It is useful to use index-free notation to encode different tensor structures. Let us write
\be
T(w,x) &=& w_\mu w_\nu T^{\mu\nu}(x),
\ee
where $w_\mu$ is an auxiliary polarization vector. Because $T^{\mu\nu}$ is traceless, we can take $w_\mu$ to be null, $w^2=0$. We can recover $T^{\mu\nu}$ as
\be
T^{\mu\nu}(x) &= D^\mu_w D^\nu_w T(w,x), 
\ee
where $D^\mu_w$ is the Todorov operator~\cite{Dobrev:1975ru}
\be
\label{eq:todorov}
D^\mu_w &=& \p{\frac{d-2}{2} + w\.\pdr{}{w}}\pdr{}{w_\mu} - \frac 1 2 w^\mu \frac{\ptl^2}{\ptl w\.\ptl w},
\ee
with $d=3$ the spacetime dimension.
Note that the Todorov operator preserves the ideal generated by $w^2$,
\be
D_w^\mu (w^2 f(w)) = w^2 (\dots),
\ee
so it is well-defined even though $w$ is constrained to be null.

\subsubsection{Conformal invariance}
\label{sec:fourptstructures}

To study the above properties, it is useful to fix a conformal frame and use representation theory of stabilizer groups to classify tensor structures, following \cite{Kravchuk:2016qvl}. This approach makes it easy to deal with degeneracies between tensor structures in low spacetime dimensions, and will also help us understand regularity conditions on the $z=\bar z$ line. We work in Euclidean signature throughout.

Using conformal transformations we can place the four operators in the 1-2 plane in the following configuration:
\be
\label{eq:conformalframe}
g(z,\bar z, w_i) &=& \<T(w_1,0) T(w_2, z) T(w_3, 1) T(w_4, \oo)\>.
\ee
We have $z=x^1 + i x^2$ and $\bar z = x^1 - i x^2$, with the direction perpendicular to the plane being $x^3$. For brevity, we have written only the holomorphic coordinate of each operator. 

We define the operator at infinity in a non-standard way, where we do not act with an inversion on the polarization vector, 
\be
T(w,\oo) \equiv \lim_{L\to\oo} L^{2\De_T} T(w,L),\quad \Delta_T=3\ .
\ee
The virtue of this convention is that the polarization vectors are treated more symmetrically, so it will be easier to understand the action of permutations.

We will consider parity-preserving theories, so the group of spacetime symmetries is $O(4,1)$. The points $0,z,1,\oo$ are stabilized by an $O(1)=\Z_2$ subgroup of $O(4,1)$ consisting of reflections in the $x^3$ direction (perpendicular to the plane). The 4-point function $g(z,\bar z, w_i)$ must be invariant under this stabilizer subgroup or ``little-group." Little-group invariance then guarantees that $g(z,\bar z, w_i)$ can be extended to an $O(4,1)$-invariant function for arbitrary configurations of the $T(w_i,x_i)$.

Let $\Bell^\pm$ denote the parity-even/odd spin-$\ell$ representation of $O(3)$, and let $\bullet^\pm$ denote the even and odd representations of $O(1)$. Each operator $T(w,x)$ transforms in the representation $\mathbf{2}^+$ of $O(3)$.
Little-group invariants are $O(1)$ singlets in
\be
\p{\mathrm{Res}^{O(3)}_{O(1)} \mathbf{2}^+}^{\otimes 4}
= \p{3\,{\bullet^{+}} \oplus\, 2\,{\bullet^{-}}}^{\otimes 4}
= 313\,{\bullet^{+}} \oplus 312\,{\bullet^{-}},
\ee
where $\Res{}^G_H \rho$ denotes the restriction of a representation $\rho$ of $G$ to a representation of $H\subseteq G$. In particular, there are $313$ parity-even tensor structures (and 312 parity-odd tensor structures).

These structures are easy to enumerate.  Define components of the polarization vectors
\be
\w &=& w^z = w^1 + i w^2\nn\\
\bar \w &=& w^{\bar z} = w^1 - i w^2\nn\\
\omega^0 &=& w^3.
\ee
For each ``helicity" $h\in \{-2,-1,0,1,2\}$, we can construct a unique monomial $[h]$ with degree $2$ and charge $h$ under rotation in the $z$-plane,
\be\label{eq:helicitybasis}
[-2]=\bar\omega^2,\quad [-1] = \bar \omega \omega^0, \quad [0] = \omega \bar \omega, \quad [1]=\omega \omega^0,\quad [2] = \omega^2.
\ee
(Using the fact that $w_\mu w^\mu = (\omega^0)^2 + \omega \bar \omega = 0$, we can ensure that the degree in $\omega^0$ is at most one.) Let $[h_1h_2h_3 h_4]$ denote a product of the corresponding monomials for each polarization vector $w_i^\mu$.\footnote{This definition differs from the one based on spinor polarizations in \cite{Kravchuk:2016qvl} by a numerical factor.}  It is easy to verify that there are 313 structures $[h_1 h_2 h_3 h_4]$ which are even under parity $\w^0\to -\w^0$, i.e.\ such that $\sum_i h_i \equiv 0\!\! \mod 2$.  The 4-point function is a linear combination of these structures, with coefficients that are functions of $z$ and $\bar z$,
\be
\label{eq:decompositionintostructures}
g(z,\bar z, w_i) &=& \sum_{\sum_i h_i \textrm{ even}} [h_1 h_2 h_3 h_4] g_{[h_1h_2 h_3 h_4]}(z,\bar z).
\ee

Using rotations around the $x_1$ axis, we can relate the point $(z,\bar z)$ to its reflection in the imaginary direction $(\bar z, z)$.  Invariance of the full correlator under this transformation implies
\be
\label{eq:zparity}
g_{[h_1 h_2 h_3 h_4]}(z,\bar z) &=& g_{[-h_1,-h_2,-h_3,-h_4]}(\bar z, z).
\ee
Meanwhile, reality\footnote{Reality of $\<TTTT\>$ follows from a combination of space parity and Euclidean Hermitian conjugation.} of $g$ implies
\be
g_{[h_1 h_2 h_3 h_4]}(z,\bar z) &=& \bar g_{[-h_1,-h_2,-h_3,-h_4]}(\bar z, z),
\ee
where we used the notation $\bar f(\bar z, z)\equiv (f(z,\bar z))^*$,
from which it follows that
\be
g_{[h_1 h_2 h_3 h_4]}(z,\bar z) &=& \bar g_{[h_1 h_2 h_3 h_4]}(z,\bar z).
\ee
In other words, the functions $g_{[h_1 h_2 h_3 h_4]}(z,\bar z)$ must have real coefficients in a Taylor series expansion in powers of $z$ and $\bar z$.

\subsubsection{Permutation invariance}
\label{subsec:permutation}

The 4-point function $\<T(w_1,x_1)\cdots T(w_4,x_4)\>$ must be invariant under permutations of the four operators. Permutations that change the cross-ratios $z,\bar z$ lead to nontrivial crossing equations that we explore later. However, permutations that leave $z,\bar z$ invariant, which we call ``kinematic permutations," give constraints on tensor structures alone \cite{Dymarsky:2013wla,Kravchuk:2016qvl}. In our case, the group of kinematic permutations is (in cycle notation)
\be
\Pi^\mathrm{kin} = \{\mathrm{id},(12)(34),(13)(24),(14)(23)\} = \Z_2 \x \Z_2.
\ee

As shown in \cite{Kravchuk:2016qvl}, $\Pi^\mathrm{kin}$-invariant tensor structures are in one-to-one correspondence with
\be
\label{eq:permsymcounting}
\p{\bigotimes_{i=1}^4\mathrm{Res}^{O(3)}_{O(1)} \mathbf{2}^+}^{\Pi^\mathrm{kin}},
\ee
where $\Pi^\mathrm{kin}$ acts on tensor factors in the natural way, and $(\rho)^G$ denotes the $G$-invariant subspace of $\rho$. These can be counted using
\be
\label{eq:z2z2invariants}
(\rho^{\otimes 4})^{\Z_2\x \Z_2} &=& \rho^4 \ominus 3(\wedge^2 \rho \otimes \mathrm{S}^2 \rho),
\ee
where $\ominus$ represents the formal difference in the character ring. Plugging in $\rho = 3\,{\bullet^{+}} \oplus\, 2\,{\bullet^{-}}$ to (\ref{eq:z2z2invariants}), we find
\be
((3\,{\bullet^{+}} \oplus\, 2\,{\bullet^{-}})^{\otimes 4})^{\Z_2\x \Z_2} &=& 97\,{\bullet^{+}} \oplus\, 78\,{\bullet^{-}},
\ee
so there are $97$ permutation-invariant parity-even structures.

\begin{table}[ht!]
\begin{center}
\begin{tabular}{r || c | c | c | c }
			& $r_1$ 		& $r_2$ 		& $r_3$ 		& $r_4$ 		\\
\hline\hline
id			& $1$ 		& $1$ 		& $1$		&	 $1$			\\\hline
$(12)(34)$	& $-(1-z)$	& $-(1-\bar z)$		& $-(1-\bar z)$ 	&	$-(1-z)$	\\\hline
$(13)(24)$	& $\bar z(1-z)$		& $\bar z(1-\bar z)$		& $z(1-\bar z)$ 		&	$z(1-z)$			\\\hline
$(14)(23)$	& $-\bar z$		& $-\bar z$ 		& $-z$		&	$-z$		\\\hline
\end{tabular}
\end{center}
\caption{Permutation phases for a 4-point function of identical operators, computed in \cite{Kravchuk:2016qvl}.}
\label{tab:permphases}
\end{table}

To write the structures explicitly, we must be more specific about the action of permutations on polarization vectors. A permutation $\pi \in \Pi^\mathrm{kin}$ acts on a monomial $[h_i]$ as
\be
\pi: [h_i] &\mto& n(r_i(\pi))^{h_i}[h_{\pi(i)}],
\ee
where $n(x) = \sqrt{x/\bar x}$ is a phase and the $r_i(\pi)$ are given in the table~\ref{tab:permphases}.
Permutation-invariant structures are given by symmetrizing with respect to this action:
\be
\<h_1h_2h_3h_4\>_z &\equiv& \frac{1}{m_{h_1h_2h_3h_4}}\Big([h_1 h_2 h_3 h_4]\nn\\
&& + n(1-z)^{-h_1+h_2+h_3-h_4} [h_2 h_1 h_4 h_3]\nn\\
&& + n(z)^{h_1+h_2-h_3-h_4} [h_4 h_3 h_2 h_1]\nn\\
&& + n(z)^{h_1+h_2-h_3-h_4} n(1-z)^{-h_1+h_2+h_3-h_4}[h_3 h_4 h_1 h_2]\Big),
\ee
where $m_{h_1h_2h_3h_4}$ is the number of elements $\Pi^\mathrm{kin}$ which stabilize $[h_1h_2h_3h_4]$. We have also added an index $z$ to the symmetric tensor structures to indicate that they depend on~$z$ and~$\bar z$. Here, it's clear that independent $\Pi^\mathrm{kin}$-invariant structures are in one-to-one correspondence with orbits of $\Z_2\x\Z_2$ when acting on quadruples $[h_1 h_2 h_3 h_4]$. Making a choice of representative for each of the 97 parity-even orbits, we can write
\be
g(z,\bar z, w_i) &=& \sum_{\substack{h_i / \Z_2^2 \\ \sum_i h_i \textrm{ even}}} \<h_1 h_2 h_3 h_4\>_z\, g_{[h_1 h_2 h_3 h_4]}(z,\bar z).
\ee
Note that the functions $g_{[h_1 h_2 h_3 h_4]}(z,\bar z)$ are the same as those appearing in~(\ref{eq:decompositionintostructures}).

\subsubsection{Conservation}
\label{subsec:conservation}

Imposing conservation of $T^{\mu\nu}(x)$ gives nontrivial differential equations relating the functions $g_{[h_1 h_2 h_3 h_4]}(z,\bar z)$. These equations can be solved up to some undetermined functions of $z,\bar z$ that we call ``functional degrees of freedom." Conversely, after imposing conservation, the functional degrees of freedom fix the entire correlator (modulo boundary terms that we discuss below). Thus, an independent set of crossing-symmetry equations should make reference to functional degrees of freedom alone.

In \cite{Dymarsky:2013wla}, it was shown that there are 5 functional degrees of freedom in a 4-point function of stress tensors in 3d. We can obtain the number $5$ with a simple group-theoretic rule from \cite{Kravchuk:2016qvl}. To account for conservation, we simply replace
\be
\Res{}^{O(3)}_{O(1)} \mathbf{2}^+ &\to& \Res{}^{O(2)}_{O(1)}\mathbf{2} = \bullet^+ \oplus \bullet^-
\ee
in~(\ref{eq:permsymcounting}).  Here, $O(2)$ can be interpreted as the little group of a massless particle in 4 dimensions, and $\mathbf{2}$ on the right-hand side of the arrow represents the spin-2 representation of $O(2)$. Plugging $\rho=\bullet^+\oplus \bullet^-$ into (\ref{eq:z2z2invariants}), we find $5\bullet^+\oplus\,\, 2\,\bullet^-$, so there are indeed 5 parity-even functional degrees of freedom.

Let us see more explicitly how these 5 degrees of freedom come about. Because the permutation group $\Pi^\mathrm{kin}$ acts freely on the four points, it suffices to impose conservation at one of the points, say $x_2$. The conservation equation is
\be
D_{w_2} \. \pdr{}{x_2} \<T(w_2,x_2) \cdots\> &=& 0,
\ee
where $D_w$ is the Todorov operator~(\ref{eq:todorov}). Restricting to the conformal frame configuration~(\ref{eq:conformalframe}), this gives\footnote{The Todorov operator in the first two terms simplifies because of our choice of tensor structures~\eqref{eq:helicitybasis}, which is at most linear in $\omega^0$.}
\be
\p{\p{\frac 3 2-\w\ptl_\w}\ptl_{\bar \w} \ptl_z + \p{\frac 3 2 -\bar \w \ptl_{\bar \w}}\ptl_\w \ptl_{\bar z} + \frac{i D_w^3 \cL_{23}}{z-\bar z}} g(z,\bar z, w_i) &=& 0,
\label{eq:conservationeq}
\ee
where
\be
\cL_{23} &=& i \sum_k \p{ \w_k^0 \p{\ptl_{\w_k} - \ptl_{\bar \w_k}} + \frac 1 2 (\w_k - \bar \w_k)\ptl_{\w_k^0}}
\ee
is the generator of rotations in the 2-3 plane acting on polarization vectors.  In (\ref{eq:conservationeq}), $\w,\bar \w,\w^0$ refer to $\w_2, \bar \w_2, \w^0_2$, respectively. The last term in the conservation equation is naively singular at $z=\bar z$. However, the singularity will be cancelled by zeros in the action of $\cL_{23}$. These complications stem from the fact that $z=\bar z$ is a locus of enhanced symmetry, where the little group becomes $O(2)$ instead of $O(1)$. We will study these issues in more detail below.

Following \cite{Dymarsky:2013wla}, we can solve (\ref{eq:conservationeq}) by thinking of one of the directions in the $z$-$\bar z$ plane as ``time" $t$ and the other as ``space'' $\xi$ and integrating away from a constant time slice. The conservation equation then has the structure 
\be
\label{eq:timespace}
(A \ptl_t + B \ptl_\xi + C) g &=& 0,
\ee
where $A,B,C$ are linear operators on the space of tensor structures. The number of functional degrees of freedom is the dimension of the kernel of $A$.

In our case, it is convenient to choose $z$ as the time direction, with $\bar z$ as the space direction. The operator $A$ is then $A_z=\p{\frac 3 2-\w_2\ptl_{\w_2}}\ptl_{\bar \w_2}$, which vanishes on any structure that is independent of $\bar \w_2$. This restricts the helicity $h_2$ to be either $1$ or $2$. Because permutations $\Pi^\mathrm{kin}$ act freely, all helicities must be either $1$ or $2$, so the kernel of $A_z$ is spanned by the five structures
\be
\<2222\>_z,\quad
\<1111\>_z,\quad
\<1212\>_z,\quad
\<1122\>_z,\quad
\<2112\>_z.
\label{eq:conservedstructures}
\ee

When integrating the conservation equation, we can set the coefficients of these structures to anything we like. In practice, it will be useful to use a slightly different basis of functional degrees of freedom.  Let
\be
\<h_1h_2 h_3 h_4\>^\pm_z &=& \frac 1 2 \p{\<h_1 h_2 h_3 h_4\>_z \pm \<-h_1,-h_2,-h_3,-h_4\>_z},
\ee
and define the corresponding coefficient functions
\be
g^\pm_{[h_1h_2h_3h_4]}(z,\bar z) &=& g_{[h_1h_2h_3h_4]}(z,\bar z) \pm g^\pm_{[-h_1,-h_2,-h_3,-h_4]}(z,\bar z).
\ee
Equation~(\ref{eq:zparity}) implies
\be
g^\pm_{[h_1h_2h_3h_4]}(\bar z, z) &=& \pm g^\pm_{[h_1h_2h_3h_4]}(z, \bar z).
\ee
We will take the functions $g^+_{[h_1h_2h_3h_4]}(z,\bar z)$ as our functional degrees of freedom. Fixing these functions is sufficient to remove ambiguities when integrating the conservation equation in the $z$-direction. By working in a Taylor expansion in $z,\bar z$, it is easy to argue that fixing $g^+_{[h_1h_2h_3h_4]}(z,\bar z)$ removes ambiguities when integrating in any direction. In particular, later we will integrate the conservation equation in the $x_2 = \Im z$ direction.

As explained in \cite{Dymarsky:2013wla}, in order to consistently integrate (\ref{eq:timespace}) away from a spatial slice, the initial data might need to satisfy additional constraints. Suppose $N$ is a matrix such that $NA=0$. Acting with $N$ on (\ref{eq:timespace}), we obtain
\be
\label{eq:initialconstraint}
(NB\ptl_\xi + NC) g &=& 0.
\ee
This constraint turns out to be first class, meaning that we only need to impose it on the initial data. Our initial slice will be the line $z=\bar z$. Because this is a locus of enhanced symmetry, we must take care while analyzing the conservation equation around it. 

\subsubsection{Regularity and boundary conditions}
\label{subsec:regularity}

For numerical bootstrap applications, we would like to write the crossing equations in a Taylor series expansion around the point $z=\bar z = \frac 1 2$. The line $z=\bar z$ corresponds to the four points $x_i$ becoming collinear, which means the stabilizer group is enhanced from $O(1) \to O(2)$. Since the tensor structures have to be invariant under the stabilizer group, we can see that there are boundary conditions at $z=\bar z$ which the functions $g_{[h_1h_2h_3h_4]}$ have to satisfy in a well-defined correlator. As we will now show, smoothness of the correlator places further constraints on the Taylor expansion of $g_{[h_1h_2h_3h_4]}$ around this locus.

Consider the 4-point function after fixing $x_1,x_3,x_4$, but before rotating $x_2$ into the 1-2 plane,
\be
g(x_2,w_i) &=& \<T(w_1,0) T(w_2,x_2) T(w_3,e) T(w_4,\oo)\>.
\ee
Here, $e=(1,0,0)$ is a unit vector in the 1-direction. We want the correlator to be smooth in $x_2$. In particular, it should have a Taylor expansion in the directions orthogonal to $e$,
\be\label{eq:yexpansion}
g(x_2, w_i) &=& \sum_{n=0,\ell=0}^\oo g^{\mu_1\cdots\mu_\ell}_n(w_i, x) y_{\mu_1} \cdots y_{\mu_\ell}y^{2n},
\ee
where $y_\mu=(x_2)_\mu - e_\mu(x_2\.e)$ is the projection of $x_2$ onto the directions orthogonal to $e$, and $x=e\.x_2$. The coefficient functions $g^{\mu_1\cdots\mu_\ell}_n(w_i,x)$ are symmetric tensors of the stabilizer group $O(2)$, built out of polarization vectors. Let us count them. Let $\mathbf{0}^\pm$ denote the parity-even/odd scalar of $O(2)$, and let $\Bell$ denote the spin-$\ell$ representation of $O(2)$. Each operator transforms in the representation
\be
\rho = \Res{}^{O(3)}_{O(2)} \mathbf{2}^+ = \mathbf{2} \oplus \mathbf{1} \oplus \mathbf{0}^+.
\ee
Although $\Z_2\x\Z_2$ permutations act in a way that depends on $x$ and $y_\mu$, the leading-order in $y$ action is simply the obvious permutation of polarization vectors, because the phases $n(r_i(\pi))$ are trivial on the line $z=\bar z$.\footnote{In fact, as shown in~\cite{Kravchuk:2016qvl}, we can define polarization vectors $\tilde w_i=w_i+O(y)$, which permute with trivial phases to all orders in $y$. We can then use these polarization vectors in \eqref{eq:yexpansion}.} Thus, for the sake of counting new permutation-invariant tensor structures at each order in $y_\mu$, we can use (\ref{eq:z2z2invariants}), which gives
\be
\label{eq:o2decomposition}
(\rho^{\otimes 4})^{\Z_2\x\Z_2} &=& 22\,\mathbf{0}^+ \oplus 3\,\mathbf{0}^- \oplus \dots. 
\ee

Equation~\eqref{eq:yexpansion} implies that a polarization structure transforming in $\Bell$ of $O(2)$ can appear starting at order $\ell$ in the $y$-expansion. From~(\ref{eq:o2decomposition}) we see that at zeroth order in $y$, there are 22 parity-even permutation-invariant structures that can appear (out of 97 total).\footnote{Incidentally, 22 is also the number of functional degrees of freedom in a 4-point function of stress tensors in 4d. This is because the stabilizer group of a generic configuration of 4-points in 4d is $O(2)$, while the little group for massless particles in 5d is $O(3)$. Thus, the representation theory computation is the same as the one here (see \cite{Dymarsky:2013wla,Kravchuk:2016qvl}).} In order for the 4-point function to be well-defined at $z=\bar z$, only the coefficients of these 22 structures can be nonzero.

It turns out that thanks to the conservation equation, this is the only condition that we have to worry about. In general, since \eqref{eq:o2decomposition} gives $O(2)$ spins up to $\mathbf8$, in the absence of the conservation equation we would have to write similar conditions for the first $8$ orders in $\Im z$. However, as the derivation above shows, these constraints follow from $O(2)$ invariance. In particular, the conservation equation is compatible with~\eqref{eq:yexpansion} in the sense that it produces a recursion relation for the coefficients $g_n$. Therefore, as long as the zeroth order constraints are satisfied, higher orders follow automatically.\footnote{One should make sure that the choice of independent two-variable degrees of freedom does not contradict the regularity constraints. Or, equivalently, that these degrees of freedom are indeed independent from the point of view of the recursion relation for~\eqref{eq:yexpansion}. We have checked that it is true for our choice of two-variable degrees of freedom.} We have explicitly verified this by working order-by order in a Taylor expansion in $\Im z$.

Thus, our initial conditions include 22 undetermined functions of a single variable $\Re z$.  We can take $5$ of these to be the restrictions of our two-variable degrees of freedom to the $z=\bar z$ line, $g_{[h_1h_2h_3h_4]}^+(\Re z, \Re z)$ where the $h_i$ are given in~(\ref{eq:conservedstructures}). Even though the structures $\<h_1 h_2 h_3 h_4\>^+_z$ do not lie in the 22-dimensional subspace of $O(2)$ singlets, we can choose the coefficients of other structures to cancel the non-$O(2)$-invariant parts.  The projection of the 5 bulk structures onto the $O(2)$-invariant subspace at $\Im z = 0$ is five-dimensional. Thus, there are exactly $22-5=17$ remaining one-variable degrees of freedom.

Finally, the constraints (\ref{eq:initialconstraint}) give $8$ independent first-order equations that these univariate functions must satisfy. Thus, in addition to 5 two-variable degrees of freedom, we have 9 one-variable degrees of freedom and 8 integration constants. We are free to choose these however we like, as long as the projection of the corresponding structures to the $O(2)$-invariant subspace is 22-dimensional.

\subsubsection{Summary and crossing equations}

Altogether, we choose the following functions as our undetermined degrees of freedom.

\begin{itemize}
\item Two-variable degrees of freedom:
\begin{align}
\label{eq:twovardegs}
g^+_{[2222]}(z,\bar z), \quad &g^+_{[1111]}(z,\bar z), \quad g^+_{[1212]}(z,\bar z),\nn\\
g^+_{[1122]}(z,\bar z), \quad &g^+_{[2112]}(z,\bar z). 
\end{align}
\item One-variable degrees of freedom:
\begin{align}
\label{eq:onevardegs}
g^+_{[0000]}(z),\quad &g^+_{[0101]}(z),\quad g^+_{[0202]}(z),\nn\\
g^+_{[0112]}(z),\quad &g^+_{[1012]}(z),\nn\\
g^+_{[0011]}(z),\quad &g^+_{[1001]}(z),\nn\\
g^+_{[0,0,-1,1]}(z),\quad &g^+_{[-1,0,0,1]}(z).
\end{align}
\item Integration constants:
\begin{align}
\label{eq:zerovardegs}
g^+_{[0022]}(1/2),\quad &g^+_{[2002]}(1/2),\nn\\
g^+_{[0,1,-1,2]}(1/2),\quad &g^+_{[-1,1,0,2]}(1/2),\nn\\
g^+_{[0,-1,1,2]}(1/2),\quad &g^+_{[1,-1,0,2]}(1/2),\nn\\
g^+_{[1,-1,-1,1]}(1/2),\quad &g^+_{[-1,-1,1,1]}(1/2).
\end{align}
\end{itemize}

The statement of crossing symmetry is simply
\be
g_{[h_1h_2h_3h_4]}^+(z,\bar z) &=& g_{[h_3h_2h_1h_4]}^+(1-z,1-\bar z).
\ee
We have chosen the set of helicities in our 
independent degrees of freedom (\ref{eq:twovardegs}), (\ref{eq:onevardegs}), and (\ref{eq:zerovardegs}) to be invariant under $h_1 \leftrightarrow h_3$. Thus, crossing symmetry becomes a constraint on these degrees of freedom alone.

As usual, we Taylor-expand the crossing equations around $z=\bar z$ to obtain the following system, parametrized by $n\leq \bar n$, $n+\bar n\leq \Lambda$.
\begin{itemize}
	\item Two-variable equations:
	\be\label{eq:twovareqns}
	\partial_z^n\partial_{\bar z}^{\bar n}g^+_{[2222]}(1/2,1/2)&=&0, \qquad(n+\bar n\text{ odd}),\nn\\
	\partial_z^n\partial_{\bar z}^{\bar n}g^+_{[1111]}(1/2,1/2)&=&0, \qquad(n+\bar n\text{ odd}),\nn\\
	\partial_z^n\partial_{\bar z}^{\bar n}g^+_{[1212]}(1/2,1/2)&=&0, \qquad(n+\bar n\text{ odd}),\nn\\
	\partial_z^n\partial_{\bar z}^{\bar n}g^+_{[1122]}(1/2,1/2)&=&(-)^{n+\bar n}\partial_z^n\partial_{\bar z}^{\bar n}g^+_{[2112]}(1/2,1/2).
	\ee
	\item One-variable equations
	\be\label{eq:onevareqns}
	\partial_z^n g^+_{[0000]}(1/2)&=&0,\qquad (n\text{ odd}),\nn\\
	\partial_z^n g^+_{[0101]}(1/2)&=&0,\qquad (n\text{ odd}),\nn\\
	\partial_z^n g^+_{[0202]}(1/2)&=&0,\qquad (n\text{ odd}),\nn\\
	\partial_z^n g^+_{[0112]}(1/2)&=&(-)^n\partial_z^n g^+_{[1102]}(1/2),\nn\\
	\partial_z^n g^+_{[0011]}(1/2)&=&(-)^n\partial_z^n g^+_{[1001]}(1/2),\nn\\
	\partial_z^n g^+_{[0,0,-1,1]}(1/2)&=&(-)^n\partial_z^n g^+_{[-1,0,0,1]}(1/2).
	\ee
	\item Integration constants
	\be\label{eq:zerovareqns}
	g^+_{[0022]}(1/2)&=&g^+_{[2002]}(1/2),\nn\\
	g^+_{[0,1,-1,2]}(1/2)&=&g^+_{[-1,1,0,2]}(1/2),\nn\\
	g^+_{[0,-1,1,2]}(1/2)&=&g^+_{[1,-1,0,2]}(1/2),\nn\\
	g^+_{[1,-1,-1,1]}(1/2)&=&g^+_{[-1,-1,1,1]}(1/2).
	\ee
\end{itemize}
Note that the analysis of the conservation constraints was necessary to make sure that the crossing equations we write are independent. We have explicitly verified that this indeed is the case by Taylor expanding to some finite order $\Lambda$ and checking that, modulo the conservation equation, the full set of crossing equations is indeed equivalent to~\eqref{eq:twovareqns}-\eqref{eq:zerovareqns} and that there are no linear dependencies among the equations~\eqref{eq:twovareqns}-\eqref{eq:zerovareqns}. 

\section{Conformal blocks}
\label{sec:conformalblocks}

We compute the conformal blocks for $\<TTTT\>$ using the approach of~\cite{Costa:2011dw}. In this approach, the conformal blocks for external operators with large spins are obtained by acting with differential operators on simpler conformal blocks, known as seed blocks, exchanging the same intermediate representation. Since in our case we only need the conformal blocks for the exchange of traceless symmetric operators, we can take the scalar blocks as our seeds. This is exactly the case studied in~\cite{Costa:2011dw}.

Consider the contribution of a single primary state $|\cO^\alpha\>$ and its descendants $P^{\{A\}}|\cO^\alpha\>$ to the 4-point function,
\be\label{eq:Ocontribution}
\sum_{\{A\},\{B\}}\<T(w_4,x_4)T(w_3,x_3)P^{\{B\}}|\cO^\beta\>Q_{\beta\{B\},\alpha\{A\}}\<\cO^\alpha|K^{\{A\}}T(w_2,x_2)T(w_1,x_1)\>.
\ee
Here $\alpha$ and $\beta$ are indices in the $SO(3)$ irrep of $\cO$, $\{A\}$ and $\{B\}$ are multi-indices such that
\be
 P^{\{A\}}=P^{A_1}\cdots P^{A_n},
\ee
and $Q_{\alpha\{A\},\beta\{B\}}$ is the matrix inverse to $\<\cO^\beta|K^{\{B\}}P^{\{A\}}|\cO^\alpha\>$. The inner products in~\eqref{eq:Ocontribution} are derivatives of the 3-point functions
\be
\<\cO^\beta|T(w_2,x_2)T(w_1,x_1)\>&=&\lambda_{TT\cO}^{(a)}\<\cO^\beta|T(w_2,x_2)T(w_1,x_1)\>_{(a)},\\
\<T(w_4,x_4)T(w_3,x_3)|\cO^\alpha\>&=&\p{\lambda_{TT\cO}^{(a)}}^*{}_{(a)}\<T(w_4,x_4)T(w_3,x_3)|\cO^\alpha\>,
\ee
where $\lambda$ are the OPE coefficients and the objects multiplying them are the tensor structures. We choose our tensor structures so that the OPE coefficients $\lambda_{TT\cO}$ are real. The sum over contributions~\eqref{eq:Ocontribution} can be then written as
\be
\<T(w_4,x_4)T(w_3,x_3)T(w_2,x_2)T(w_1,x_1)\>=\sum_{\cO}\lambda^{(a)}_{TT\cO}\lambda^{(b)}_{TT\cO}G_{\cO,ab}(w_i,x_i),
\ee
where we defined the conformal block
\begin{multline}
G_{\cO,ab}(w_i,x_i)\equiv\\\sum_{\{A\},\{B\}}{}_{(b)}\<T(w_4,x_4)T(w_3,x_3)P^{\{B\}}|\cO^\beta\>Q_{\beta\{B\},\alpha\{A\}}\<\cO^\alpha|K^{\{A\}}T(w_2,x_2)T(w_1,x_1)\>_{(a)}.
\end{multline}
Note that if $\cO$ is parity-even then both $a$ and $b$ should correspond to parity-even structures, and if $\cO$ is parity-odd then both $a$ and $b$ should correspond to parity-odd structures. The corresponding conformal blocks will have different properties in what follows, and we hence refer to these cases as even-even and odd-odd respectively.

The main observation in~\cite{Costa:2011dw} was that one can find conformally-invariant differential operators $\cD^{(a)}_{ij}(w_i,w_j)$ acting on a pair of points such that\footnote{The existence of the $\cD_{ij}^{(a)}$ can be understood in terms of ``weight-shifting operators" \cite{Karateev:2017jgd}.}
\be\label{eq:diffbasisdefn}
\<\cO^\alpha|T(w_2,x_2)T(w_1,x_1)\>_{(a)}&=&\cD^{(a)}_{12}(w_1,w_2)\<\cO^\alpha|\phi_2(x_2)\phi_1(x_1)\>,\nn\\
{}_{(b)}\<T(w_4,x_4)T(w_3,x_3)|\cO^\beta\>&=&\cD^{(b)}_{34}(w_3,w_4)\<\phi_4(x_4)\phi_3(x_3)|\cO^\beta\>.
\ee
Here in the right-hand side the operators act on some standard scalar 3-point functions,\footnote{Of course, this relation is purely kinematical (i.e between tensor structures), and the operators $\phi_i$ do not actually exist in the physical theory.}
which we choose to be, in the formalism of~\cite{Costa:2011mg},
\be
	\<\phi_1\phi_2\cO_3\>\equiv \frac{V_3^{\ell_3}}{
		X_{12}^{\frac{\De_1+\De_2-\De_3-\ell_3}{2}}
		X_{23}^{\frac{\De_2+\De_3-\De_1+\ell_3}{2}}
		X_{31}^{\frac{\De_3+\De_1-\De_2+\ell_3}{2}}
	},\quad X_{ij}=-2X_i\cdot X_j.
\ee

Conformal invariance of these differential operators means that the same relations~\eqref{eq:diffbasisdefn} hold even if we insert $P^{\{B\}}$ or $K^{\{A\}}$ in these 3-point functions. We thus find
\be\label{eq:spinseedrelation}
	G_{a,b}(w_i,x_i)=\cD^{(a)}_{12}(w_1,w_2)\cD^{(b)}_{34}(w_3,w_4)G_\mathrm{scalar}(x_i),
\ee
where the scalar block is given by
\be
G_\mathrm{scalar}(w_i,x_i)=\sum_{\{A\},\{B\}}\<\phi_4(x_4)\phi_3(x_3)P^{\{B\}}|\cO^\beta\>Q_{\beta\{B\},\alpha\{A\}}\<\cO^\alpha|K^{\{A\}}\phi_2(x_2)\phi_1(x_1)\>.
\ee
The problem of calculating conformal blocks then reduces to three subproblems:
\begin{enumerate}
	\item Construction of the conformally-invariant differential operators $\cD^{(a)}_{ij}$ which satisfy~\eqref{eq:diffbasisdefn}.
	\item Computation of the scalar conformal blocks $G_\mathrm{scalar}$.
	\item Performing the differentiation in the right-hand side of~\eqref{eq:spinseedrelation}.
\end{enumerate}

\subsection{Differential basis}

Construction of the differential operators $\cD_{ij}^{(a)}$ has been discussed in~\cite{Costa:2011dw}. Let us first consider the operators $\cD^{(a)}_{12}$ and restrict ourselves to parity-even structures. They are constructed as products of the basic operators
\be\label{eq:basicoperators}
	D_{11},\, D_{12},\, D_{21},\, D_{22},\, H_{12},
\ee
where the first order operators $D_{ij}$ increase spin at position $i$ by 1 while decreasing the scaling dimension at position $j$ by 1. The operator $H_{12}$ is just multiplication by the structure $H_{12}$ and it increases the spin and the scaling dimension by 1 at both positions. These operators do not commute, but their algebra closes, so that one can consider the following general ansatz,
\be\label{eq:diffopansatz}
	\cD^{(a)}_{12}=\sum_{n_{ij},m_k}c_{n_{12},n_{23},n_{13},m_1,m_2}^{(a)} H_{12}^{n_{12}}D_{12}^{n_{13}}D_{21}^{n_{23}}D_{11}^{m_1}D_{22}^{m_2}\Sigma_{1}^{n_{12}+n_{23}+m_1}\Sigma_{2}^{n_{12}+n_{13}+m_2},
\ee
where the parameters in the sum are constrained so that the resulting operator increases spin by 2 at both points. Here $\Sigma_i$ is a formal operator which increases the scaling dimension at position $i$ by $1$. This is needed because various terms in the sum change the scaling dimensions by different amounts. Accordingly,~\eqref{eq:spinseedrelation} should actually contain several types of scalar blocks differing by the scaling dimensions of the external operators. We will return to this issue when we discuss the calculation of these scalar blocks.

One can check that the differential basis ansatz~\eqref{eq:diffopansatz} contains 14 different operators. This is the same as the number of \emph{algebraic} (not yet conserved or symmetric) tensor structures for $\<TT\cO_\ell\>$ one can build out of $H_{ij}$ and $V_i$ for $\ell\geq 4$. We can therefore find a change of basis between the algebraic and differential bases.

We can then easily formulate the conservation and the permutation symmetry constraints for $\<TT\cO_\ell\>$ in the algebraic basis and then translate these constraints to the differential basis. This results in a system of linear equations for the coefficients $c$,
\be\label{eq:diffbasisconstraints}
	\sum_{n_{ij},m_k}M^{\alpha}_{n_{ij},m_k}(\De)c^{(a)}_{n_{ij},m_k}=0.
\ee
The coefficients in this equation are rational functions of the dimension $\De$ of the exchanged primary $\cO$, and thus the solutions are rational functions of $\De$ as well. Consistently with the discussion in section~\ref{sec:threeptstructures}, we find that there exist 2 solutions for even $\ell\geq 4$.
To simplify the numerical evaluation of~\eqref{eq:spinseedrelation}, we choose a basis of the solutions $c^{(a)}_{n_{ij},m_k}$ which is polynomial in $\De$ of the lowest possible degree. These degrees are $6$ and $4$ for the two solutions.

In the above discussion we have glossed over a slight subtlety that in the algebraic basis in 3d, there is one tensor structure~\eqref{eq:redundant_tensor_structure} which is redundant and can be expressed in terms of other structures, so the number of independent structures is actually $13$. There is also a corresponding relation in the differential basis. If we were to ignore this relation, we would find more solutions to the conservation constraints. Taking it into account, we can use it to simplify the form of the solutions $c^{(a)}_{n_{ij},m_k}$.

A similar procedure works for $\ell\leq 4$, the only difference being that there appear new relations in the differential basis (while the algebraic basis simply becomes smaller). These relations are easily controlled by the transformation matrix which expresses the differential basis structures in terms of the algebraic ones. We then use these relations to find the simplest form of the non-redundant solutions of~\eqref{eq:diffbasisconstraints}. 

The parity-odd structures can be treated in a similar way, except that we generally find more redundancies than in the parity-even case. We describe the construction of parity-odd differential basis in appendix~\ref{app:tensor}, together with the explicit expressions for the coefficients $c^{(a)}_{n_{ij},m_k}$. In both the parity-even and the parity-odd cases the operators $\cD^{(a)}_{34}$ can be obtained by applying a simple permutation to the operators $\cD^{(a)}_{12}$. 

\subsection{Computing the scalar blocks}

Since~\eqref{eq:diffopansatz} involves the formal dimension-shifting operators $\Sigma_{1,2}$, there are several scalar conformal blocks entering~\eqref{eq:spinseedrelation}, which differ by the dimensions $\De_i$ of the external scalars. 

Let us analyze the dimensions of the scalar at positions 1 and 2. The exponents in~\eqref{eq:diffopansatz} are constrained by the spins of the stress tensors
\be
	n_{12}+n_{13}+m_1=n_{12}+n_{23}+m_2=2.
\ee
On the other hand, the dimensions of the scalar operators in each term are given by
\be
	\De_1&=\De_T+n_{12}+n_{23}+m_1,\\
	\De_2&=\De_T+n_{12}+n_{13}+m_2.
\ee
It follows that the sum
\be
\De_1+\De_2=2\De_T+4=10
\ee
is the same for all the terms. On the other hand, the difference is
\be
\De_{12}=\De_1-\De_2=n_{23}-n_{13}+m_1-m_2=2(m_1-m_2),
\ee
and one can see that it takes all even values $-4\leq \De_{12}\leq 4$. The same is true for $\De_{34}$.

The analysis for parity-odd operators is similar, with the result that $\De_1+\De_2=9$, while $\De_{12}$ assumes all odd values $-3\leq \De_{12}\leq 3$. The same is true for $\De_{34}$.

Note that the scalar blocks essentially depend only on the differences $\De_{12}$ and $\De_{34}$. Furthermore, there is a $\Z_2\times\Z_2$ group of permutations of the external operators which preserves the OPE $s$-channel and the cross-ratios,\footnote{Of course, we can also use the permutations which change the cross-ratios, but in practice it is easier to have all scalar blocks with the same arguments.} and thus acts in a simple way on the conformal blocks. The elements of this group change the scaling dimensions of the scalar blocks according to
\be
	(12)(34):&\quad \De_{12}\to-\De_{12},\,\De_{34}\to-\De_{34},\\
	(13)(24):&\quad \De_{12}\leftrightarrow\De_{34},\\
	(14)(23):&\quad \De_{12}\leftrightarrow-\De_{34}.
\ee
We thus only need to compute the scalar blocks with $\De_{12}$ and $\De_{34}$ in a fundamental domain for these transformations, and then all the other blocks can be easily inferred. It is easy to check that a fundamental domain is given by
\be\label{eq:fundamentaldims}
	\De_{12}\geq |\De_{34}|.
\ee
\begin{figure}[ht!]
	\begin{center}
		\includegraphics[width=0.7\textwidth]{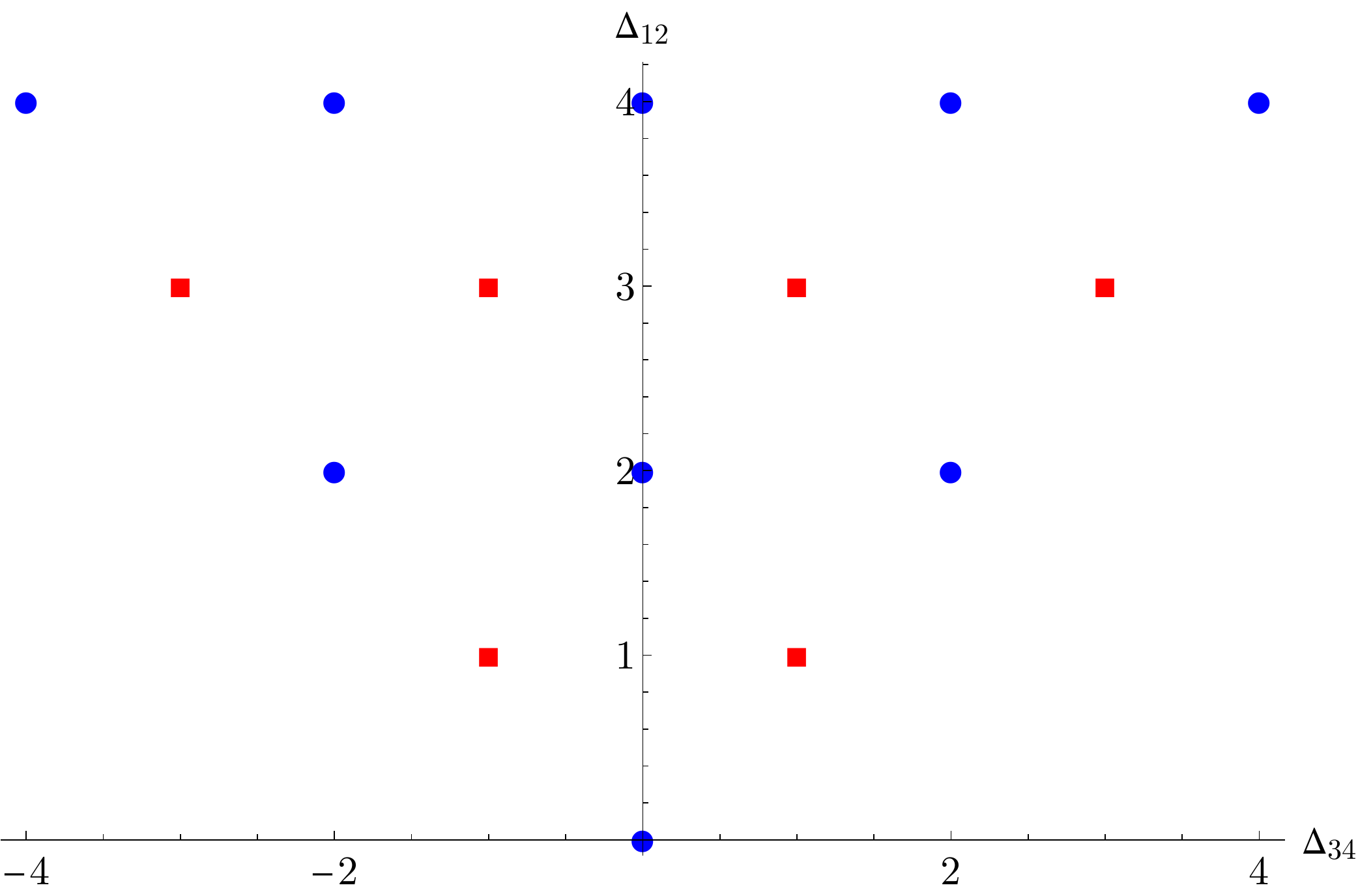}
	\end{center}
	\caption{Parameters of scalar conformal blocks for the even-even (blue dots) and odd-odd (red squares) cases.}\label{fig:scalarblockdims}
\end{figure}
The resulting fundamental set of the parameters $\De_{12},\,\De_{34}$ for the scalar blocks is shown in figure~\ref{fig:scalarblockdims}. There are $9$ scalar blocks required for the computation of even-even $\<TTTT\>$ blocks, and $6$ scalar blocks required for the computation of odd-odd $\<TTTT\>$ blocks.\footnote{Note that by using the dimension-shifting differential operators~\cite{Dolan:2011dv,Karateev:2017jgd} we can reduce this set to just one scalar conformal block for each parity.} In practice we compute them efficiently using the pole expansion of~\cite{Kos:2014bka,Penedones:2015aga} evaluated on the diagonal $z=\bar{z}$ combined with the recursion relation implied by the Casimir equation to evaluate scalar block derivatives away from the diagonal.

\subsection{Applying the differential operators}

To finish the calculation of the stress-tensor conformal blocks, it is necessary to apply the differential operators $\cD^{(a)}_{ij}$ to the scalar blocks. The embedding-space definition of these operators, given in~\cite{Costa:2011dw}, seems inadequate for this purpose because the embedding-space 4-point tensor structures in 3d contain many degeneracies. Therefore, it is convenient to reformulate these operators directly in the conformal frame basis constructed in section~\ref{sec:fourptstructures}.

The first step is to convert the embedding-space expression for the differential operators to explicit expressions in 3 dimensions. For this purpose, we consider an explicit uplift of 3 dimensional primary operators to embedding space operators,
\be
	\cO(Z,X)=\frac{1}{(X^+)^{\De}}\cO\p{Z^\mu-Z^{+}\frac{X^\mu}{X^+},\frac{X^\mu}{X^+}},
\ee
where on the right-hand side we have the 3d operator $\cO(w,x)$. Applying embedding-space differential operators to this expression, we reproduce on the right-hand side the corresponding differential operators in 3 dimensions. Choosing a different uplift will yield the same result due to the consistency conditions imposed on the embedding space differential operators.

With the 3-dimensional expressions at hand, we can understand the action of the differential operators in the conformal frame. In the conformal frame, some of the operators are placed at fixed positions. In order to apply derivatives in these constrained directions, we simply solve the equations
\be
	\sum_{k=1}^4 L_{k\,AB}\<TTTT\>=0
\ee
for these derivatives. Here $L_k$ are the conformal generators acting on point $k$. As a result, we can write for any 3d differential operator $D$
\be
	D \p{[h_1h_2h_3h_4]g_{[h_1h_2h_3h_4]}(z,\bar z)}=\sum_{h'_i}[h_1'h_2'h_3'h_4']D^{[h_1'h_2'h_3'h_4']}_{[h_1h_2h_3h_4]}g_{[h_1h_2h_3h_4]}(z,\bar z),
\ee
where $D_{[h_i]}^{[h'_i]}$ are differential operators in $z$ and $\bar z$. In this equation, we can keep the spins $\ell_i$ and the parameters $h_i$ as variables, in which case $h'_i$ differ from $h_i$ by finite shifts. Using in place of $D$ the basic differential operators~\eqref{eq:basicoperators} and their parity-odd analogs, we obtain their counterparts in the conformal frame.

This allows us to efficiently compute the more complicated compositions~\eqref{eq:diffopansatz} directly in conformal frame without encountering any redundancies in tensor structures in intermediate steps. In the end, we find expressions for the $\<TTTT\>$ blocks of the form
\be
	\p{G_{\De,\ell,ab}}_{[h_1h_2h_3h_4]}(z,\bar z)=\sum_{i=1}^{N_\text{scalar}}\sum_{m,n}a^{i,mn,ab}_{[h_1h_2h_3h_4]}(\De,\ell,z,\bar z)\ptl_z^m\ptl_{\bar z}^n G_{\De,\ell}^{\De_{12}^{(i)},\De_{34}^{(i)}}(z,\bar z),
\ee
where $a$ are some rational functions of $z,\bar z,\ell$, and polynomial in $\De$,\footnote{Because of our polynomial choice of the solutions $c_{n_{ij},m_k}^{(a)}$ to~\eqref{eq:diffbasisconstraints}.} while $\De_{12}^{(i)}$ and $\De_{34}^{(i)}$ are the parameters of the scalar conformal blocks from the fundamental region~\eqref{eq:fundamentaldims}. The derivative order is $m+n\leq 8$ for even-even blocks and $m+n\leq 10$ for odd-odd blocks; $N_\text{scalar}$ is $9$ and $6$ respectively.

The functions $a$ contain powers of $(z-\bar z)$ in their denominators, but these get canceled when one takes into account that the scalar blocks are symmetric under $z\leftrightarrow \bar z$. For example, if we rewrite the above expression in coordinates $z+\bar z$ and $(z-\bar z)^2$, then the functions $a$ manifestly have only the OPE singularities. This is to be expected, since the functions entering the decomposition~\eqref{eq:decompositionintostructures} must have the same singularities as the physical correlator. Therefore, we can take further derivatives directly in this expression, and then evaluate it at $z=\bar z=1/2$ to find the derivatives of $\<TTTT\>$ blocks in terms of linear combinations of the derivatives of scalar blocks with coefficients polynomial in $\De$. Substituting rational approximations for the derivatives of the scalar blocks then immediately yields rational approximations for $\<TTTT\>$ blocks suitable for use in \texttt{SDPB}~\cite{Simmons-Duffin:2015qma}.

\section{Numerical bounds}

In this section we discuss how to use the crossing equations and conformal blocks derived in the previous sections to compute numerical bounds on the OPE coefficients and scaling dimensions appearing in the $T \times T$ OPE. Further details of our numerical implementation are given in appendix~\ref{app:numerics}.

\subsection{Initial comments: $C_T$ and $\theta$}

To begin, let us return to the conformal block decomposition of the stress-tensor 4-point function in a general 3d CFT,
\be
	\<TTTT\> = \lambda_{TT\mathbf{1}}^2 G_{\mathbf{1}}+\frac{1}{C_T}\lambda_{TTT}^{(a)}\lambda_{TTT}^{(b)}G_{T,ab}+\sum_{\cO}\lambda^{(a)}_{TT\cO}\lambda^{(b)}_{TT\cO}G_{\cO,ab},
\ee
where we have explicitly separated the contribution of the identity operator and the stress tensor itself. We have also assumed that the CFT in question possesses a unique stress tensor. The factor $\frac{1}{C_T}$ comes from the fact that $C_T$ enters the 2-point function of the canonically-normalized stress tensor $T$.

The OPE coefficient $\lambda_{TT\mathbf{1}}$ of the identity operator is just the coefficient in the 2-point function $\<TT\>$, and thus is essentially the central charge $C_T$. At the same time, the OPE coefficients for the stress tensor itself are given by $\lambda^{(1)}_{TTT}=n_B$ and $\lambda^{(2)}_{TTT}=n_F$. Due to the Ward identity constraint~\eqref{eq:wardidentity}, these three coefficients are not independent. It is therefore convenient to introduce the following parametrization,\footnote{Another, perhaps more natural, parametrization would be $n_B = C_T \cos^2\theta'$, $n_F = C_T \sin^2\theta'$. However this parametrization doesn't allow us to numerically test negative values of $n_B$ and $n_F$ so we adopt the one in the text in order to probe the conformal collider bounds.}
\be
\label{thetadef}
	n_B&=C_T\frac{\cos\theta}{\sin\theta+\cos\theta},\\
	n_F&=C_T\frac{\sin\theta}{\sin\theta+\cos\theta}.
\ee
Note that $\theta = \tan^{-1}(n_F/n_B)$ is $\pi$-periodic, so we can assume that $\theta\in(-\pi/4,3\pi/4)$, where the denominators are positive. We also renormalize the 4-point function $\<TTTT\>$ so that $C_T$ appears only in one of the terms,
\be\label{eq:TTTTbootstrapNorm}
	C_T^{-2}\<TTTT\>&= G_{\mathbf{1}}+\frac{1}{C_T}\Theta^{ab}G_{T,ab}+\sum_{\cO}\hat\lambda^{(a)}_{TT\cO}\hat\lambda^{(b)}_{TT\cO}G_{\cO,ab}\nn\\
	&=G_{\mathbf{1}}+\frac{1}{C_T}\Theta^{ab}G_{T,ab}+\sum_{\De,\rho}M^{ab}_{\De,\rho} G_{\De,\rho,ab},
\ee
where $\hat\lambda_{TT\cO}^{(a)}=C_T^{-2}\lambda_{TT\cO}^{(a)}$ and the positive-semidefinite matrix $\Theta^{ab}$ is given by
\be\label{eq:Thetadefn}
	\Theta=\frac{1}{(\sin\theta+\cos\theta)^2}\begin{pmatrix}
		\cos^2\theta & \cos\theta\sin\theta\\
		\cos\theta\sin\theta & \sin^2\theta
	\end{pmatrix}.
\ee
We have also defined the positive-semidefinite OPE matrix $M^{ab}_{\De,\rho}$ to be the sum of $\hat\lambda^{(a)}_{TT\cO}\hat\lambda^{(b)}_{TT\cO}$ over the operators $\cO$ with scaling dimension $\De$ and in the $O(3)$ representation $\rho$. Of course, the operators appearing in the $T\times T$ OPE are singlets of global symmetries and we generically do not expect there to be any degeneracies. Therefore, we expect that all matrices $M_{\De,\rho}$ have rank $1$. However, without additional assumptions the operators are allowed to have arbitrarily close scaling dimensions, which is numerically indistinguishable from a degeneracy in the spectrum. In other words, even if we had a way of constraining all $M_{\De,\rho}$ to have rank $1$, numerically this would make no difference unless we also input assumptions about gaps between operators. The stress-tensor four-point function written in the form~\eqref{eq:TTTTbootstrapNorm} is suitable for numerical analysis using the standard methods which we review in appendix~\ref{app:numerics}. Here, let us make some initial comments about our assumptions and on the kind of bounds we can expect to find.

Note that $C_T^{-1}\Theta$ is essentially a special case of the OPE matrices $M_{\De,\rho}$. We only consider the theories with a unique  spin-$\mathbf{2}^+$ conserved operator, and this is reflected in the fact that we explicitly assume $\Theta$ to have rank 1 by writing~\eqref{eq:Thetadefn}. Unlike in the case of generic $M_{\De,\rho}$, this constraint matters. Indeed, parity-even spin-2 operators \textit{strictly above} the unitarity bound only have a single OPE coefficient and thus are clearly distinguishable from $T$ even if their scaling dimension is arbitrarily close to $3$. It is therefore more appropriate to think about $T$ as an isolated operator.\footnote{Although not completely appropriate --- there is still a direction in the 3-dimensional space of symmetric matrices $\Theta$ which can be ``altered'' by spin-$\mathbf{2}^+$ operators with $\De=3+\epsilon$. This direction, however, coincides with~\eqref{eq:Thetadefn} only if $\theta\to-\pi/4+\pi k$.}

It is important to note that although this assumption on the form of $\Theta$ is non-trivial, it does not necessarily imply that this CFT has a unique conserved spin-$\mathbf{2}^+$ operator. Indeed, consider a decoupled system of any number $N\geq 2$ of CFTs, all of which satisfy~\eqref{eq:Thetadefn} with the same value of $\theta$. If the stress tensors in these theories are $T_i$, then the stress tensor of the full system is 
\be
	T=\sum_{i=1}^N T_i.
\ee
We also have $C_{T}=\sum_i C_{T_i}$. It is easy to check that $\<TTTT\>$ in this system satisfies~\eqref{eq:TTTTbootstrapNorm} and~\eqref{eq:Thetadefn}, even though each $T_i$ is a distinct conserved spin-$\mathbf{2}^+$ operator. 

This also shows that for any value of $\theta$ which is allowed by the crossing symmetry of~\eqref{eq:TTTTbootstrapNorm} the central charge $C_T$ is unbounded from above -- we can simply take $N$ copies of the same CFT for arbitrarily large $N$. In the limit $N\to \infty$, the corresponding four-point function approaches that of the mean field theory (MFT). The stress-tensor 4-point function in MFT is dual to the 4-point scattering of free spin-2 massless particles in $\mathrm{AdS}_4$ and is given by Wick's theorem,
\be\label{eq:TTTTinMFT}
\<TTTT\>=\<TT\>\<TT\>+\<TT\>\<TT\>+\<TT\>\<TT\>.
\ee
In this theory $C_T$ is formally infinite. In other words, it gives a unitary solution to crossing symmetry for which the second term in~\eqref{eq:TTTTbootstrapNorm} vanishes. In particular, its existence shows that any value of $\theta$ is formally allowed unless one excludes $C_T=\infty$.

From the above discussion it follows that we cannot put upper bounds on $C_T$ or constrain $\theta$ without extra assumptions which go beyond unitarity, parity invariance, crossing symmetry and existence of a unique stress tensor. Importantly, this is not a technical obstruction of the associated semidefinite problem. As we noted, $T$ is effectively an isolated operator and thus there is no a-priori problem with such bounds. The problem is more physical in nature and ultimately due to existence of the MFT. We will repeatedly see that as soon as MFT is excluded by additional assumptions, these bounds become possible.

\subsection{General theories}
\label{sec:generaltheories}

Given that MFT has infinite central charge, we can hope to exclude some values of $\theta$ by assuming that $C_T$ is finite.  One way this can be possible is if there exists a $\theta$-dependent lower bound on $C_T$ which diverges for some values of $\theta$. Of course, numerically we might not reproduce the divergence but instead see a finite bound which grows as we improve our numerical approximation (i.e.\ increase the derivative order $\Lambda$). 

\begin{figure}[ht!]
\begin{center}
\includegraphics[width=0.78\textwidth]{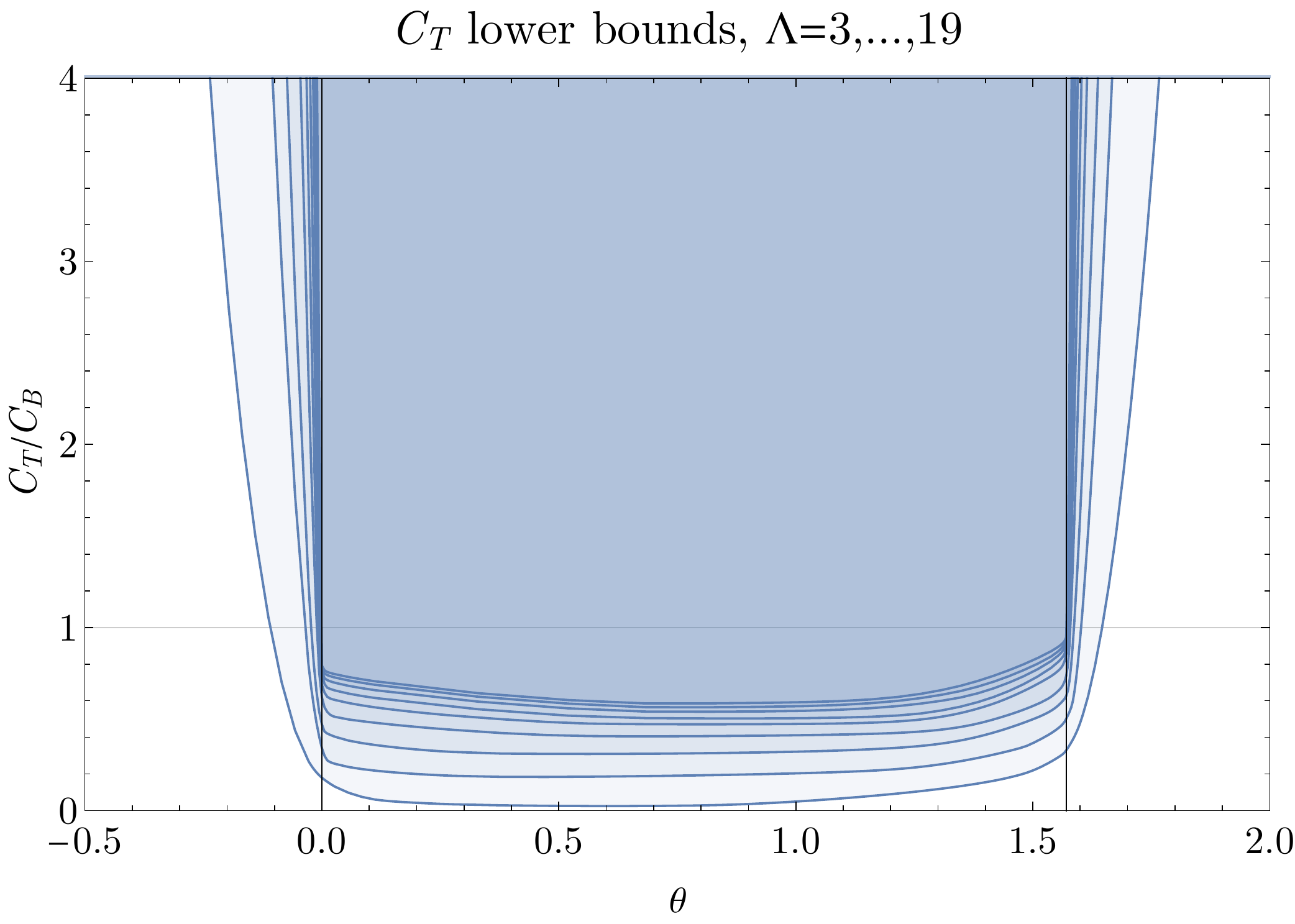}
\end{center}
\caption{A series of lower bounds on $C_T$ as a function of $\theta$, valid in any unitary parity-preserving 3d CFT. The shaded region is allowed.
} \label{fig:hofmanmaldacena}
\end{figure}

This is indeed what happens. In figure~\ref{fig:hofmanmaldacena} we show a series of lower bounds on $C_T$ as a function of $\theta$ for derivative orders $\Lambda=3,\ldots,19$, with no assumptions beyond unitarity, crossing symmetry, parity conservation, and the existence of a unique stress tensor. The behavior of the bound differs dramatically depending on whether $\theta\in [0,\pi/2]$ or not. For $\theta\in [0,\pi/2]$, the bound appears to converge to a finite value. Strikingly, for $\theta<0$ or $\theta>\pi/2$ the bound diverges with growing $\Lambda$.

These numerical results strongly suggest that for unitary parity-preserving theories with finite $C_T$, $\theta$ necessarily lies in the interval $[0,\pi/2]$. Note that $\theta\in[0,\pi/2]$ corresponds to $n_B,n_F\geq 0$, which is equivalent to the conformal collider bounds~\cite{Hofman:2008ar,Buchel:2009sk}. We have thus essentially recovered the stress-tensor conformal collider bounds using the numerical bootstrap.\footnote{Similar conformal collider bounds for OPE coefficients of conserved currents were recovered numerically in \cite{Dymarsky:2017xzb}.} Note that the recent analytical proof \cite{Hofman:2016awc} of the conformal collider bounds uses the lightcone limit of the crossing equation. The analysis of \cite{Simmons-Duffin:2016wlq} suggests that numerical bootstrap techniques at high derivative order can probe the lightcone limit of the crossing equation (despite the fact that the numerical bootstrap usually involves expanding the crossing equation around a Euclidean point). Thus, it is perhaps unsurprising that we make contact with analytical results at large $\L$.

When the conformal collider bounds are saturated ($n_F=0$ or $n_B=0$), the theory is expected to be free \cite{Zhiboedov:2013opa}. Our lower bounds at $\th=0,\pi/2$ are consistent with the existence of the free boson theory ($\th=0$) and the free fermion theory ($\th=\pi$), though they are not yet saturated by those theories. However, the bounds continue to change as we increase the derivative order $\Lambda$. It is possible that at sufficiently large $\L$, our lower bound will become $C_B$ at each endpoint. We do not currently have enough data to perform a reliable extrapolation to $\L=\oo$ (as in, e.g.\ \cite{Beem:2015aoa}).

\subsection{Scalar gaps}
\label{sec:scalargaps}

\subsubsection{Parity-even scalar gaps}

\begin{figure}[ht!]
\begin{center}
\includegraphics[width=0.78\textwidth]{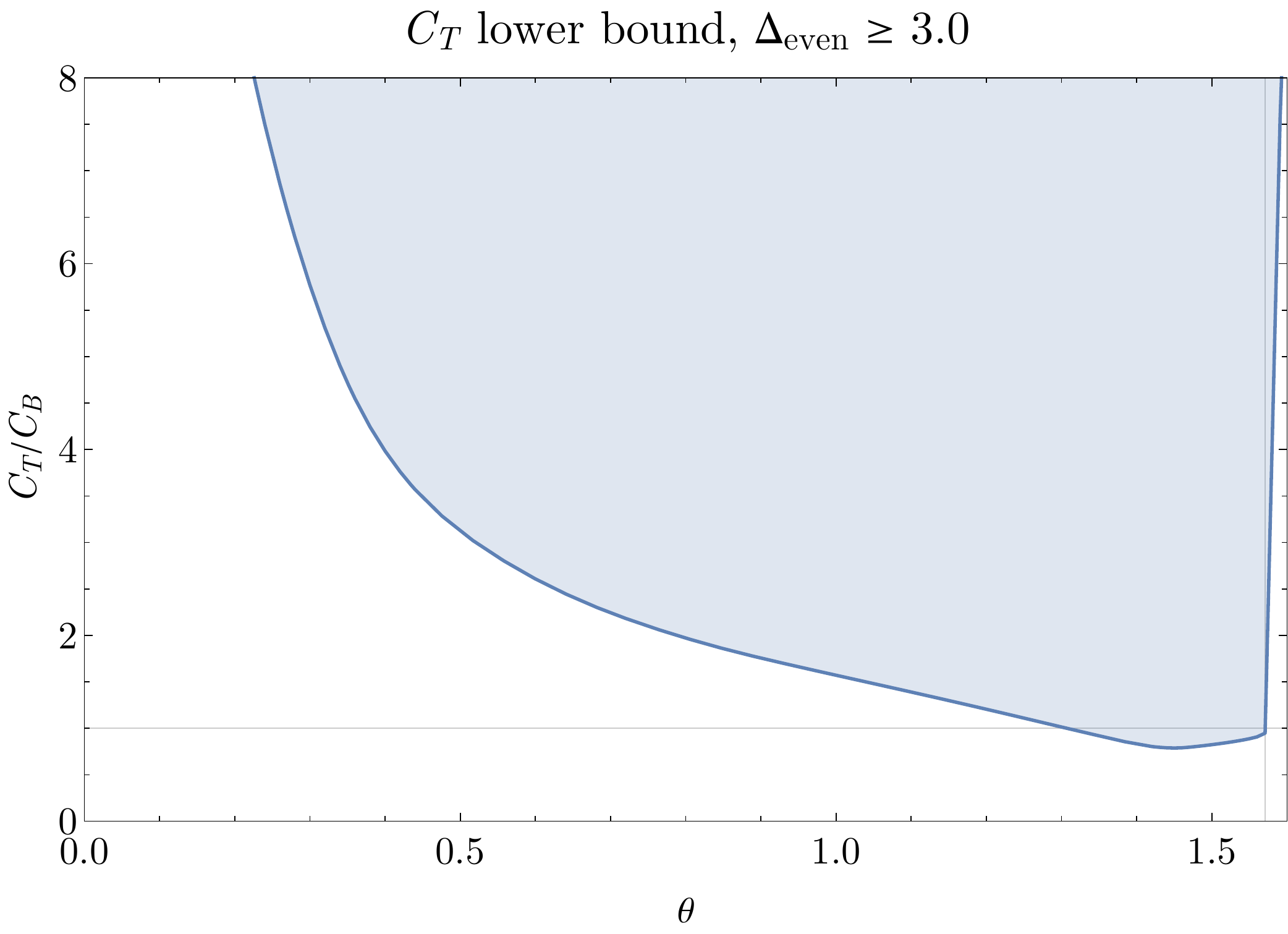}
\end{center}
\caption{A lower bound on $C_T$ as a function of $\th$ in 3d CFTs with no relevant parity-even scalars.}\label{fig:deltaEven3Plot}
\end{figure}

Let us now explore how the bounds on $C_T$ and $\th$ change when we impose further restrictions on the CFT data. It is natural to ask: what is the allowed space of $(\th,C_T)$ in theories with no relevant parity-even scalars --- i.e.\ parity-preserving ``dead-end" CFTs. Denoting the dimension of the lowest-dimension parity-even scalar by $\Delta_{\text{even}}$, we show a bound on theories with $\De_{\text{even}}\geq 3$ in figure~\ref{fig:deltaEven3Plot}. The free fermion at $\th=\pi$ is allowed (the lowest-dimension parity-even singlet in the free-fermion theory is $\psi^2 \ptl_\mu \psi^\a \ptl^\mu \psi_\a$, which has $\De=6$), whereas the free boson is of course excluded. The lower bound on $C_T$ falls quickly as $\th$ varies between $0$ and $\pi$, dipping below $C_B$ only for a small range $\th\in[1.3,\pi]$.  

\begin{figure}[ht!]
\begin{center}
\includegraphics[width=0.78\textwidth]{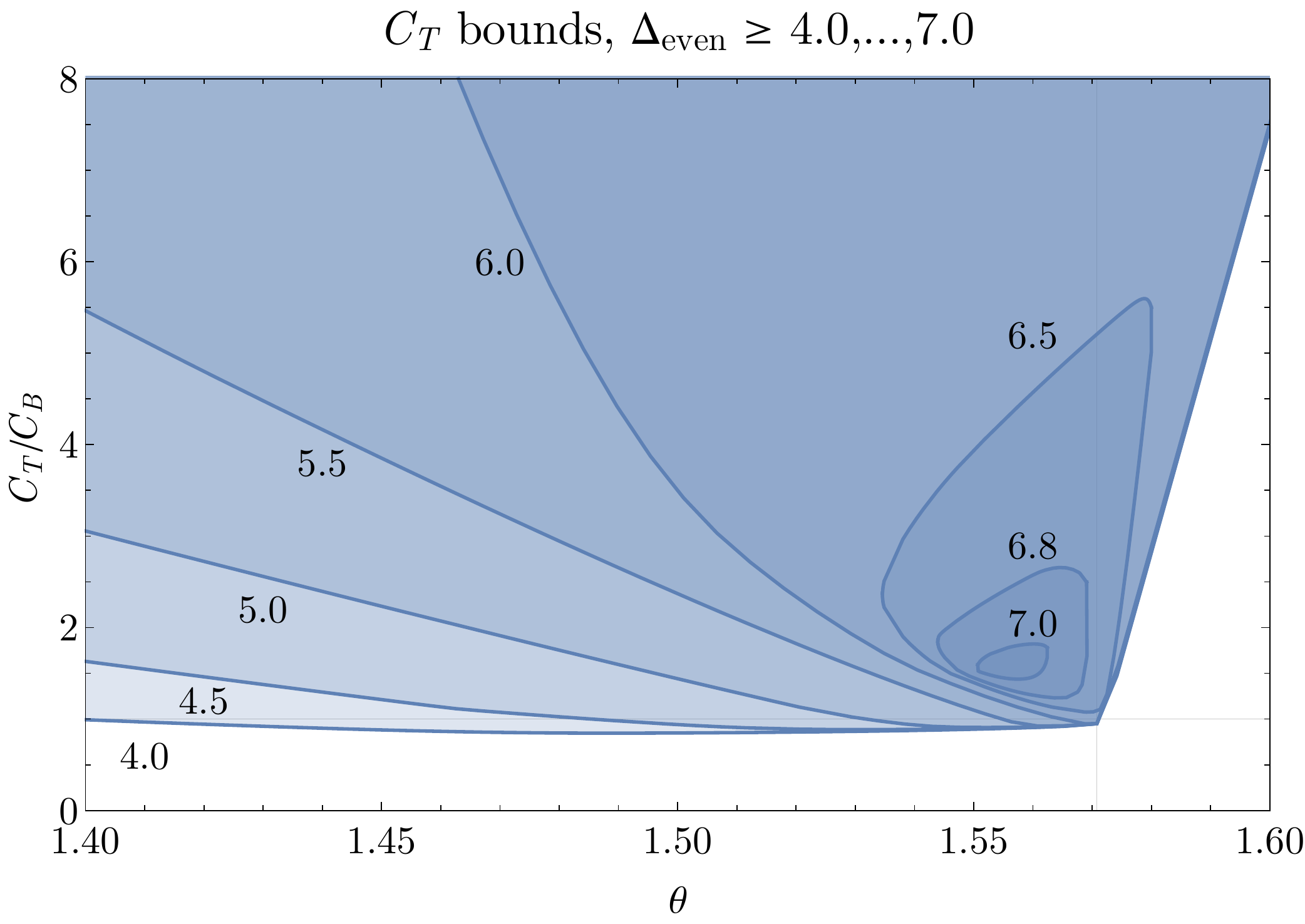}
\end{center}
\caption{Bounds on $(\th,C_T)$ with varying gaps in the parity-even scalar sector. When $\De_{\text{even}}^{\min}=4.0,\dots,6.0$, we have a series of lower bounds on $C_T$ as a function of $\th$.  When $\De_{\text{even}}^{\min}>6.0$, we have closed islands which eventually shrink to zero size.}\label{fig:deltaEvenGapsPlot}
\end{figure}

As we increase the imposed gap in the parity-even scalar sector, $\Delta_{\text{even}} \geq \Delta_{\text{even}}^{\min}$, the lower bounds on $C_T$ get stronger, while still remaining consistent with the existence of the free fermion up to $\De_{\text{even}}^{\min}=6$. We illustrate these bounds in figure~\ref{fig:deltaEvenGapsPlot}.  Note that it is not possible to place upper bounds on $C_T$ when $\De_{\text{even}}^{\min}<6$, because of the existence of MFT, which has $\De_{\text{even}}=6$ (associated with $\cO_{\text{even}}=T_{\mu\nu}T^{\mu\nu}$) and infinite $C_T$.  However, when $\De_{\text{even}}^{\min}>6$, upper bounds become possible, and indeed $C_T$ and $\th$ become confined to a small island in the vicinity of the free fermion point. For example, when $\De_{\text{even}}^{\min}=6.8$, we find $\th\in[1.54,1.57]$ and $C_T/C_B \in[1.2, 2.6]$. It is interesting to ask whether any CFT realizes these values. For even larger values of $\De_{\text{even}}^{\min}$, the allowed region disappears. 

\subsubsection{Parity-odd scalar gaps}

\begin{figure}[ht!]
	\begin{center}
		\includegraphics[width=0.78\textwidth]{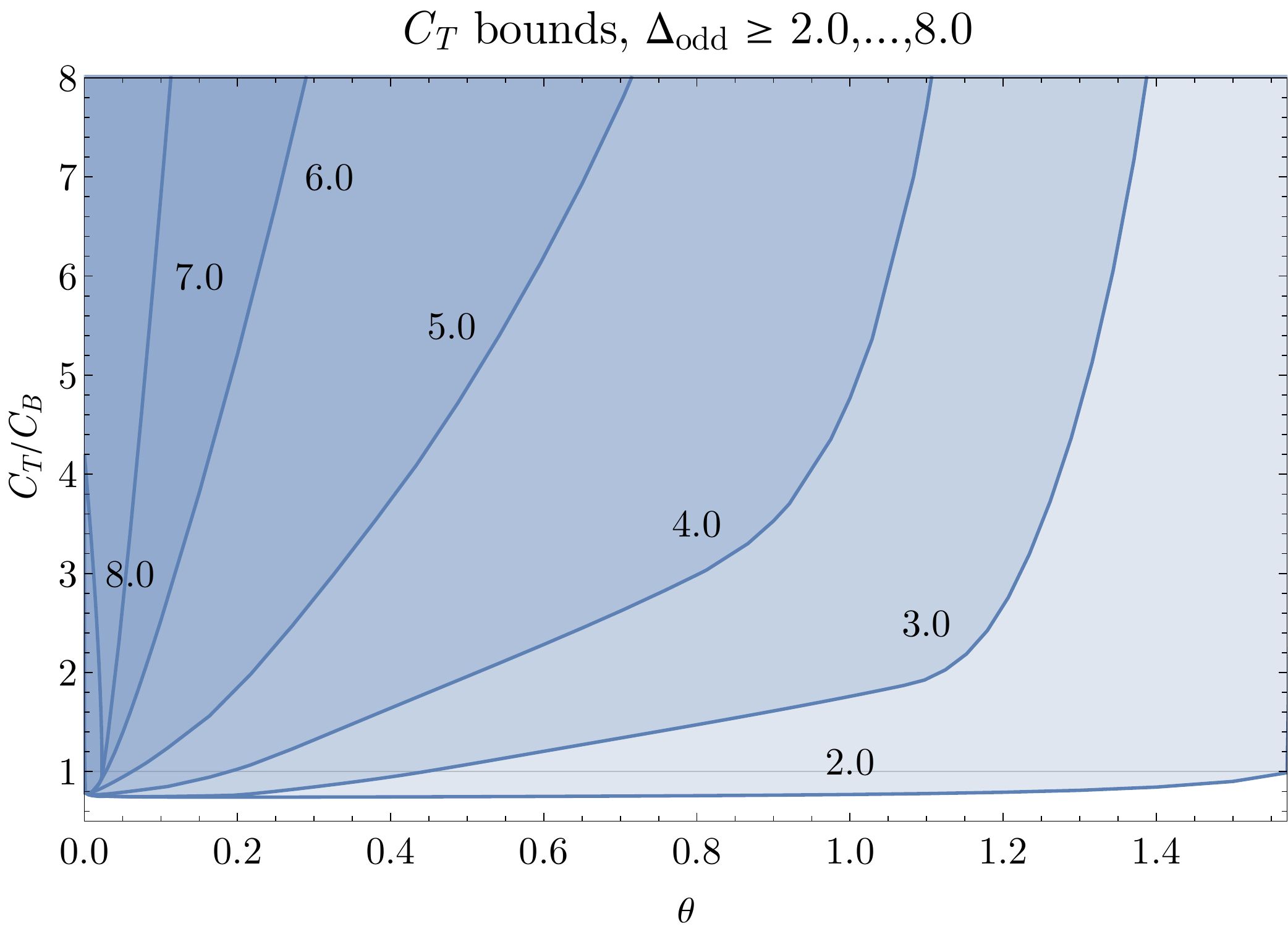}
	\end{center}
	\caption{Bounds on $(\th,C_T)$ with varying gaps in the parity-odd scalar sector. When the value of the gap $\De_{\textrm{odd}}^{\min}>7$, it becomes possible to find both upper and lower bounds on $C_T$ as.}\label{fig:deltaOddGapsLargePlot}
\end{figure}

\begin{figure}[ht!]
	\begin{center}
		\includegraphics[width=0.78\textwidth]{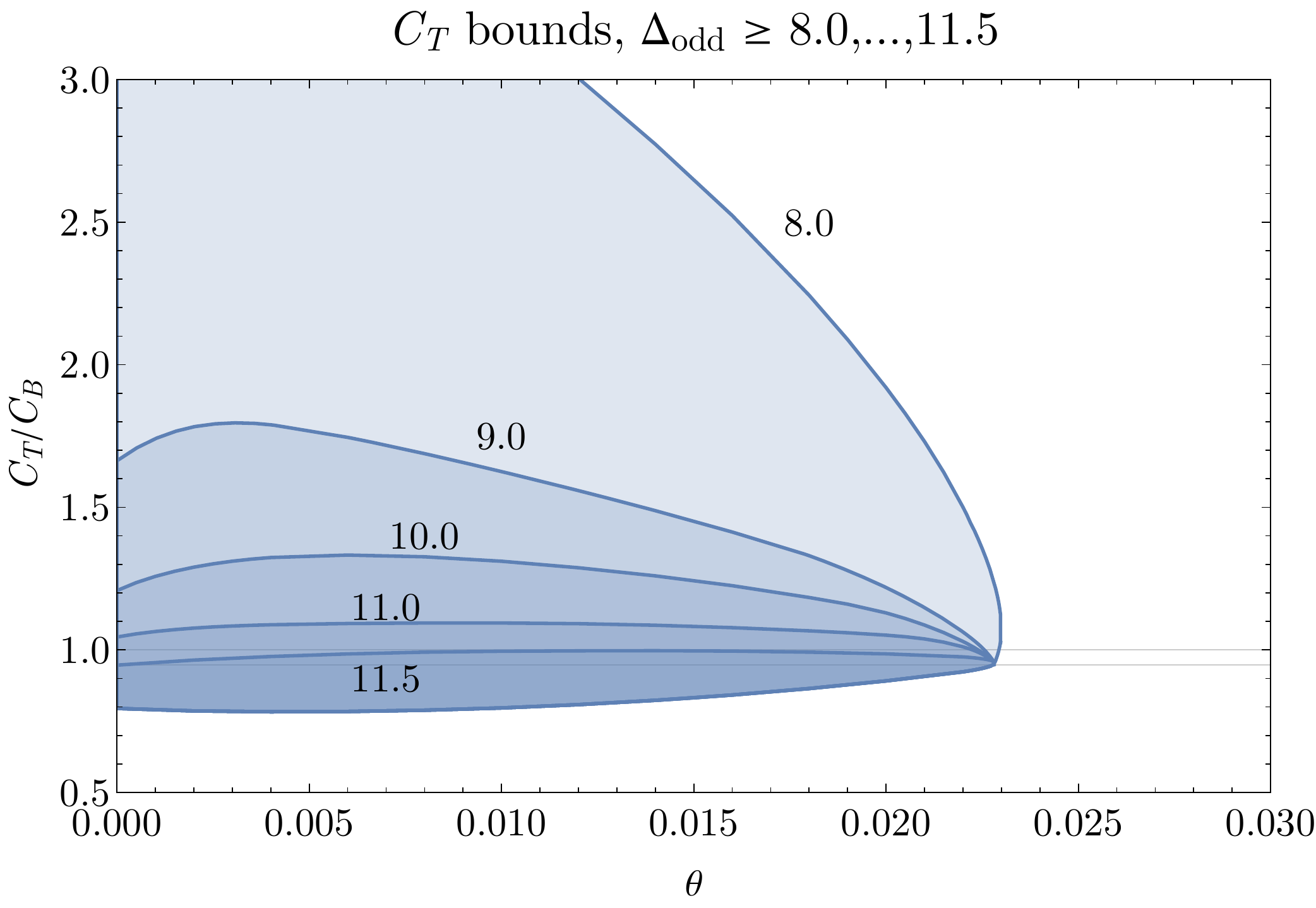}
	\end{center}
	\caption{Closed regions for $(\th,C_T)$, given various large gaps in the parity-odd scalar sector. The lower horizontal line shows the value of $C_T$ in the 3d Ising CFT.}\label{fig:deltaOddGapsSmallPlot}
\end{figure}

Next we study the effect of a gap in the parity-odd scalar operators. In figure~\ref{fig:deltaOddGapsLargePlot}, we show a series of bounds on $C_T$ as a function of $\theta$, for various gaps in the parity-odd scalar sector, $\Delta_{\text{odd}} \geq \Delta_{\text{odd}}^{\min}$.  The bounds are roughly a mirror image of those in the previous subsection. For $\De_{\textrm{odd}}^{\min}=2,\ldots,7$, we find a series of increasingly strong bounds pushing the allowed region towards smaller $\th$. When $\De_\textrm{odd}^{\min}>7$, our assumption excludes MFT (which has $\cO_\textrm{odd}=\e_{\mu\nu\rho} T^{\mu \s} \ptl^\nu T^{\rho}{}_\s$, of dimension $7$), and it becomes possible to find both upper and lower bounds on $C_T$. Indeed, we find a series of islands (figure~\ref{fig:deltaOddGapsSmallPlot}), which finally exclude the free-boson theory when $\De_\textrm{odd}\gtrsim 11$.\footnote{The lightest parity-odd scalar in the theory of a single free boson is the dimension-11 scalar $\epsilon^{\mu\nu\rho}\phi (\ptl_\alpha\ptl_{\beta_1}\ptl_{\beta_2}\ptl_\mu\phi)(\ptl^{\alpha}\ptl_\nu\phi)(\ptl^{\beta_1}\ptl^{\beta_2}\ptl_\rho\phi)+\text{desc}$.} A common corner point of these islands is very close to the $C_T$ value of the 3d Ising CFT. We return to this point in section~\ref{sec:Ising}, where we will see that further imposing known gaps in the 3d Ising CFT slightly reduces this apparent upper bound on $\theta_{\textrm{Ising}}$. 

Finally, note that these bounds imply that any CFT with a large parity-odd gap must have a stress-tensor 3-point function close to the bosonic one, with $\theta < .023$. 

\subsubsection{Scalar gaps in both sectors}

\begin{figure}[ht!]
\begin{center}
\includegraphics[width=0.78\textwidth]{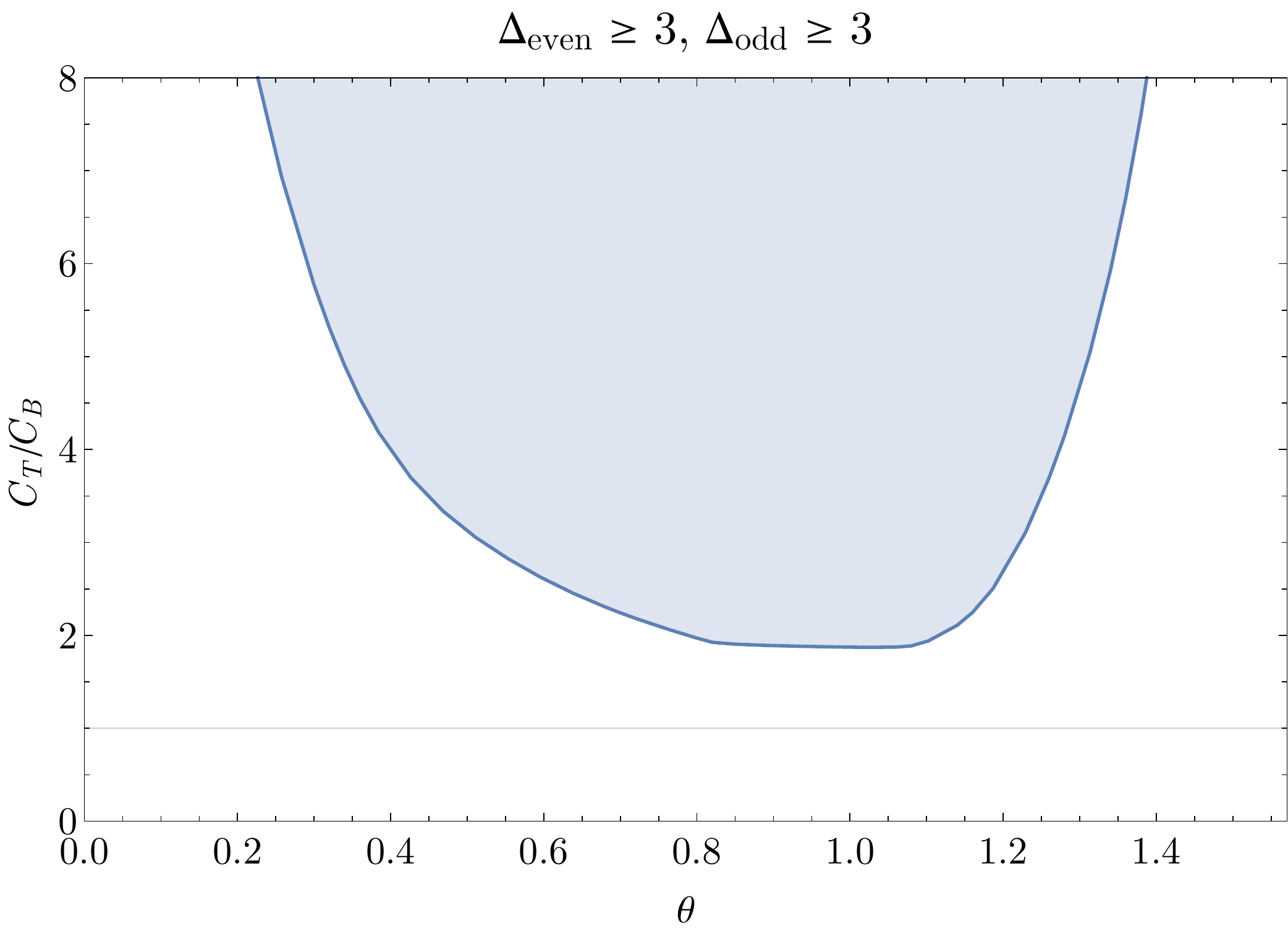}
\end{center}
\caption{Lower bound on $C_T$ as a function of $\theta$ assuming no relevant scalar operators.}\label{fig:oddGap3evenGap3Plot}
\end{figure}

In figure~\ref{fig:oddGap3evenGap3Plot}, we show a bound on the space of true ``dead-end" CFT's, i.e.\ theories with no parity-preserving or parity-breaking relevant deformations. We see from this plot that such theories must have $C_T\gtrsim 2$.  In addition, for a given $C_T$, $\th$ is constrained to lie towards the middle of the range $[0,\pi/2]$.

\begin{figure}[ht!]
\begin{center}
\includegraphics[width=0.78\textwidth]{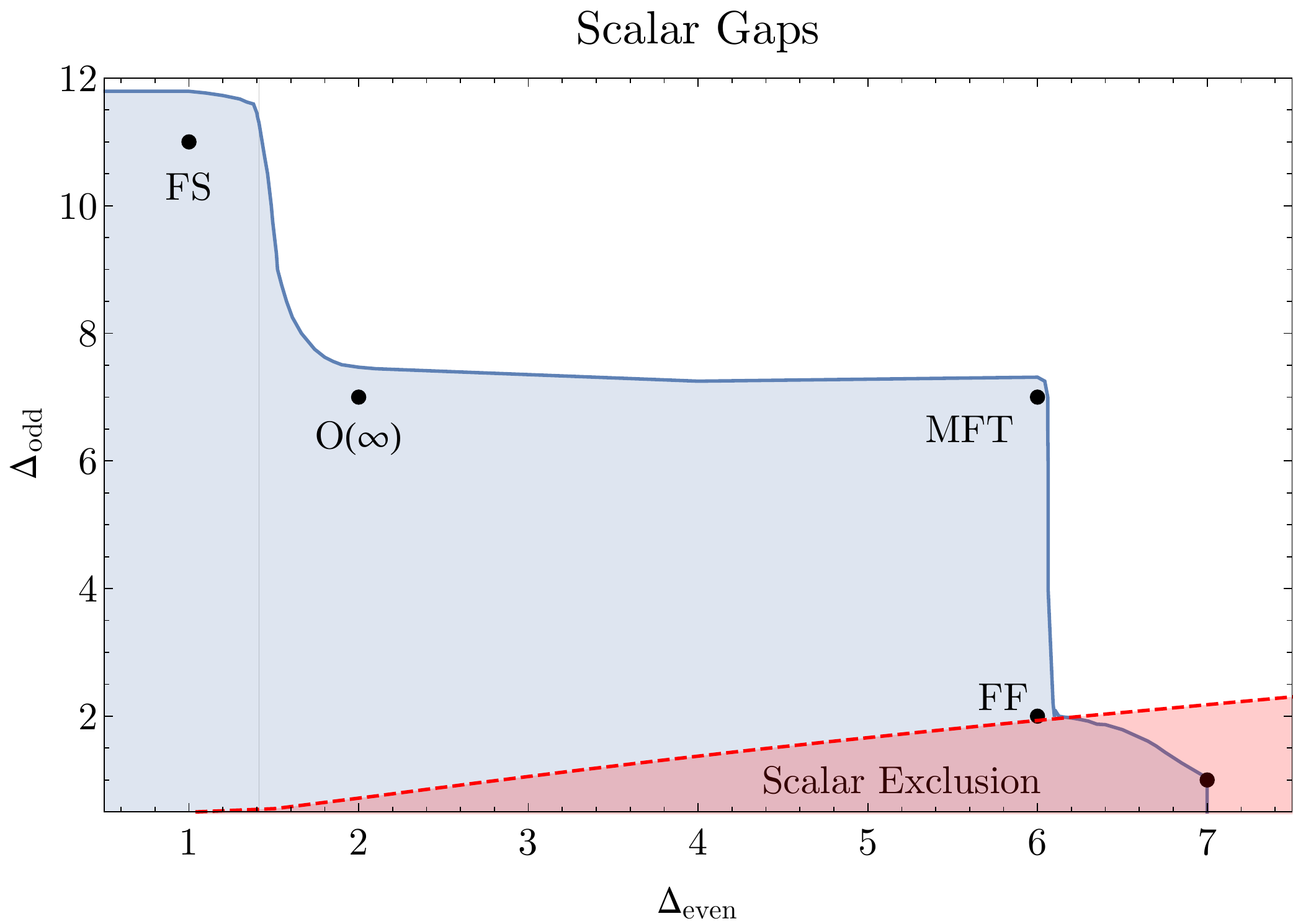}
\end{center}
\caption{Bound on the allowed gaps in parity-even and parity-odd scalar sectors (imposed simultaneously). The blue shaded region is allowed by the $\<TTTT\>$ bootstrap. The vertical grey line indicates the scaling dimension of $\epsilon$ in the Ising model. The red region is excluded from the scalar bootstrap for 4-point functions $\<\cO_{\text{odd}}\cO_{\text{odd}}\cO_{\text{odd}}\cO_{\text{odd}}\>$ assuming $\cO_{\text{even}}$ appears in both the $\cO_{\text{odd}} \times \cO_{\text{odd}}$ and $T \times T$ OPEs.}
\label{fig:scalarGapsExclusionPlot}
\end{figure}

For each of the parity-even and parity-odd sectors, we have seen that there exists a maximal gap beyond which no CFT can exist (figures~\ref{fig:deltaEvenGapsPlot} and \ref{fig:deltaOddGapsSmallPlot}). In figure~\ref{fig:scalarGapsExclusionPlot}, we show the full space of allowed gaps in the both sectors. Along the axes, this plot reproduces the gaps at which the islands disappear in figures~\ref{fig:deltaEvenGapsPlot} and \ref{fig:deltaOddGapsSmallPlot}.  The full bound shows several interesting features that approximately coincide with known theories. Notable points include MFT at $(\De_\textrm{odd},\De_\textrm{even})=(6,7)$, the free Majorana fermion at $(6,2)$, the free real scalar at $(11,1)$,\footnote{Note that the fundamental field in a free scalar theory is charged under a $\Z_2$ symmetry and thus does not appear in the $T \times T$ OPE.} and the $N=\infty$ limit of the $O(N)$ models at $(2,7)$. We also see the maximal possible gaps $\Delta_{\text{even}} \leq 7.0$ and $\Delta_{\text{odd}} \leq 11.78$. 

The known scaling dimension $\De_\epsilon=1.412625(10)$~\cite{Kos:2016ysd} of the energy operator $\epsilon$ in the 3d Ising CFT is shown in figure~\ref{fig:scalarGapsExclusionPlot} by a vertical line. We see that while most features seem to be related to free theories, there appears to be a sharp transition in the upper part of the allowed region, very close to the Ising line. We return to this point in section~\ref{sec:Ising}. 

There is also a feature near $(\De_\textrm{even},\De_\textrm{odd})=(7,1)$, which does not seem to correspond to a known theory. Such a theory, if exists, is constrained by the bound in figure~\ref{fig:deltaEvenGapsPlot} to have $C_T/C_B\sim 2$ and a value of $\theta$ very close to but lower than the free fermion value, $1.55 < \theta< 1.563$. Since this putative theory requires a very light parity-odd operator $\cO_{\text{odd}}$, such a large parity-even gap should be excluded by the bootstrap constraints for 4-point functions of $\cO_{\text{odd}}$ unless the $\cO_{\text{odd}} \times \cO_{\text{odd}}$ OPE contains an additional parity-even scalar not present in the $T \times T$ OPE. We leave it as an open question whether this can occur and if this region has any physical significance.

Note that every point which is allowed in this plot must be allowed together with a rectangular region to its lower left. Because of this, a large part of the allowed region is due to existence of MFT. It is therefore interesting to study analogous bounds under assumptions which would exclude the MFT. We leave this question for future work.

\subsection{Spin-2 gaps}
\label{Spin-2}

\begin{figure}[ht!]
\begin{center}
\includegraphics[width=0.78\textwidth]{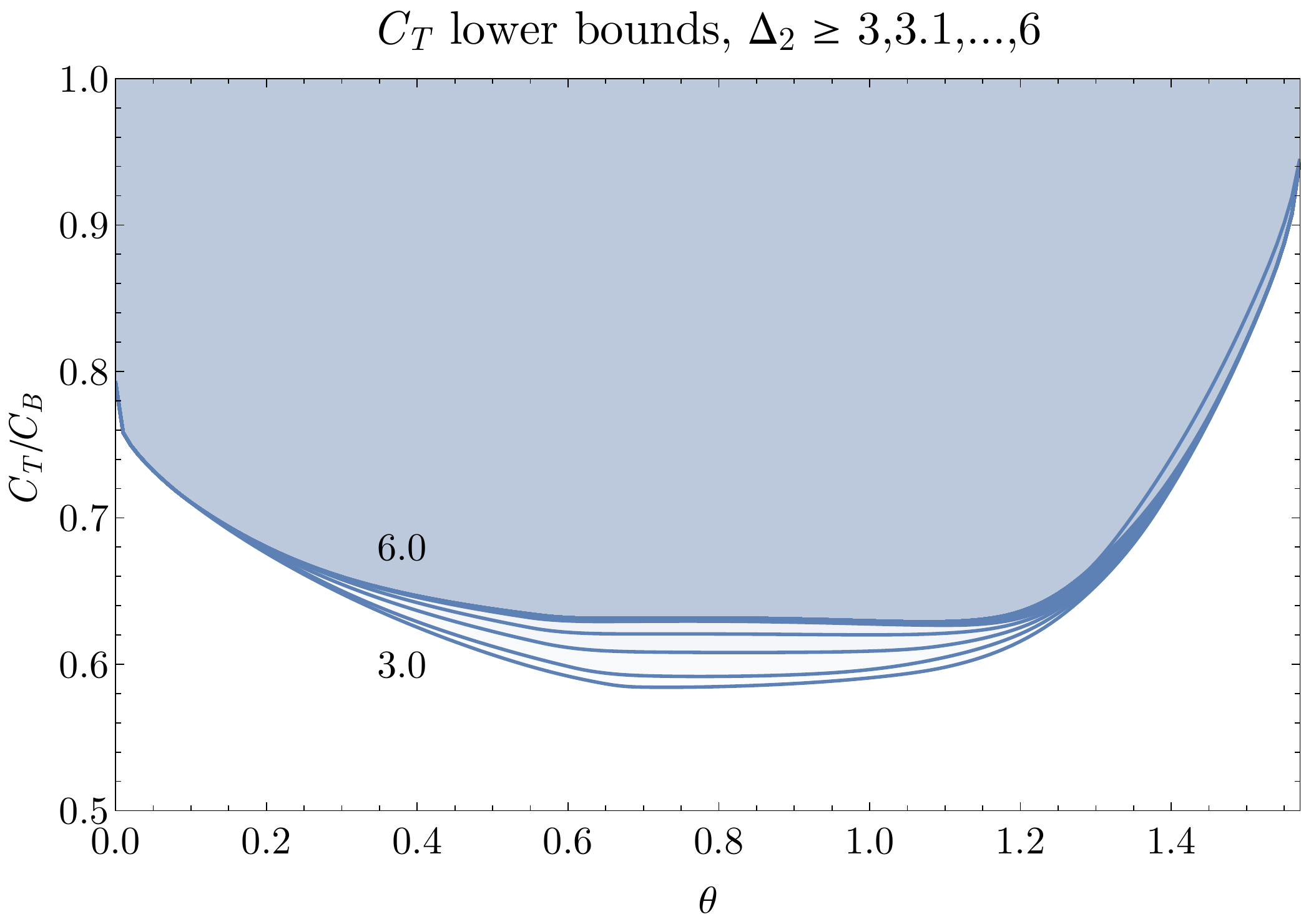}
\end{center}
\caption{Lower bounds on $C_T$ as a function of $\theta$ in 3d CFTs for different gaps between the stress tensor and the second parity-even spin-2 operator.}\label{fig:spin2Gap3to6}
\end{figure}

Next we turn to imposing gaps in the spin-2 spectrum. First we ask how the gap until the second parity-even spin-2 operator $T'$ of dimension $\Delta_2$ affects the lower bounds on $C_T$.  This is shown for gaps $\Delta_2 \geq 3,\ldots,6$ in figure~\ref{fig:spin2Gap3to6}. We can see that such gaps have a minimal effect on the lower bound. The gap $\Delta_2 = 6$ is special because this dimension occurs for the operator $T'_{\mu\nu} = T_{\mu\sigma} T^\sigma_\nu$ in a number of different CFTs, including free theories, $O(N)$ models at large $N$, and MFT. Thus it is not surprising that the full range of $\theta$ is still allowed at this gap and that the bound is not very strong.

\begin{figure}[ht!]
\begin{center}
\includegraphics[width=0.78\textwidth]{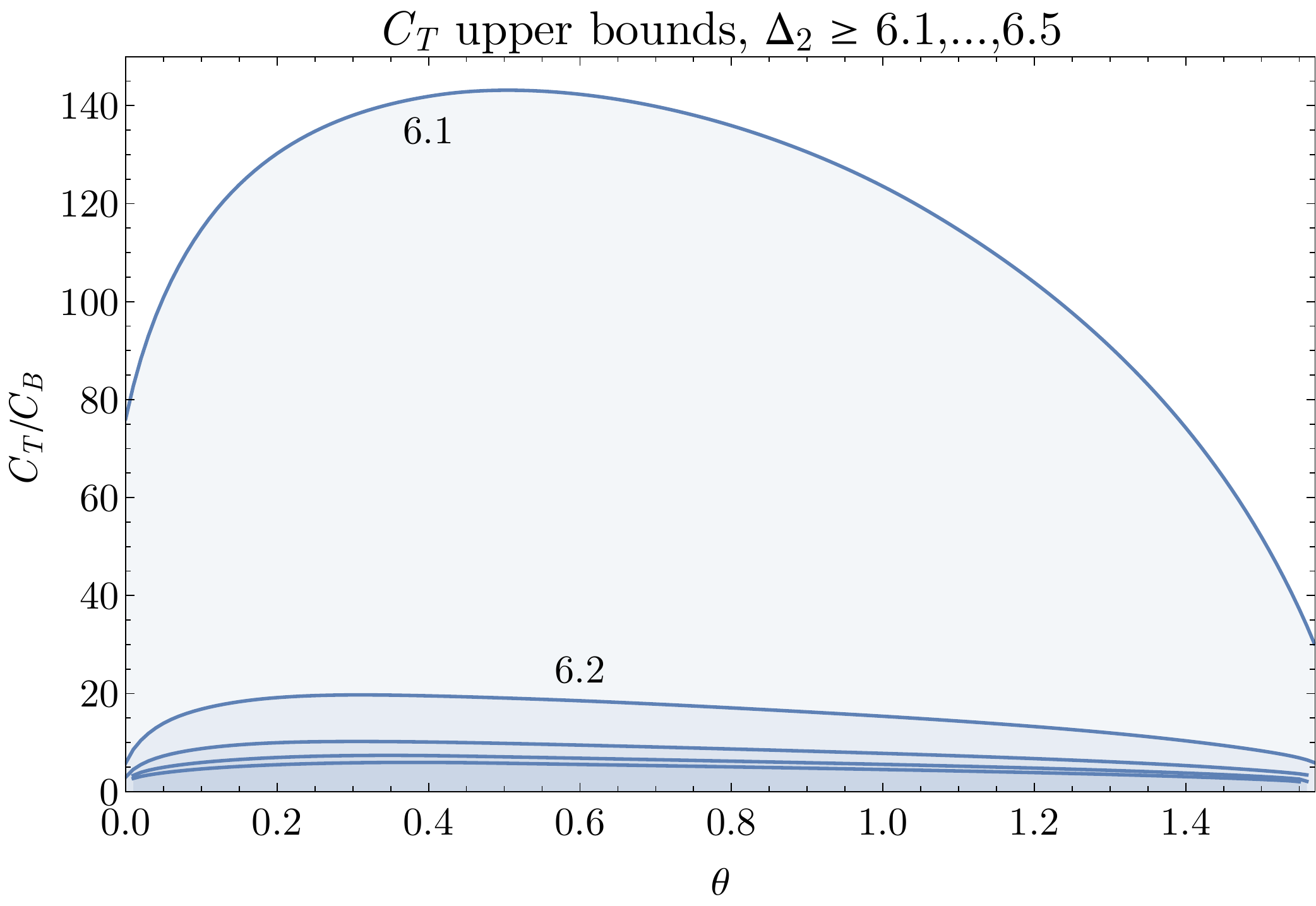}
\end{center}
\caption{Upper bounds on $C_T$ as a function of $\theta$ in 3d CFTs for different gaps between the stress tensor and the second parity-even spin-2 operator.}\label{fig:spin2Gap61to65}
\end{figure}

\begin{figure}[ht!]
\begin{center}
\includegraphics[width=0.78\textwidth]{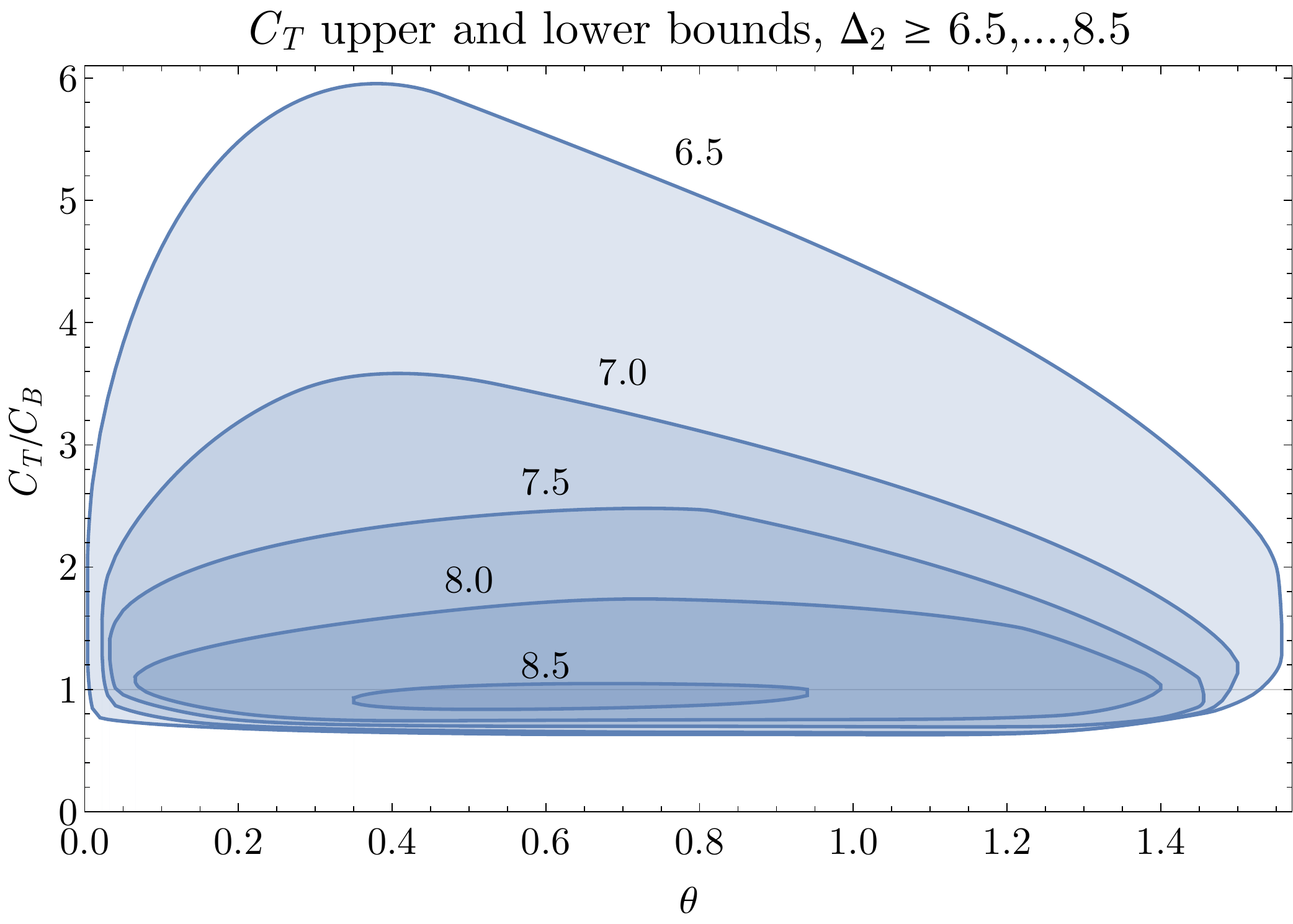}
\end{center}
\caption{Upper and lower bounds on $C_T$ as a function of $\theta$ in 3d CFTs for different gaps between the stress tensor and the second parity-even spin-2 operator.}\label{fig:spin2Gap65to85}
\end{figure}

However, we expect that if the $\Delta_2^{\min}$ is raised above 6, then we may be able to start excluding MFT and large $N$ theories by obtaining an upper bound on $C_T$. This is because the ``double-trace" operator $T_{\mu\sigma} T^\sigma_\nu$ in large $C_T$ theories will have a dimension $\Delta_2 = 6 + \cO(1/C_T)$, so imposing a gap above 6 will exclude some set of these theories. This is realized in figures~\ref{fig:spin2Gap61to65} and~\ref{fig:spin2Gap65to85}, where for gaps slightly above 6 the upper bound is fairly weak, but as it is raised further it becomes very strong and for gaps near 8.5 the closed region shrinks to a small island around $C_T/C_B \sim 1$ and $.4 \lesssim \theta \lesssim .9$. It is interesting to ask if there is a unitary CFT with such a large spin-2 gap and $\theta \approx \pi/4$ which lives inside of this allowed region.

\subsection{Spin-4 gaps}
\label{Spin-4}
In this section we move on to considering the constraints resulting from imposing a bound on the dimension of lightest spin-four operator $\Delta_4$. Consistency of crossing with the OPE in Minkowski space when two operators are light-like separated imposes a number of non-trivial constraints on the spectrum of ``intermediate" operators. In particular the ``Nachtmann theorem" stipulates that the leading twist, defined as the twist of the lightest primary of spin $\ell$ appearing in the OPE $\cO \times \cO$,
\bea
\tau_\ell=\Delta_\ell -\ell\ ,
\eea
is a monotonically non-decreasing convex function of $\ell$ which asymptotes to $2\tau_{\cO}$ \cite{Nachtmann:1973mr,Komargodski:2012ek,Fitzpatrick:2012yx,Li:2015itl,Costa:2017twz}. So far this has been rigorously established for scalar $\cO$ and even $\ell$, although the result is expected to hold more generally, for primary $\cO$ of any spin. Applying this to the stress tensor one finds that the dimension of the lightest operator of spin $\ell$ should not exceed $\ell+2$. For the leading spin-4 operator this implies inconsistency of unitary theories with $\Delta_4>6$. Moreover, when $\Delta_4=6$, the lightest operators of spin $\ell>4$ must have dimensions exactly equal to $\ell+2$. The corresponding theory is a MFT dual to pure gravity in AdS${}_4$ with Newton's constant taken to zero. The operators in question  are double-trace operators, schematically  $T\partial^{\ell-4}T$, where we omit indices for simplicity. 

When $\Delta_4$ approaches $6$ from below, by convexity all higher spin operators must approach $\ell+2$. This is exactly the behavior expected for a theory dual to weakly coupled gravity in AdS${}_4$. The double-trace anomalous dimensions $\Delta_\ell-\ell-2$ are due to graviton exchange in the bulk, which is proportional to Newton's constant $G_N\sim 1/C_T$. This picture suggests that imposing a gap $\Delta_4>6-\epsilon$ should result in a numerical bound on the central charge  $C_T\geq C^*_T$,  with $C_T^*$ going to infinity as $C_T^* \sim 1/\epsilon$. 

Such behavior was observed previously in the context of the $\mathcal N=8$ numerical supersymmetric bootstrap in $3d$ \cite{Chester:2014fya}. There the lower bound on $C_T$ was studied as a function of the dimensions of spin-0 and spin-2 long multiplets, $\Delta^*_0$ and $\Delta^*_2$ respectively. When the dimensions approached the values associated with $N\rightarrow \infty$ ABJM theory, the exclusion region for $C_T$ grew accordingly, with the lower bound on $C_T$ scaling as $1/(2-\Delta^*_2)$. Another related result is in the context of numerical bootstrap of four conserved currents \cite{Dymarsky:2017xzb}. In this case imposing  $\Delta_4=6$ resulted in the lower bound on $C_T$ growing indefinitely as the numerical precision (the derivative order $\Lambda$) increased. 

\begin{figure}[ht!]
\begin{center}
\includegraphics[width=0.78\textwidth]{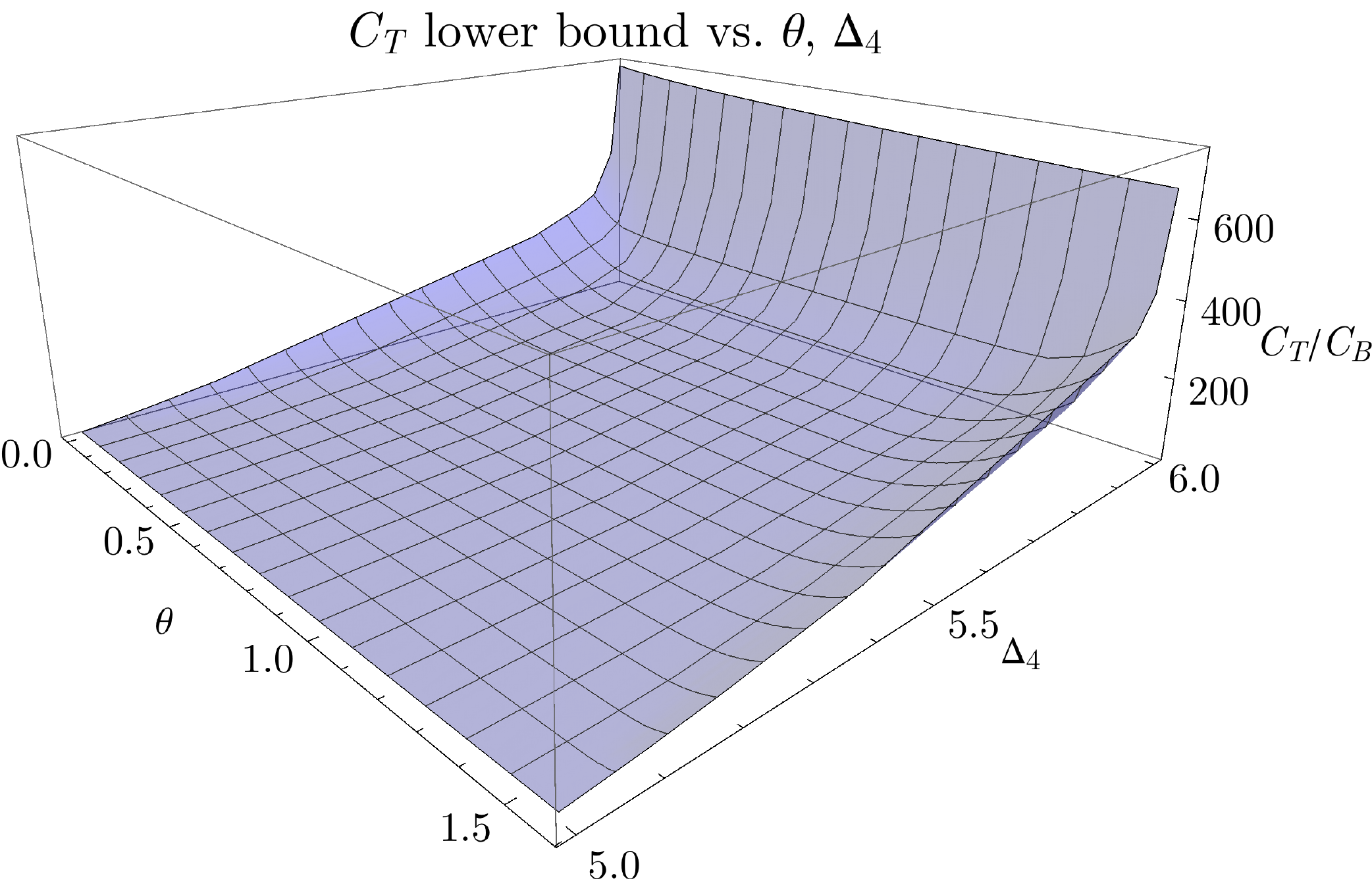}
\end{center}
\caption{Lower bounds on $C_T$ as a function of $\theta$ and the spin-4 gap $\Delta_4$.}
\label{spin4Gap3dPlot}
\end{figure}

\begin{figure}[ht!]
\begin{center}
\includegraphics[width=0.78\textwidth]{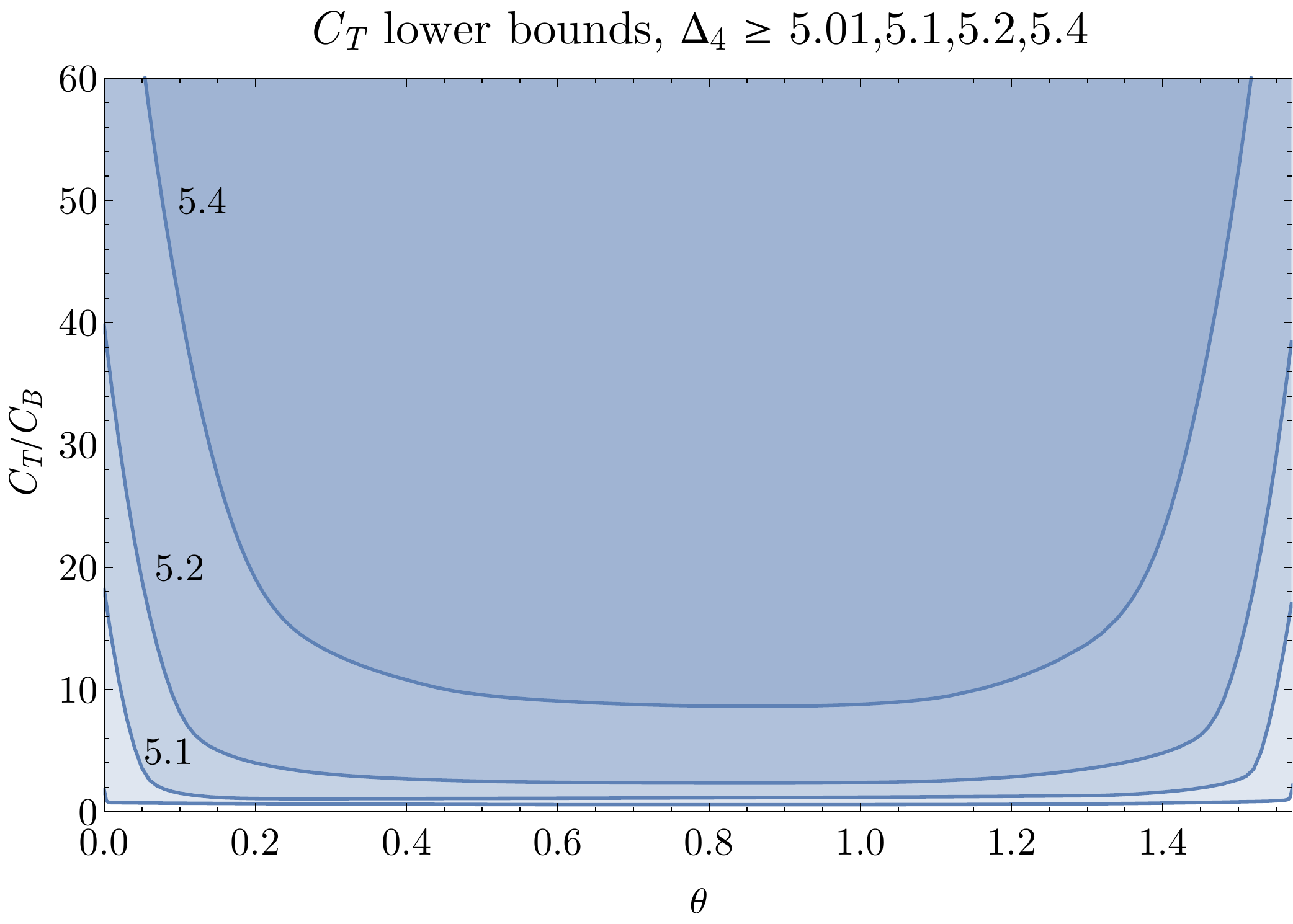}
\end{center}
\caption{Lower bounds on $C_T$ as a function of $\theta$ for spin-4 gaps $\Delta_4 \geq 5.01, 5.1, 5.2, 5.4$.}
\label{fig:spin4GapPlot}
\end{figure}

The numerical results of imposing a gap on $\Delta_4$ are shown in figure~\ref{spin4Gap3dPlot}, with some projections at smaller values of $\Delta_4$ shown in figure~\ref{fig:spin4GapPlot}. For each value of $\Delta_4$ and $0\leq \theta\leq \pi/2$ we find a minimal allowed value of $C_T$. This value is quite sensitive to $\theta$, generally reaching maximal values for $\theta\rightarrow 0,\pi/2$ and remaining relatively small around $\theta\approx \pi/4$. At the same time when $\Delta_4$ approaches $6$ the bound rapidly grows for all value of $\theta$, and seems to diverge (numerically we see bounds of $\cO(600-700)$) as $\Delta_4\rightarrow 6$, consistent with the Nachtmann theorem. Our bounds do not seem to show sufficient convergence to read off the expected $1/\epsilon$ scaling, but it will be interesting to study this divergent behavior more closely in future work.

\subsection{Ising-like spectrum}
\label{sec:Ising}

Next we focus our attention on what can be learned about the 3d Ising model from the $\<TTTT\>$ bootstrap. In earlier numerical bootstrap work~\cite{El-Showk:2014dwa}, a precise determination of the central charge $C_T^{\textrm{Ising}}/C_B = 0.946534(11)$ was found. As far as we are aware, no determinations of the $\<TTT\>$ 3-point function in the 3d Ising model have been made previously. 

The Ising model has a $\mathbb{Z}_2$ global symmetry, but only $\mathbb{Z}_2$-even operators appear in the $T \times T$ OPE. Such operators can be either even or odd under spacetime parity.  The scaling dimensions of the leading parity-even operators in the 3d Ising spectrum have been computed to high precision using numerical bootstrap methods (see table 2 of~\cite{Simmons-Duffin:2016wlq} for a summary). However, as far as we are aware very little is known about the parity-odd spectrum.

\begin{figure}
\begin{center}
\includegraphics[width=0.78\textwidth]{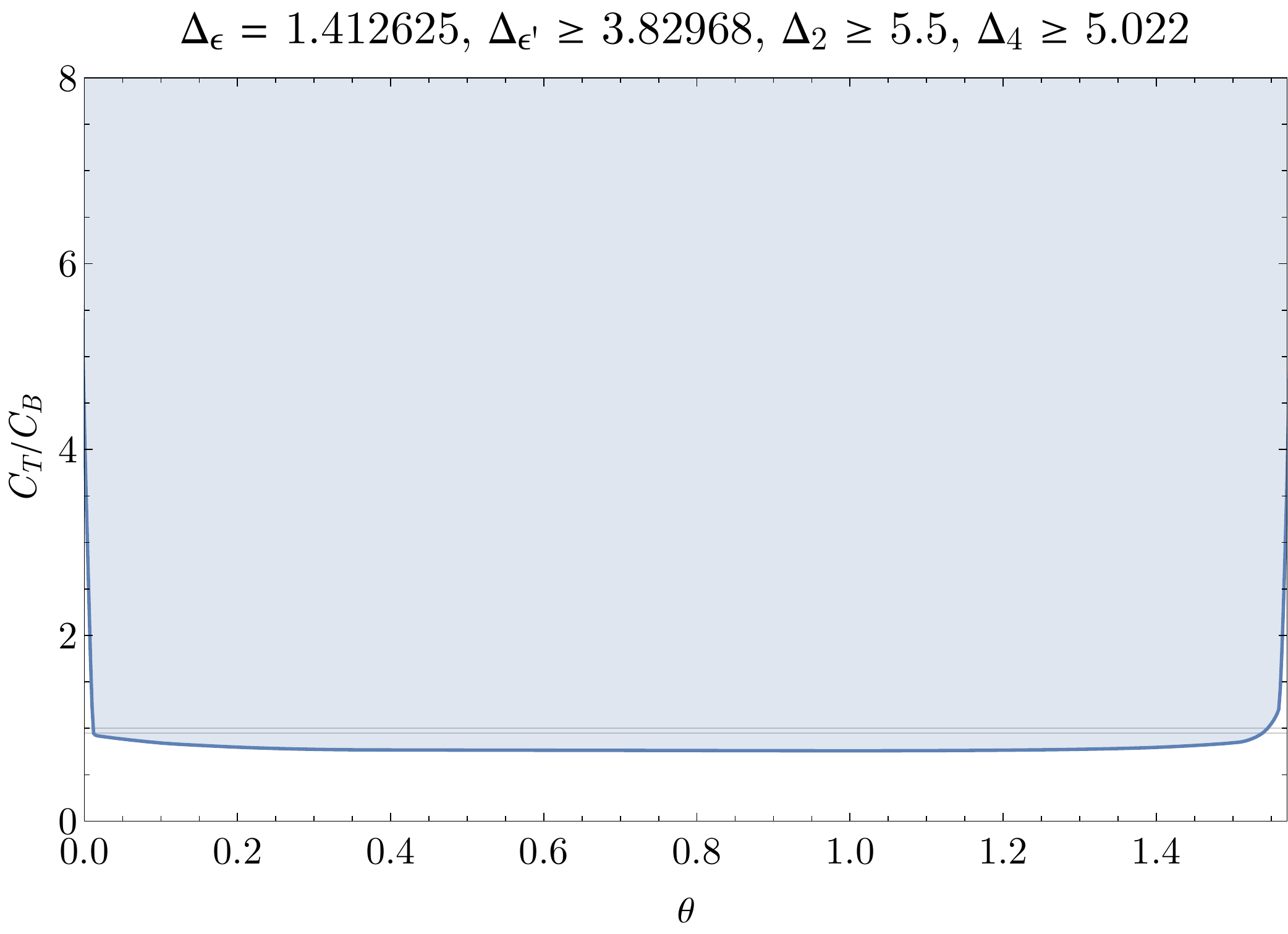}
\end{center}
\caption{Lower bound on $C_T$ as a function of $\theta$ assuming known low-lying gaps in the parity-even spectrum in the 3d Ising CFT.}\label{fig:isingEvenSectorGapsPlot}
\end{figure}

In figure~\ref{fig:isingEvenSectorGapsPlot} we show the result of inputting the approximate known scaling dimensions for the leading parity-even scalars $\{\epsilon, \epsilon'\}$, the second spin-2 operator $T'$, and the leading spin-4 operator. The horizontal lines show the 3d Ising value of $C_T$ as well as the free scalar value. Regions very close to $\theta=0$ and $\theta=\pi/2$ are excluded (primarily due to the spin-4 gap) but otherwise this data does not place a very strong constraint.

\begin{figure}
\begin{center}
\includegraphics[width=0.78\textwidth]{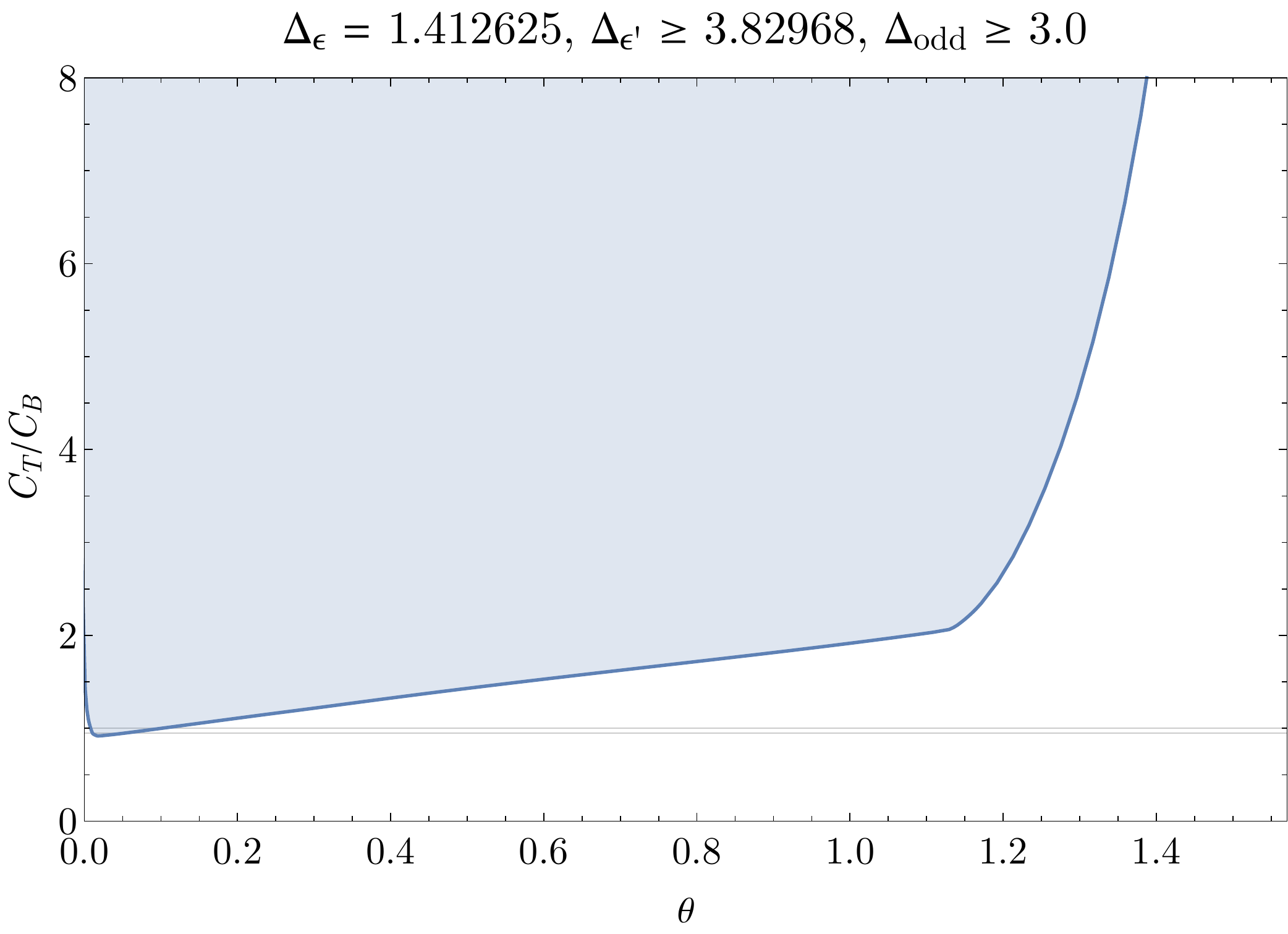}
\end{center}
\caption{Lower bound on $C_T$ as a function of $\theta$ assuming known low-lying gaps in the parity-even scalar spectrum in the 3d Ising CFT, combined with the assumption that the leading parity-odd scalar is irrelevant.}\label{fig:Isingoddgap3}
\end{figure}

\begin{figure}
\begin{center}
\includegraphics[width=0.78\textwidth]{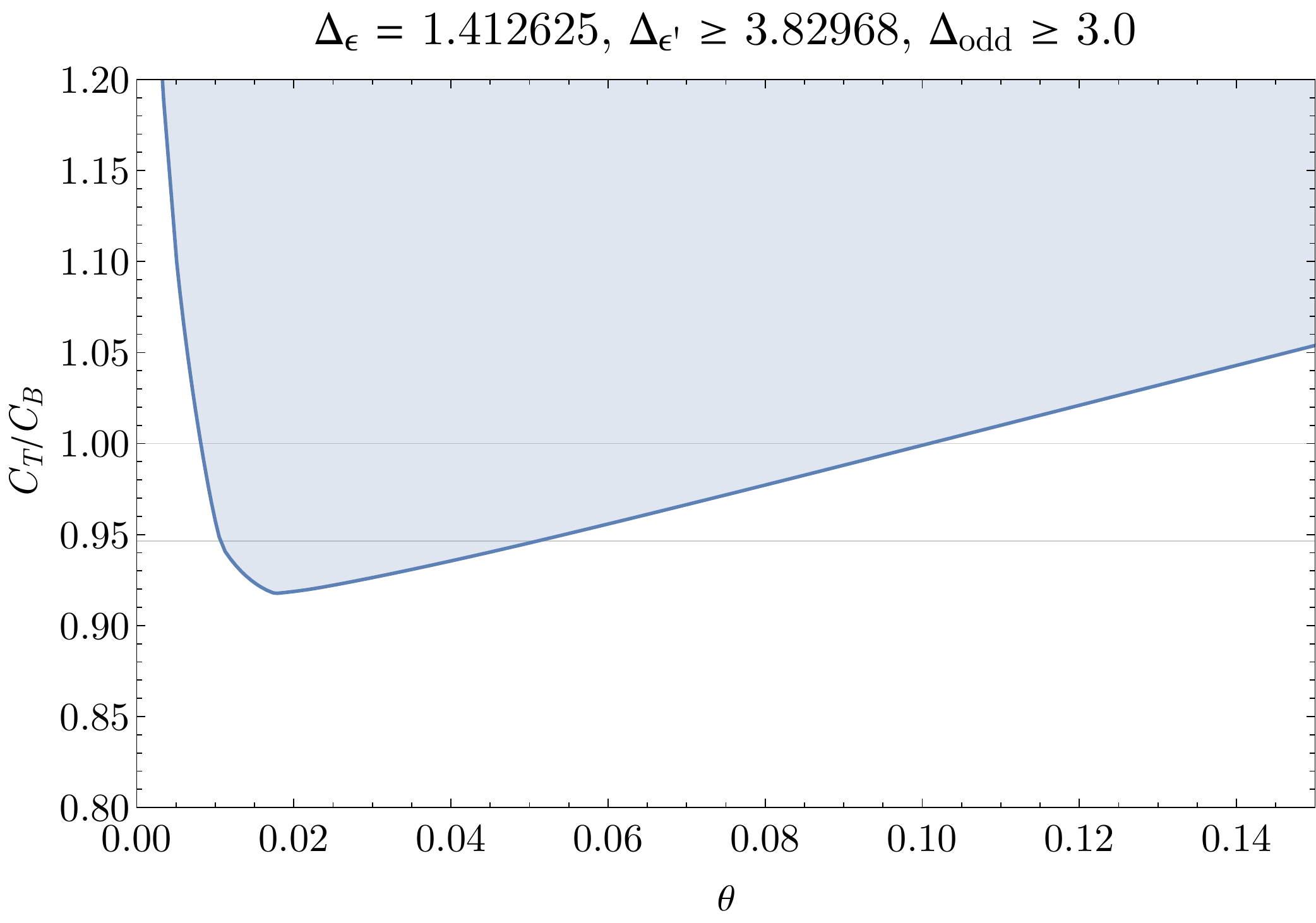}
\end{center}
\caption{Lower bound on $C_T$ as a function of $\theta$ assuming known low-lying gaps in the parity-even scalar spectrum in the 3d Ising CFT, combined with the assumption that the leading parity-odd scalar is irrelevant.}\label{fig:Isingoddgap3zoom}
\end{figure}

On the other hand, we find that imposing a parity-odd gap places a very strong constraint on the allowed region. In figure~\ref{fig:Isingoddgap3} we show the effect of inputting the expectation (overwhelmingly supported by experiment, simulation, and other theoretical techniques) that the leading parity-odd scalar is irrelevant, in addition to inputting the leading parity-even scalar dimensions. Only a tiny window at small $\theta$ is compatible with the 3d Ising value of $C_T$. We show a zoom of this region in figure~\ref{fig:Isingoddgap3zoom}, where it can be seen that these assumptions imply $.01 < \theta < .05$.

In fact, it is likely that the parity-odd scalar gap in the 3d Ising model is significantly larger than 3. E.g., it may be close to the free scalar value $\Delta_{\textrm{odd}} = 11$. This large gap is also plausible given figure~\ref{fig:scalarGapsExclusionPlot}, where it can be seen that a sharp transition in the allowed region occurs near the Ising value of $\Delta_{\textrm{even}}$. In light of this plot, if the gap is maximal we see that it may be as large as $\Delta_{\textrm{odd}} \lesssim 11.2$. 

\begin{figure}
\begin{center}
\includegraphics[width=0.78\textwidth]{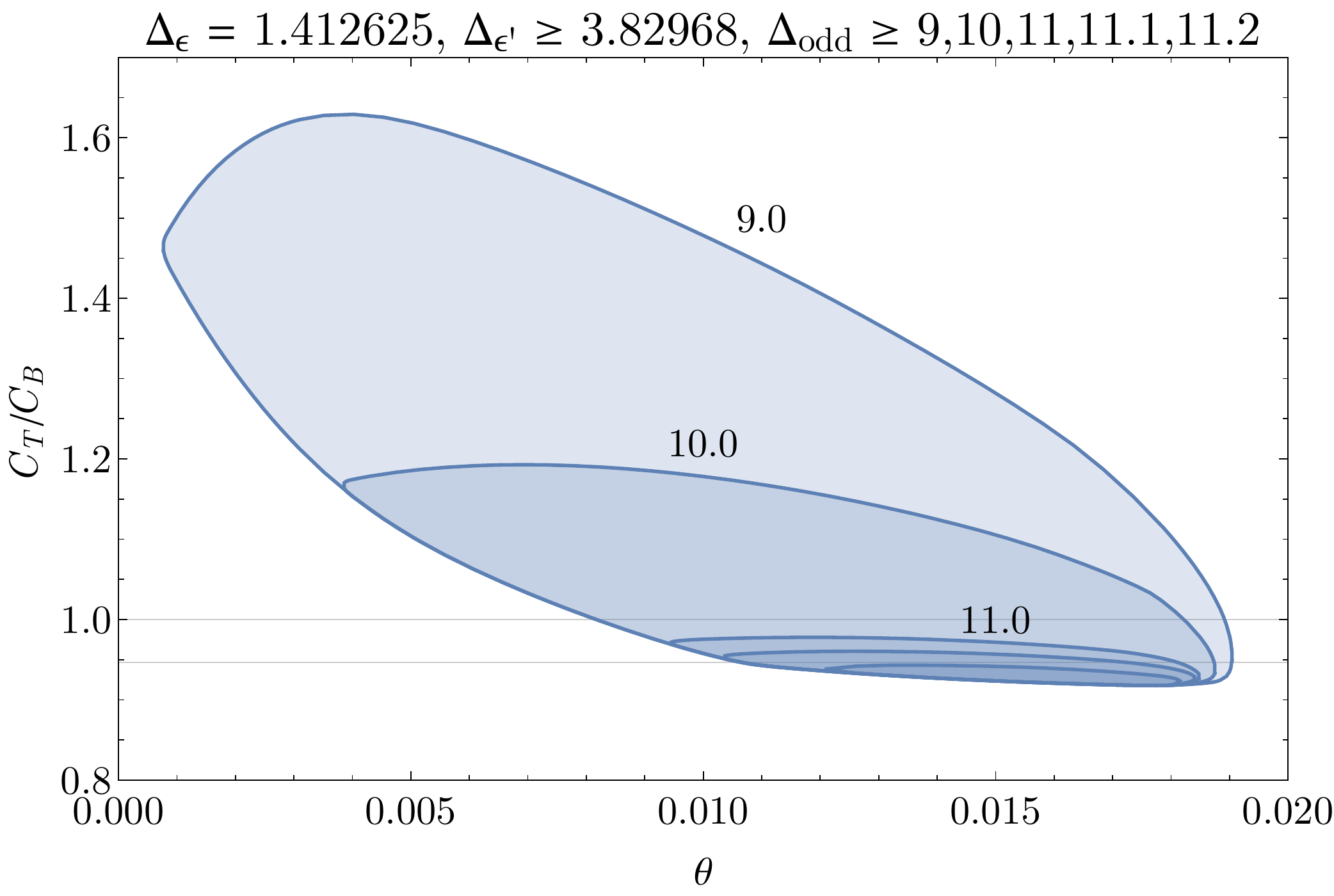}
\end{center}
\caption{Lower and upper bounds on $(\theta, C_T)$ assuming known low-lying gaps in the parity-even scalar spectrum in the 3d Ising CFT, combined with various larger gaps in the parity-odd spectrum. A gap $\Delta_{\text{odd}}=11.1$ is compatible with $C_T^{\text{Ising}}$ (shown as the lower horizontal line) but a gap $\Delta_{\text{odd}}=11.2$ is not.}\label{fig:IsingparityOddGap91011Plot}
\end{figure}

Previously in figure~\ref{fig:deltaOddGapsSmallPlot} we saw that a parity-odd gap close to this value on its own imposes a robust restriction $\theta < .023$, with an allowed region compatible with $C_T^{\text{Ising}}$. In figure~\ref{fig:IsingparityOddGap91011Plot} we show the result on the allowed region of additionally imposing the known values of $\Delta_{\epsilon}$ and $\Delta_{\epsilon'}$, combined with the sequence of assumptions $\Delta_{\textrm{odd}} \geq 9, 10, 11, 11.1, 11.2$. These assumptions lead to closed islands and if the gap is close to being saturated allow us to make the tighter determination $.01 < \theta < .018-.019$, with the precise upper bound depending on the gap.

\newpage
\section{Discussion}
\label{Discussion}
In this work we used the numerical conformal bootstrap to study the space of unitary parity-preserving CFTs in three dimensions. Assuming the existence of a unique stress tensor (conserved spin-2 current) and imposing crossing symmetry of its four-point correlation function, we found a number of universal bounds on CFT data. One striking discovery is the necessity of both light parity-even ($\Delta_{\rm even}\leq 7$) and parity-odd ($\Delta_{\rm odd}\leq 11.78$) scalars in the spectrum of {\it any\/} consistent local unitary CFT, see figure~\ref{fig:scalarGapsExclusionPlot}. Among other universal results are those limiting  the value of the central charge $C_T$ modulo additional assumptions. For example, in hypothetical ``dead-end" CFTs without any relevant scalars $C_T$ is constrained to be larger than roughly twice the central charge of a free 3d scalar or Majorana fermion. These, and other similar findings presented in this paper are of a new kind, in the sense that they cannot be derived (as far as we know) using any theoretical tools other than the numerical bootstrap. 

There is another class of discoveries presented in this paper which further support and extend previously established theoretical results. Our numerical results reproduce the ``conformal collider" bounds, see figure~\ref{fig:hofmanmaldacena}. Imposing scalar or spin-2 gaps above the values they take in holographic theories further allows us to place upper bounds on $C_T$. Similarly, imposing a gap on the dimension of the lightest spin-4 operator discussed in section \ref{Spin-4}, $\Delta_4\geq 6-\epsilon$, $\epsilon\rightarrow 0$, forces the CFT in question to have an apparently diverging central charge and a spectrum likely dual to weakly coupled gravity in AdS${}_4$, in full consistency with the Nachtmann theorem \cite{Nachtmann:1973mr,Komargodski:2012ek,Fitzpatrick:2012yx,Costa:2017twz,Li:2015itl}. Reproducing these results is a strong consistency check on our numerical setup. 

Many exclusion plots in this work exhibit characteristic features potentially signaling the existence of an underlying theory saturating the corresponding bounds. The scalar exclusion plot in figure~\ref{fig:scalarGapsExclusionPlot} has a kink that tentatively corresponds to the 3d Ising model, in addition to reassuring corners that coincide with other known free or mean-field solutions. This gives hope to extend our results to further elucidate precise properties of particular theories. The first few steps in this direction for the 3d Ising model were already undertaken in section  \ref{sec:Ising}, where known dimensions of light scalar operators\footnote{Assuming that the lightest parity-odd scalar is irrelevant.} were used to obtain a strong bound $0.01<\theta<0.05$ on the OPE coefficient controlling the 3pt function of stress tensors \eqref{thetadef}. By assuming larger gaps in the parity-odd scalar sector this window can be reduced down to $0.010 < \theta < 0.019$. We also find closed islands in Figs.~\ref{fig:deltaEvenGapsPlot} and~\ref{fig:spin2Gap65to85} which may indicate new nontrivial solutions to the bootstrap equations and could be interesting to study further.

Our work paves the way for many future investigations. Below we briefly describe only some of the possible directions, which we find particularly interesting and important. A substantial extension of this work would be to combine stress tensors with other operators, such as scalars, fermions, or global symmetry currents, using a larger mixed correlator bootstrap. In this way one should be able to isolate e.g.~theories with global $O(N)$ symmetry and obtain a host of new constraints pertaining to such theories. One can also extend our work to CFTs with varying amounts of supersymmetry, requiring additional computation of the necessary superconformal blocks. From the technical point of view these generalizations are relatively straightforward and only require combining previously developed ingredients. 

Yet another natural generalization is to extend the analysis of this paper to parity-breaking theories. This direction is interesting in part because it would help us gain a better understanding of the large family of Chern-Simons-matter theories in three dimensions, recently understood to be interconnected by a large web of RG flows and dualities (e.g.~\cite{Aharony:2015mjs,Hsin:2016blu,Aharony:2016jvv}). From the technical point of view such an extension would require the straightforward task of generalizing the analysis of sections \ref{sec:conformalstructures} and \ref{sec:conformalblocks} to additional parity-breaking structures.

Finally, the numerical analysis performed in this paper, and the theoretical developments which it required, constitute significant progress in the development of the conformal bootstrap in $d=3$ dimensions. It would be very interesting to generalize the current analysis to higher dimensions, first to $d=4$. The needed conformal blocks in four dimensions were recently calculated implicitly in a number of works \cite{Elkhidir:2014woa, Echeverri:2015rwa, Echeverri:2016dun, Costa:2016hju, Cuomo:2017wme, Karateev:2017jgd}. Accordingly, the bootstrap for the stress tensor and other operators with spin in four dimensions is now accessible in principle, although it still represents a substantial technical challenge. We hope to address this problem in the future. This research program can also be potentially extended to arbitrary $d$ yielding universal constraints on CFTs in $d=5,6$ and beyond. We hope this study will eventually yield new non-trivial results contributing to our understanding of interacting CFTs, or their absence, in $d>6$.

\section*{Acknowledgements}

We are grateful to Clay C\'ordova, Daliang Li, David Meltzer, Jo\~ao Penedones, Eric Perlmutter, Slava Rychkov, Marco Serone, Emilio Trevisani, Alessandro Vichi, and Alexander Zhiboedov for discussions. We also thank Revant Nayar for collaboration in the initial stages of this work. Many thanks to the organizers and participants of the bootstrap collaboration workshops at Yale, Princeton, and ICTP S\~ao Paulo where part of this work was completed. AD is supported by NSF grant PHY-1720374. DSD is supported by DOE grant DE-SC0009988, a William D. Loughlin Membership at the Institute for Advanced Study, and Simons Foundation grant 488657 (Simons Collaboration on the Nonperturbative Bootstrap). PK is supported by DOE grant DE-SC0011632. DP is supported by NSF grant PHY-1350180 and Simons Foundation grant 488651. The computations in this paper were run on the Omega and Grace computing clusters supported by the facilities and staff of the Yale University Faculty of Arts and Sciences High Performance Computing Center, on the Hyperion computing cluster supported by the School of Natural Sciences Computing Staff at the Institute for Advanced Study and on the computing clusters of the National Energy Research Scientific Computing Center, a DOE Office of Science User Facility supported by the Office of Science of the U.S. Department of Energy under Contract No. DE-AC02-05CH11231.

\newpage
\appendix

\section{Tensor structures}
\label{app:tensor}
In this section we give the explicit expressions for the three-point tensor structures in the differential basis as required for the computation of conformal blocks in section~\ref{sec:conformalblocks}.

\subsection{Parity-even structures in differential basis}
For a given spin $\ell$, we define the basis of parity-even differential operators for $\<TT\cO_\ell\>$ as
\be
	\cD_{n_{23},n_{13},n_{12}}=H_{12}^{n_{12}}D_{12}^{n_{13}}D_{21}^{n_{23}}D_{11}^{m_1}D_{22}^{m_2}\Sigma_{1}^{n_{12}+n_{23}+m_1}\Sigma_{2}^{n_{12}+n_{13}+m_2},
\ee
where $m_1=2-n_{12}-n_{13}$ and $m_2=2-n_{12}-n_{23}$.
\paragraph{Structures for $\<TT\cO_0\>$}
There exists a single parity-even tensor structure for $\<TT\cO_0\>$, given by the differential operator
\begin{align}
	\cD^{(1)}_{\mathbf{0}^+}=-\mathcal{D}_{0, 0, 0}+(\Delta -5) (\Delta +2)\mathcal{D}_{0, 0, 1}-\frac{1}{8} (\Delta -5) (\Delta -3) \Delta  (\Delta +2)\mathcal{D}_{0, 0, 2}.
\end{align}
\paragraph{Structures for $\<TT\cO_2\>$}
There exists a single parity-even tensor structure for $\<TT\cO_2\>$, with $\De>3$, given by the differential operator
\begin{align}
	\cD^{(1)}_{\mathbf{2}^+}=&-8 \left(7 \Delta ^2-13 \Delta +30\right)\mathcal{D}_{0, 0, 0}+16 (\Delta +2) (5 \Delta -11)\mathcal{D}_{1, 0, 0}\nn\\
	&-16 (\Delta +2) (\Delta +4)\mathcal{D}_{2, 0, 0}+16 (\Delta +2) (5 \Delta -11)\mathcal{D}_{0, 1, 0}\nn\\
	&-32 \Delta  (2 \Delta -5)\mathcal{D}_{1, 1, 0}-16 (\Delta +2) (\Delta +4)\mathcal{D}_{0, 2, 0}+8 \Delta  \left(\Delta ^2+29 \Delta -78\right)\mathcal{D}_{0, 0, 1}\nn\\
	&-8 (\Delta -3) (\Delta +2) \left(\Delta ^2-2 \Delta -2\right)\mathcal{D}_{1, 0, 1}-8 (\Delta -2) (\Delta +2) \left(\Delta ^2-3 \Delta +8\right)\mathcal{D}_{0, 1, 1}\nn\\
	&+8 (\Delta -2)^2 (\Delta -1) \Delta\mathcal{D}_{1, 1, 1}+(\Delta -2) (\Delta -1) \Delta  \left(\Delta ^3-6 \Delta ^2-25 \Delta +78\right)\mathcal{D}_{0, 0, 2}.
\end{align}
\paragraph{$\<TTT\>$ structures}
There exist two parity-even tensor structures for $\<TTT\>$, one realized in the theory of a single free scalar field, and the other in the theory of single free Majorana fermion. They are given by the following differential operators
\begin{align}
	\cD_{T}^{(B)}=&-\frac{9}{128 \pi ^3}\mathcal{D}_{0, 0, 0}+\frac{35}{256 \pi ^3}\mathcal{D}_{1, 0, 0}-\frac{245}{1024 \pi ^3}\mathcal{D}_{2, 0, 0}+\frac{35}{256 \pi ^3}\mathcal{D}_{0, 1, 0}-\frac{33}{512 \pi ^3}\mathcal{D}_{1, 1, 0}\nn\\
	&-\frac{245}{1024 \pi ^3}\mathcal{D}_{0, 2, 0}+\frac{153}{1024 \pi ^3}\mathcal{D}_{0, 0, 1}-\frac{35}{256 \pi ^3}\mathcal{D}_{1, 0, 1}-\frac{159}{1024 \pi ^3}\mathcal{D}_{0, 1, 1}-\frac{63}{1024 \pi ^3}\mathcal{D}_{1, 1, 1},\\
%****************************
	\cD_{T}^{(F)}=&-\frac{9}{64 \pi ^3}\mathcal{D}_{0, 0, 0}+\frac{5}{16 \pi ^3}\mathcal{D}_{1, 0, 0}-\frac{35}{64 \pi ^3}\mathcal{D}_{2, 0, 0}+\frac{5}{16 \pi ^3}\mathcal{D}_{0, 1, 0}-\frac{9}{64 \pi ^3}\mathcal{D}_{1, 1, 0}-\frac{35}{64 \pi ^3}\mathcal{D}_{0, 2, 0}\nn\\
	&+\frac{45}{128 \pi ^3}\mathcal{D}_{0, 0, 1}-\frac{5}{16 \pi ^3}\mathcal{D}_{1, 0, 1}-\frac{39}{128 \pi ^3}\mathcal{D}_{0, 1, 1}-\frac{9}{64 \pi ^3}\mathcal{D}_{1, 1, 1}.
\end{align}

\paragraph{Structures for $\<TT\cO_\ell\>$}
There exists two parity-even tensor structure for $\<TT\cO_{\ell}\>$ for even $\ell\geq4$, given by the differential operators
\begin{align}
\cD^{(1)}_{\Bell^+}=&(\Delta ^4-6 \Delta ^3+43 \Delta ^2-102 \Delta +3 \ell ^4+6 \ell ^3-4 \Delta ^2 \ell ^2\nn\\&\quad+12 \Delta  \ell ^2-35 \ell ^2-4 \Delta ^2 \ell +12 \Delta  \ell -38 \ell +184)\mathcal{D}_{0, 0, 0}\nn\\
&-2 (-\Delta +\ell +1) (\Delta +\ell ) \left(-\Delta ^2+3 \Delta +\ell ^2+\ell -14\right)\mathcal{D}_{1, 0, 0}\nn\\
&+(-\Delta +\ell -1) (-\Delta +\ell +1) (\Delta +\ell ) (\Delta +\ell +2)\mathcal{D}_{2, 0, 0}\nn\\
&-2 (-\Delta +\ell +1) (\Delta +\ell ) \left(-\Delta ^2+3 \Delta +\ell ^2+\ell -14\right)\mathcal{D}_{0, 1, 0}\nn\\
&-4 \left(-\Delta ^4+6 \Delta ^3-13 \Delta ^2+12 \Delta +\ell ^4+2 \ell ^3-7 \ell ^2-8 \ell +44\right)\mathcal{D}_{1, 1, 0}\nn\\
&+2 (-\Delta +\ell +1) (\Delta +\ell ) \left(\Delta ^2-3 \Delta +\ell ^2+\ell -10\right)\mathcal{D}_{2, 1, 0}\nn\\
&+(-\Delta +\ell -1) (-\Delta +\ell +1) (\Delta +\ell ) (\Delta +\ell +2)\mathcal{D}_{0, 2, 0}\nn\\
&+2 (-\Delta +\ell +1) (\Delta +\ell ) \left(\Delta ^2-3 \Delta +\ell ^2+\ell -10\right)\mathcal{D}_{1, 2, 0}\nn\\
&+(\Delta ^4-6 \Delta ^3-5 \Delta ^2+42 \Delta +\ell ^4+2 \ell ^3-\ell ^2-2 \ell +40)\mathcal{D}_{2, 2, 0}\nn\\
&-2 (\ell -1) (\ell +2) \left(12 \Delta ^2-36 \Delta +\ell ^4+2 \ell ^3-\Delta ^2 \ell ^2+3 \Delta  \ell ^2-13 \ell ^2\right.\nn\\&\quad\quad\left.-\Delta ^2 \ell +3 \Delta  \ell -14 \ell +72\right)\mathcal{D}_{0, 0, 1}\nn\\
&-12 \left(\ell ^2+\ell -4\right) (-\Delta +\ell +1) (\Delta +\ell )\mathcal{D}_{1, 0, 1}\nn\\
&-8 \ell  (\ell +1) (-\Delta +\ell +1) (\Delta +\ell )\mathcal{D}_{0, 1, 1}-8 (\ell -1) \ell  (\ell +1) (\ell +2)\mathcal{D}_{1, 1, 1}\nn\\
&+\frac{1}{4} (\ell -1) \ell  (\ell +1) (\ell +2) \left(-\Delta ^4+6 \Delta ^3+5 \Delta ^2-42 \Delta +\ell ^4+2 \ell ^3-17 \ell ^2\right.\nn\\&\quad\quad\left.-18 \ell +104\right)\mathcal{D}_{0, 0, 2},\\
%****************************
\cD^{(2)}_{\Bell^+}=&(-\Delta ^2+3 \Delta -\ell ^2-\ell +36)\mathcal{D}_{0, 0, 0}+2 (-\Delta +\ell +1) (\Delta +\ell )\mathcal{D}_{1, 0, 0}\nn\\
&+2 (-\Delta +\ell +1) (\Delta +\ell )\mathcal{D}_{0, 1, 0}+4 \left(\Delta ^2-3 \Delta +\ell ^2+\ell -6\right)\mathcal{D}_{1, 1, 0}\nn\\
&+(\Delta ^4-6 \Delta ^3-5 \Delta ^2+42 \Delta +\ell ^4+2 \ell ^3-17 \ell ^2-18 \ell +72)\mathcal{D}_{0, 0, 1}\nn\\
&+2 (-\Delta +\ell +1) (\Delta +\ell )\mathcal{D}_{1, 0, 1}\nn\\
&+\frac{1}{8} \left(-\Delta ^6+9 \Delta ^5-13 \Delta ^4-57 \Delta ^3+86 \Delta ^2+120 \Delta -\ell ^6-3 \ell ^5-\Delta ^2 \ell ^4+3 \Delta  \ell ^4\right.\nn\\&\quad\quad\left.+15 \ell ^4-2 \Delta ^2 \ell ^3+6 \Delta  \ell ^3+35 \ell ^3-\Delta ^4 \ell ^2+6 \Delta ^3 \ell ^2+6 \Delta ^2 \ell ^2-45 \Delta  \ell ^2-54 \ell ^2\right.\nn\\&\quad\quad\left.-\Delta ^4 \ell +6 \Delta ^3 \ell +7 \Delta ^2 \ell -48 \Delta  \ell -72 \ell \right)\mathcal{D}_{0, 0, 2}.
\end{align}

\subsection{Parity-odd structures in differential basis}
To construct the differential operators for parity-odd tensor structures, we use the differential operators derived in~\cite{Costa:2011dw},
\begin{align}
	Q_1&=\epsilon\left(Z_1,Z_2,X_1,X_2,\pdr{}{X_1}\right),\\
	Q_2&=\epsilon\left(Z_1,Z_2,X_1,X_2,\pdr{}{X_2}\right),\\
	\widetilde{D}_1&=\epsilon\left(Z_1,X_1,\pdr{}{X_1},X_2,\pdr{}{X_2}\right),\\
	\widetilde{D}_2&=\epsilon\left(Z_2,X_2,\pdr{}{X_2},X_1,\pdr{}{X_1}\right).
\end{align}
Note that the operators $\widetilde{D}_i$ satisfy all consistency conditions of~\cite{Costa:2011dw} only when operators $1$ and $2$ have spin $0$.\footnote{In~\cite{Costa:2011dw} these operators are defined with extra terms containing derivatives in polarizations. However, even with that definition $\widetilde{D}_1$ does not commute with $X_1\cdot \pdr{}{Z_1}$ and one needs to add extra terms to ensure full consistency for action on generic operators.}

Using these, we can define the operators
\begin{align}
	E_{13}&=\widetilde{D}_1,\\
	E_{23}&=\widetilde{D}_2,\\
	E_{12}&=\frac{1}{2}\left(Q_1\Sigma_1^1+Q_2\Sigma_2^1\right).
\end{align}
We define the basis of parity-odd differential operators for $\<TT\cO_\ell\>$ as
\begin{align}
	\cD_{n_{23},n_{13},n_{12},1}^-&=\cD_{n_{23},n_{13},n_{12}}E_{23},\\
	\cD_{n_{23},n_{13},n_{12},2}^-&=\cD_{n_{23},n_{13},n_{12}}E_{13},\\
	\cD_{n_{23},n_{13},n_{12},3}^-&=\cD_{n_{23},n_{13},n_{12}}E_{12}.
\end{align}
Here $\cD_{n_{23},n_{13},n_{12}}$ are the parity-even differential operators with $m_1,m_2$ defined depending on which $E_{ij}$ it multiplies so that the total spins at points $1$ and $2$ agree.

\paragraph{Structures for $\<TT\cO_0\>$}
There exists a unique parity-odd tensor structure for $\<TT\cO_{0}\>$, given by the differential operator
\begin{align}
\tilde{\mathcal{D}}^{(1)}_{\mathbf{0}^-}&=-4\mathcal{D}^-_{0, 0, 0, 3}+(\Delta -4) (\Delta +1)\mathcal{D}^-_{0, 0, 1, 3}.
\end{align}
There is a slight complication in this case, since the transition matrix between the differential and algebraic bases vanishes at $\De=1$. Thus any differential basis structure with polynomial coefficients vanishes for $\De=1$, which is undesirable since we would like to have a non-zero conformal block for every $\De\geq 1/2$. We therefore in this case consider the non-polynomial solution given by
\be
 \mathcal{D}^{(1)}_{\mathbf{0}^-}=\frac{1}{\De-1}\tilde{\mathcal{D}}^{(1)}_{\mathbf{0}^-}.
\ee
In practice, we work with $\tilde{\mathcal{D}}^{(1)}_{\mathbf{0}^-}$ and only in the end divide the numerator of the resulting rational approximation to the parity-odd scalar block by $(\De-1)^2$.\footnote{We need the square since there are left and right three-point structures.} The construction guarantees that this division is possible.

\paragraph{Structures for $\<TT\cO_2\>$}
There exists a unique parity-odd tensor structure for $\<TT\cO_{2}\>$, given by the differential operator
\begin{align}
\mathcal{D}^{(1)}_{\mathbf{2}^-}=&-4\mathcal{D}^-_{0, 1, 0, 1}-2 (\Delta -2) (\Delta +3)\mathcal{D}^-_{0, 1, 0, 3}+(\Delta ^4-6 \Delta ^3-13 \Delta ^2+66 \Delta +144)\mathcal{D}^-_{0, 0, 1, 3}\nn\\
&+2 (\Delta -6) (\Delta +2)\mathcal{D}^-_{0, 1, 1, 1}-(4)\mathcal{D}^-_{1, 0, 0, 2}-2 (\Delta -2) (\Delta +3)\mathcal{D}^-_{1, 0, 0, 3}+8 (\Delta +6)\mathcal{D}^-_{1, 1, 0, 3}\nn\\
&+2 (\Delta -6) (\Delta +2)\mathcal{D}^-_{1, 0, 1, 2}.
\end{align}

\paragraph{Structures for $\<TT\cO_\ell\>$ for even $\ell$}
There exists a unique parity-odd tensor structure for $\<TT\cO_{\ell}\>$ for even $\ell\geq 4$, given by the differential operator
\begin{align}
\mathcal{D}^{(1)}_{\mathbf{\ell}^-}=&8 \left(-3 \Delta ^2+9 \Delta +\ell ^2+\ell +24\right)\mathcal{D}^-_{0, 0, 0, 1}-16 (\Delta -4) (\Delta +1)\mathcal{D}^-_{0, 0, 0, 2}\nn\\
&-16 \left(\Delta ^4-6 \Delta ^3-\Delta ^2+30 \Delta +\Delta ^2 \ell ^2-3 \Delta  \ell ^2-4 \ell ^2+\Delta ^2 \ell -3 \Delta  \ell -4 \ell \right)\mathcal{D}^-_{0, 0, 0, 3}\nn\\
&+16 \left(\ell ^2+\ell +6\right)\mathcal{D}^-_{0, 1, 0, 1}+8 (\ell -\Delta ) (\Delta +\ell +1)\mathcal{D}^-_{0, 1, 0, 2}\nn\\
&+8 \left(\Delta ^4-6 \Delta ^3-9 \Delta ^2+54 \Delta +44\right)\mathcal{D}^-_{0, 1, 0, 3}\nn\\
&+4 \left(\Delta ^4-6 \Delta ^3-7 \Delta ^2+48 \Delta +\Delta ^2 \ell ^2-3 \Delta  \ell ^2+\Delta ^2 \ell -3 \Delta  \ell +72\right)\mathcal{D}^-_{0, 2, 0, 1}\nn\\
&+4 \left(\Delta ^6-9 \Delta ^5+13 \Delta ^4+57 \Delta ^3-86 \Delta ^2-120 \Delta +\Delta ^2 \ell ^4\right.\nn\\&\quad\quad\left.-3 \Delta  \ell ^4+2 \Delta ^2 \ell ^3-6 \Delta  \ell ^3+2 \Delta ^4 \ell ^2-12 \Delta ^3 \ell ^2-11 \Delta ^2 \ell ^2\right.\nn\\&\quad\quad\left.+87 \Delta  \ell ^2+40 \ell ^2+2 \Delta ^4 \ell -12 \Delta ^3 \ell -12 \Delta ^2 \ell +90 \Delta  \ell +40 \ell \right)\mathcal{D}^-_{0, 0, 1, 1}\nn\\
&-4 (\Delta -3) \Delta  \left(\Delta ^2-3 \Delta +\ell ^2+\ell -16\right)\mathcal{D}^-_{0, 0, 1, 2}+8 (\ell -\Delta ) (\Delta +\ell +1)\mathcal{D}^-_{0, 0, 1, 3}\nn\\
&+16 \left(\Delta ^2-3 \Delta +\ell ^2+\ell -10\right)\mathcal{D}^-_{0, 1, 1, 1}+8 \left(\Delta ^2-3 \Delta +\ell ^2+\ell -16\right)\mathcal{D}^-_{1, 0, 0, 1}.
\end{align}
\paragraph{Structures for $\<TT\cO_\ell\>$ for odd $\ell$}
There exists a unique parity-odd tensor structure for $\<TT\cO_{\ell}\>$ for odd $\ell\geq 5$, given by the differential operator
\begin{align}
\mathcal{D}^{(1)}_{\mathbf{\ell}^-}=&-4 (\Delta -2) (\Delta -1) \left(\Delta ^2-3 \Delta -3 \ell ^2-3 \ell +32\right)\mathcal{D}^-_{0, 0, 0, 1}\nn\\
&+8 (\ell -3) (\ell -1) (\ell +2) (\ell +4)\mathcal{D}^-_{0, 0, 0, 2}\nn\\
&+8 \ell  (\ell +1) \left(-6 \Delta ^2+18 \Delta +\ell ^4+2 \ell ^3+\Delta ^2 \ell ^2-3 \Delta  \ell ^2-11 \ell ^2\right.\nn\\&\quad\quad\left.+\Delta ^2 \ell -3 \Delta  \ell -12 \ell +12\right)\mathcal{D}^-_{0, 0, 0, 3}\nn\\
&-8 \left(-\Delta ^4+6 \Delta ^3-25 \Delta ^2+48 \Delta +\ell ^4+2 \ell ^3+\Delta ^2 \ell ^2-3 \Delta  \ell ^2\right.\nn\\&\quad\quad\left.-11 \ell ^2+\Delta ^2 \ell -3 \Delta  \ell -12 \ell -4\right)\mathcal{D}^-_{0, 1, 0, 1}\nn\\
&+4 (\Delta -2) (\Delta -1) (\ell -\Delta ) (\Delta +\ell +1)\mathcal{D}^-_{0, 1, 0, 2}\nn\\
&-4 (\Delta -2) (\Delta -1) \left(\ell ^4+2 \ell ^3-21 \ell ^2-22 \ell +84\right)\mathcal{D}^-_{0, 1, 0, 3}\nn\\
&-2 \left(\ell ^6+3 \ell ^5+\Delta ^2 \ell ^4-3 \Delta  \ell ^4-15 \ell ^4+2 \Delta ^2 \ell ^3-6 \Delta  \ell ^3-35 \ell ^3-17 \Delta ^2 \ell ^2\right.\nn\\&\quad\quad\left.+51 \Delta  \ell ^2+54 \ell ^2-18 \Delta ^2 \ell +54 \Delta  \ell +72 \ell -144\right)\mathcal{D}^-_{0, 2, 0, 1}\nn\\
&-2 \ell  (\ell +1) \left(-2 \Delta ^4+12 \Delta ^3+82 \Delta ^2-300 \Delta +\ell ^6+3 \ell ^5+2 \Delta ^2 \ell ^4-6 \Delta  \ell ^4\right.\nn\\&\quad\quad\left.-13 \ell ^4+4 \Delta ^2 \ell ^3-12 \Delta  \ell ^3-31 \ell ^3+\Delta ^4 \ell ^2-6 \Delta ^3 \ell ^2-23 \Delta ^2 \ell ^2+96 \Delta  \ell ^2\right.\nn\\&\quad\quad\left.+20 \ell ^2+\Delta ^4 \ell -6 \Delta ^3 \ell -25 \Delta ^2 \ell +102 \Delta  \ell +36 \ell +64\right)\mathcal{D}^-_{0, 0, 1, 1}\nn\\
&+2 (\ell -3) (\ell -2) (\ell +3) (\ell +4) \left(\Delta ^2-3 \Delta +\ell ^2+\ell \right)\mathcal{D}^-_{0, 0, 1, 2}\nn\\
&-4 (\Delta -2) (\Delta -1) (\ell -\Delta ) (\Delta +\ell +1)\mathcal{D}^-_{0, 0, 1, 3}\nn\\
&-8 (\Delta -2) (\Delta -1) \left(\Delta ^2-3 \Delta +\ell ^2+\ell -10\right)\mathcal{D}^-_{0, 1, 1, 1}\nn\\
&+4 (\Delta -2) (\Delta -1) \left(\Delta ^2-3 \Delta +\ell ^2+\ell -16\right)\mathcal{D}^-_{1, 0, 0, 1}.
\end{align}

\section{Details on the numerics}
\label{app:numerics}
In this appendix we give specific details on how the bounds in this paper are obtained from the crossing equations~\eqref{eq:twovareqns}-\eqref{eq:zerovareqns} and the conformal block decomposition~\eqref{eq:TTTTbootstrapNorm}. 

First, we organize the crossing equations~\eqref{eq:twovareqns}-\eqref{eq:zerovareqns} in a single vector equation 
\be
	\vec F_{TTTT}=0.
\ee
The conformal block decomposition~\eqref{eq:TTTTbootstrapNorm} then induces a decomposition of the vector $\vec F_{TTTT}$,
\be\label{eq:cbequation}
	\vec F_{TTTT}=\vec F_{\mathbf{1}}+\frac{1}{C_T}\Theta^{ab}\vec F_{T,ab}+\sum_{(\De,\rho)\in S}M^{ab}_{\De,\rho} \vec F_{\De,\rho,ab}=0.
\ee
Here we have explicitly specified that the summation is over some assumed set of dimensions and spins $S$. This equation has to be satisfied in any theory whose spectrum of operators is a subset of $S$. For example, when we say that we impose a gap $\De_\text{even}^\text{min}$ in the parity-even scalar sector, we choose
\begin{align}
	S=&\{(\De,\Bell^+)|\De\geq \ell+1,\ell=2k\geq 2\}\,\cup\nn\\
	&\{(\De,\Bell^-)|\De\geq \ell+1,\ell\geq 4\}\,\cup\nn\\
	&\{(\De,\mathbf{2}^-)|\De\geq 3\}\,\cup\nn\\
	&\{(\De,\mathbf{0}^+)|\De\geq \De_\text{even}^\text{min}\}\,\cup\nn\\
	&\{(\De,\mathbf{0}^-)|\De\geq \tfrac{1}{2}\}.	
\end{align}
Given a choice of $S$, we then study two questions:
\begin{enumerate}
	\item \qquad {\bf Feasibility:} Does the system~\eqref{eq:cbequation} have a solution for some $\theta$?
	\item \qquad {\bf Optimization:} What is the minimal (maximal) value of $C_T$ for a given $\theta$?
\end{enumerate} 

\paragraph{Feasibility:}
To answer the feasibility question, we look for a vector $\vec\alpha$ such that
\begin{align}
	\vec\alpha\cdot \vec F_{\mathbf{1}}&=1,\label{eq:funcIdconstr}\\
	\vec\alpha\cdot \vec F_{T}&\succeq 0,\\
	\vec\alpha\cdot \vec F_{\De,\rho}&\succeq 0,\quad \forall(\De,\rho)\in S.	\label{eq:funcSconstr}
\end{align}
Clearly, if such $\vec\alpha$ is found, then there cannot be a solution to~\eqref{eq:cbequation}, since positive-semidefiniteness of $M_{\De,\rho}$, $\Theta$ and $C_T>0$ imply
\be\label{eq:Fcontradiction}
	\vec\alpha\cdot \vec F_{TTTT}\geq 1.
\ee
We then conclude that CFTs with the spectral assumption $S$ do not exist. As usual, this conclusion is rigorous for any $\Lambda$, given that the equations~\eqref{eq:funcIdconstr}-\eqref{eq:funcSconstr} are satisfied to a sufficient precision. If such an $\vec\alpha$ cannot be found, we cannot conclude anything and the spectral assumption $S$ is formally ``allowed'' by our bounds.

\paragraph{Optimization:}
Let us start with the case that we want to find a lower bound on $C_T$ for a given $\theta$. Suppose that we have found a vector $\vec\alpha$ such that 
\begin{align}\label{eq:funcMin1}
\vec\alpha\cdot \vec F_{\mathbf{1}}&=-1,\\
\vec\alpha\cdot \vec F_{\De,\rho}&\succeq 0,\quad \forall(\De,\rho)\in S.\label{eq:funcMin2}
\end{align}
It then follows from $\vec F_{TTTT}=0$ that
\be
	 -1+\frac{1}{C_T}\vec\alpha\cdot(\Theta^{ab}\vec F_{T,ab})\leq 0,
\ee
and thus
\be
	C_T\geq \vec\alpha\cdot(\Theta^{ab}\vec F_{T,ab}).
\ee
We then search for an $\vec\alpha$ which maximizes 
\be\label{eq:objective}
	\vec\alpha\cdot(\Theta^{ab}\vec F_{T,ab})
\ee
subject to~\eqref{eq:funcMin1} and \eqref{eq:funcMin2} in order to find the optimal bound. Again, the bounds are rigorous for every $\Lambda$.

If our goal is to find an upper bound on $C_T$, we replace~\eqref{eq:funcMin1} with
\be
\vec\alpha\cdot \vec F_{\mathbf{1}}=+1,
\ee
which then analogously implies
\be
C_T\leq -\vec\alpha\cdot(\Theta^{ab}\vec F_{T,ab}).
\ee
We again look for such $\vec\alpha$ which maximizes~\eqref{eq:objective} in order to find the optimal bound.

\paragraph{Numerical implementation:}
To search for the vectors $\alpha$ we use the semidefinite solver \texttt{SDPB}~\cite{Simmons-Duffin:2015qma}. In section~\ref{sec:conformalblocks} we explained how to obtain rational approximations of the $\<TTTT\>$ conformal blocks required by \texttt{SDPB} starting from rational approximations of scalar conformal blocks arising from their pole expansions~\cite{Kos:2014bka,Penedones:2015aga}.

These approximations are controlled by the integral parameter~$\kappa$ defined in~\cite{Simmons-Duffin:2015qma}. The blocks become exact in the limit~$\kappa\to\infty$; the convergence is exponential. In practice we use a finite value of $\kappa$ and check that our results don't change if $\kappa$ is increased. Another approximation that we have to make is the truncation to a finite range of spins in constraints~\eqref{eq:funcSconstr} and~\eqref{eq:funcMin2}. Again, we choose a sufficiently large cutoff and check that the results are independent of it.

Below we list $\kappa$, the spin cutoff, and the relevant \texttt{SDPB} parameters that we used in calculations for various values of $\Lambda$ (all figures except figure~\ref{fig:hofmanmaldacena} correspond to $\Lambda=19$):
\begin{center}
\begin{tabular}{l|c|c|c|c|c}
	$\Lambda$ 								& $\leq 11$ & $13$ & $15$ & $17$ & $19$ \\
	\hline
	$\kappa$								& 20 & 24 & 24 & 24 & 24 \\
	spins									& $\leq 25$ & $\leq 30$ & $\leq 36$ & $\leq 42$ & $\leq 42$ \\
	\texttt{precision} 						& 832 & 832 & 832 & 832 & 1024 \\
	\texttt{findPrimalFeasible}				& False& False& False& False& False\\
	\texttt{findDualFeasible}				& False& False& False& False& False\\
	\texttt{detectPrimalFeasibleJump}		& False& False& False& False& False\\
	\texttt{detectDualFeasibleJump}			& False& False& False& False& False\\
	\texttt{dualityGapThreshold}			& $10^{-10}$& $10^{-10}$& $10^{-10}$& $10^{-10}$& $10^{-10}$\\
	\texttt{primalErrorThreshold}			& $10^{-30}$& $10^{-30}$& $10^{-30}$& $10^{-30}$& $10^{-30}$\\
	\texttt{dualErrorThreshold}				& $10^{-30}$& $10^{-30}$& $10^{-30}$& $10^{-30}$& $10^{-30}$\\
	\texttt{initialMatrixScalePrimal}		& $10^{20}$& $10^{20}$& $10^{20}$& $10^{20}$& $10^{20}$\\
	\texttt{initialMatrixScaleDual}			& $10^{20}$& $10^{20}$& $10^{20}$& $10^{20}$& $10^{20}$\\
	\texttt{feasibleCenteringParameter}		& $0.1$& $0.1$& $0.1$& $0.1$& $0.1$\\
	\texttt{infeasibleCenteringParameter}	& $0.3$ & $0.3$ & $0.3$ & $0.3$ & $0.3$ \\
	\texttt{stepLengthReduction}			& $0.7$& $0.7$& $0.7$& $0.7$& $0.7$\\
	\texttt{choleskyStabilizeThreshold}		& $10^{-120}$ & $10^{-120}$ & $10^{-120}$ & $10^{-120}$ & $10^{-180}$ \\
	\texttt{maxComplementarity}				& $10^{100}$& $10^{100}$& $10^{100}$& $10^{100}$& $10^{100}$
\end{tabular}
\end{center}
The exclusion plot in figure~\ref{fig:scalarGapsExclusionPlot} requires testing only feasibility so we set \texttt{findPrimalFeasible}	and \texttt{findDualFeasible} to True. For the scalar bound in figure~\ref{fig:scalarGapsExclusionPlot} we used the parameters of~\cite{Simmons-Duffin:2015qma} with $\Lambda=35$. The stress-tensor conformal blocks as well as the code used for their generation and setting up~\texttt{SDPB} are available upon request.

\bibliography{Biblio}

\providecommand{\href}[2]{#2}\begingroup\raggedright\begin{thebibliography}{10}

\bibitem{Ferrara:1973yt}
S.~Ferrara, A.~F. Grillo, and R.~Gatto, ``{Tensor representations of conformal
  algebra and conformally covariant operator product expansion},''
\href{http://dx.doi.org/10.1016/0003-4916(73)90446-6}{{\em Annals Phys.}
  {\bfseries 76} (1973) 161--188}.
%%CITATION = APNYA,76,161;%%.

\bibitem{Polyakov:1974gs}
A.~M. Polyakov, ``{Nonhamiltonian approach to conformal quantum field
  theory},''
{\em Zh. Eksp. Teor. Fiz.} {\bfseries 66} (1974) 23--42.
%%CITATION = ZETFA,66,23;%%.

\bibitem{Mack:1975jr}
G.~Mack, ``{Duality in quantum field theory},''
\href{http://dx.doi.org/10.1016/0550-3213(77)90238-3}{{\em Nucl. Phys.}
  {\bfseries B118} (1977) 445--457}.
%%CITATION = NUPHA,B118,445;%%.

\bibitem{Rattazzi:2008pe}
R.~Rattazzi, V.~S. Rychkov, E.~Tonni, and A.~Vichi, ``{Bounding scalar operator
  dimensions in 4D CFT},''
  \href{http://dx.doi.org/10.1088/1126-6708/2008/12/031}{{\em JHEP} {\bfseries
  12} (2008) 031},
\href{http://arxiv.org/abs/0807.0004}{{\ttfamily arXiv:0807.0004 [hep-th]}}.
%%CITATION = 0807.0004;%%.

\bibitem{Rychkov:2016iqz}
S.~Rychkov, \href{http://dx.doi.org/10.1007/978-3-319-43626-5}{{\em {EPFL
  Lectures on Conformal Field Theory in D>= 3 Dimensions}}}.
\newblock SpringerBriefs in Physics. 2016.
\newblock \href{http://arxiv.org/abs/1601.05000}{{\ttfamily arXiv:1601.05000
  [hep-th]}}.
\newblock
\url{http://inspirehep.net/record/1415968/files/arXiv:1601.05000.pdf}.
\newblock
%%CITATION = ARXIV:1601.05000;%%.

\bibitem{Simmons-Duffin:2016gjk}
D.~Simmons-Duffin, \href{http://dx.doi.org/10.1142/9789813149441_0001}{``{The
  Conformal Bootstrap},''} in {\em {Proceedings, Theoretical Advanced Study
  Institute in Elementary Particle Physics: New Frontiers in Fields and Strings
  (TASI 2015): Boulder, CO, USA, June 1-26, 2015}}, pp.~1--74.
\newblock 2017.
\newblock \href{http://arxiv.org/abs/1602.07982}{{\ttfamily arXiv:1602.07982
  [hep-th]}}.
\newblock
\url{http://inspirehep.net/record/1424282/files/arXiv:1602.07982.pdf}.
\newblock
%%CITATION = ARXIV:1602.07982;%%.

\bibitem{Poland:2016chs}
D.~Poland and D.~Simmons-Duffin, ``{The conformal bootstrap},''
\href{http://dx.doi.org/10.1038/nphys3761}{{\em Nature Phys.} {\bfseries 12}
  no.~6, (2016) 535--539}.
%%CITATION = NPAHA,12,535;%%.

\bibitem{ElShowk:2012ht}
S.~El-Showk, M.~F. Paulos, D.~Poland, S.~Rychkov, D.~Simmons-Duffin, and
  A.~Vichi, ``{Solving the 3D Ising Model with the Conformal Bootstrap},''
  \href{http://dx.doi.org/10.1103/PhysRevD.86.025022}{{\em Phys.Rev.}
  {\bfseries D86} (2012) 025022},
\href{http://arxiv.org/abs/1203.6064}{{\ttfamily arXiv:1203.6064 [hep-th]}}.
%%CITATION = ARXIV:1203.6064;%%.

\bibitem{El-Showk:2014dwa}
S.~El-Showk, M.~F. Paulos, D.~Poland, S.~Rychkov, D.~Simmons-Duffin, {\em
  et~al.}, ``{Solving the 3d Ising Model with the Conformal Bootstrap II.
  c-Minimization and Precise Critical Exponents},''
  \href{http://dx.doi.org/10.1007/s10955-014-1042-7}{{\em J.Stat.Phys.}
  {\bfseries 157} (2014) 869},
\href{http://arxiv.org/abs/1403.4545}{{\ttfamily arXiv:1403.4545 [hep-th]}}.
%%CITATION = ARXIV:1403.4545;%%.

\bibitem{Gliozzi:2014jsa}
F.~Gliozzi and A.~Rago, ``{Critical exponents of the 3d Ising and related
  models from Conformal Bootstrap},''
  \href{http://dx.doi.org/10.1007/JHEP10(2014)042}{{\em JHEP} {\bfseries 1410}
  (2014) 42},
\href{http://arxiv.org/abs/1403.6003}{{\ttfamily arXiv:1403.6003 [hep-th]}}.
%%CITATION = ARXIV:1403.6003;%%.

\bibitem{Kos:2014bka}
F.~Kos, D.~Poland, and D.~Simmons-Duffin, ``{Bootstrapping Mixed Correlators in
  the 3D Ising Model},'' \href{http://dx.doi.org/10.1007/JHEP11(2014)109}{{\em
  JHEP} {\bfseries 1411} (2014) 109},
\href{http://arxiv.org/abs/1406.4858}{{\ttfamily arXiv:1406.4858 [hep-th]}}.
%%CITATION = ARXIV:1406.4858;%%.

\bibitem{Kos:2016ysd}
F.~Kos, D.~Poland, D.~Simmons-Duffin, and A.~Vichi, ``{Precision Islands in the
  Ising and $O(N)$ Models},''
  \href{http://dx.doi.org/10.1007/JHEP08(2016)036}{{\em JHEP} {\bfseries 08}
  (2016) 036},
\href{http://arxiv.org/abs/1603.04436}{{\ttfamily arXiv:1603.04436 [hep-th]}}.
%%CITATION = ARXIV:1603.04436;%%.

\bibitem{Simmons-Duffin:2016wlq}
D.~Simmons-Duffin, ``{The Lightcone Bootstrap and the Spectrum of the 3d Ising
  CFT},'' \href{http://dx.doi.org/10.1007/JHEP03(2017)086}{{\em JHEP}
  {\bfseries 03} (2017) 086},
\href{http://arxiv.org/abs/1612.08471}{{\ttfamily arXiv:1612.08471 [hep-th]}}.
%%CITATION = ARXIV:1612.08471;%%.

\bibitem{Rattazzi:2010yc}
R.~Rattazzi, S.~Rychkov, and A.~Vichi, ``{Bounds in 4D Conformal Field Theories
  with Global Symmetry},''
  \href{http://dx.doi.org/10.1088/1751-8113/44/3/035402}{{\em J. Phys.}
  {\bfseries A44} (2011) 035402},
\href{http://arxiv.org/abs/1009.5985}{{\ttfamily arXiv:1009.5985 [hep-th]}}.
%%CITATION = 1009.5985;%%.

\bibitem{Kos:2013tga}
F.~Kos, D.~Poland, and D.~Simmons-Duffin, ``{Bootstrapping the $O(N)$ vector
  models},'' \href{http://dx.doi.org/10.1007/JHEP06(2014)091}{{\em JHEP}
  {\bfseries 1406} (2014) 091},
\href{http://arxiv.org/abs/1307.6856}{{\ttfamily arXiv:1307.6856 [hep-th]}}.
%%CITATION = ARXIV:1307.6856;%%.

\bibitem{Chester:2014gqa}
S.~M. Chester, S.~S. Pufu, and R.~Yacoby, ``{Bootstrapping $O(N)$ vector models
  in 4 $< d <$ 6},'' \href{http://dx.doi.org/10.1103/PhysRevD.91.086014}{{\em
  Phys. Rev.} {\bfseries D91} no.~8, (2015) 086014},
\href{http://arxiv.org/abs/1412.7746}{{\ttfamily arXiv:1412.7746 [hep-th]}}.
%%CITATION = ARXIV:1412.7746;%%.

\bibitem{Kos:2015mba}
F.~Kos, D.~Poland, D.~Simmons-Duffin, and A.~Vichi, ``{Bootstrapping the O(N)
  Archipelago},'' \href{http://dx.doi.org/10.1007/JHEP11(2015)106}{{\em JHEP}
  {\bfseries 11} (2015) 106},
\href{http://arxiv.org/abs/1504.07997}{{\ttfamily arXiv:1504.07997 [hep-th]}}.
%%CITATION = ARXIV:1504.07997;%%.

\bibitem{Iliesiu:2015qra}
L.~Iliesiu, F.~Kos, D.~Poland, S.~S. Pufu, D.~Simmons-Duffin, and R.~Yacoby,
  ``{Bootstrapping 3D Fermions},''
  \href{http://dx.doi.org/10.1007/JHEP03(2016)120}{{\em JHEP} {\bfseries 03}
  (2016) 120},
\href{http://arxiv.org/abs/1508.00012}{{\ttfamily arXiv:1508.00012 [hep-th]}}.
%%CITATION = ARXIV:1508.00012;%%.

\bibitem{Iliesiu:2017nrv}
L.~Iliesiu, F.~Kos, D.~Poland, S.~S. Pufu, and D.~Simmons-Duffin,
  ``{Bootstrapping 3D Fermions with Global Symmetries},''
\href{http://arxiv.org/abs/1705.03484}{{\ttfamily arXiv:1705.03484 [hep-th]}}.
%%CITATION = ARXIV:1705.03484;%%.

\bibitem{Beem:2013qxa}
C.~Beem, L.~Rastelli, and B.~C. van Rees, ``{The $\mathcal{N}=4$ Superconformal
  Bootstrap},'' \href{http://dx.doi.org/10.1103/PhysRevLett.111.071601}{{\em
  Phys.Rev.Lett.} {\bfseries 111} (2013) 071601},
\href{http://arxiv.org/abs/1304.1803}{{\ttfamily arXiv:1304.1803 [hep-th]}}.
%%CITATION = ARXIV:1304.1803;%%.

\bibitem{Chester:2014fya}
S.~M. Chester, J.~Lee, S.~S. Pufu, and R.~Yacoby, ``{The $ \mathcal{N}=8 $
  superconformal bootstrap in three dimensions},''
  \href{http://dx.doi.org/10.1007/JHEP09(2014)143}{{\em JHEP} {\bfseries 1409}
  (2014) 143},
\href{http://arxiv.org/abs/1406.4814}{{\ttfamily arXiv:1406.4814 [hep-th]}}.
%%CITATION = ARXIV:1406.4814;%%.

\bibitem{Beem:2014zpa}
C.~Beem, M.~Lemos, P.~Liendo, L.~Rastelli, and B.~C. van Rees, ``{The $
  \mathcal{N}=2 $ superconformal bootstrap},''
  \href{http://dx.doi.org/10.1007/JHEP03(2016)183}{{\em JHEP} {\bfseries 03}
  (2016) 183},
\href{http://arxiv.org/abs/1412.7541}{{\ttfamily arXiv:1412.7541 [hep-th]}}.
%%CITATION = ARXIV:1412.7541;%%.

\bibitem{Bobev:2015vsa}
N.~Bobev, S.~El-Showk, D.~Mazac, and M.~F. Paulos, ``{Bootstrapping the
  Three-Dimensional Supersymmetric Ising Model},''
  \href{http://dx.doi.org/10.1103/PhysRevLett.115.051601}{{\em Phys. Rev.
  Lett.} {\bfseries 115} no.~5, (2015) 051601},
\href{http://arxiv.org/abs/1502.04124}{{\ttfamily arXiv:1502.04124 [hep-th]}}.
%%CITATION = ARXIV:1502.04124;%%.

\bibitem{Bobev:2015jxa}
N.~Bobev, S.~El-Showk, D.~Mazac, and M.~F. Paulos, ``{Bootstrapping SCFTs with
  Four Supercharges},'' \href{http://dx.doi.org/10.1007/JHEP08(2015)142}{{\em
  JHEP} {\bfseries 08} (2015) 142},
\href{http://arxiv.org/abs/1503.02081}{{\ttfamily arXiv:1503.02081 [hep-th]}}.
%%CITATION = ARXIV:1503.02081;%%.

\bibitem{Chester:2015qca}
S.~M. Chester, S.~Giombi, L.~V. Iliesiu, I.~R. Klebanov, S.~S. Pufu, and
  R.~Yacoby, ``{Accidental Symmetries and the Conformal Bootstrap},''
  \href{http://dx.doi.org/10.1007/JHEP01(2016)110}{{\em JHEP} {\bfseries 01}
  (2016) 110},
\href{http://arxiv.org/abs/1507.04424}{{\ttfamily arXiv:1507.04424 [hep-th]}}.
%%CITATION = ARXIV:1507.04424;%%.

\bibitem{Beem:2015aoa}
C.~Beem, M.~Lemos, L.~Rastelli, and B.~C. van Rees, ``{The (2, 0)
  superconformal bootstrap},''
  \href{http://dx.doi.org/10.1103/PhysRevD.93.025016}{{\em Phys. Rev.}
  {\bfseries D93} no.~2, (2016) 025016},
\href{http://arxiv.org/abs/1507.05637}{{\ttfamily arXiv:1507.05637 [hep-th]}}.
%%CITATION = ARXIV:1507.05637;%%.

\bibitem{Poland:2015mta}
D.~Poland and A.~Stergiou, ``{Exploring the Minimal 4D $\mathcal{N}=1$ SCFT},''
  \href{http://dx.doi.org/10.1007/JHEP12(2015)121}{{\em JHEP} {\bfseries 12}
  (2015) 121},
\href{http://arxiv.org/abs/1509.06368}{{\ttfamily arXiv:1509.06368 [hep-th]}}.
%%CITATION = ARXIV:1509.06368;%%.

\bibitem{Lemos:2015awa}
M.~Lemos and P.~Liendo, ``{Bootstrapping $ \mathcal{N}=2 $ chiral
  correlators},'' \href{http://dx.doi.org/10.1007/JHEP01(2016)025}{{\em JHEP}
  {\bfseries 01} (2016) 025},
\href{http://arxiv.org/abs/1510.03866}{{\ttfamily arXiv:1510.03866 [hep-th]}}.
%%CITATION = ARXIV:1510.03866;%%.

\bibitem{Chester:2015lej}
S.~M. Chester, L.~V. Iliesiu, S.~S. Pufu, and R.~Yacoby, ``{Bootstrapping
  $O(N)$ Vector Models with Four Supercharges in $3 \leq d \leq4$},''
  \href{http://dx.doi.org/10.1007/JHEP05(2016)103}{{\em JHEP} {\bfseries 05}
  (2016) 103},
\href{http://arxiv.org/abs/1511.07552}{{\ttfamily arXiv:1511.07552 [hep-th]}}.
%%CITATION = ARXIV:1511.07552;%%.

\bibitem{Lin:2015wcg}
Y.-H. Lin, S.-H. Shao, D.~Simmons-Duffin, Y.~Wang, and X.~Yin, ``{$ \mathcal{N}
  $ = 4 superconformal bootstrap of the K3 CFT},''
  \href{http://dx.doi.org/10.1007/JHEP05(2017)126}{{\em JHEP} {\bfseries 05}
  (2017) 126},
\href{http://arxiv.org/abs/1511.04065}{{\ttfamily arXiv:1511.04065 [hep-th]}}.
%%CITATION = ARXIV:1511.04065;%%.

\bibitem{Beem:2016wfs}
C.~Beem, L.~Rastelli, and B.~C. van Rees, ``{More ${\mathcal N}=4$
  superconformal bootstrap},''
\href{http://arxiv.org/abs/1612.02363}{{\ttfamily arXiv:1612.02363 [hep-th]}}.
%%CITATION = ARXIV:1612.02363;%%.

\bibitem{Lemos:2016xke}
M.~Lemos, P.~Liendo, C.~Meneghelli, and V.~Mitev, ``{Bootstrapping
  $\mathcal{N}=3$ superconformal theories},''
  \href{http://dx.doi.org/10.1007/JHEP04(2017)032}{{\em JHEP} {\bfseries 04}
  (2017) 032},
\href{http://arxiv.org/abs/1612.01536}{{\ttfamily arXiv:1612.01536 [hep-th]}}.
%%CITATION = ARXIV:1612.01536;%%.

\bibitem{Li:2017ddj}
D.~Li, D.~Meltzer, and A.~Stergiou, ``{Bootstrapping mixed correlators in 4D $
  \mathcal{N} $ = 1 SCFTs},''
  \href{http://dx.doi.org/10.1007/JHEP07(2017)029}{{\em JHEP} {\bfseries 07}
  (2017) 029},
\href{http://arxiv.org/abs/1702.00404}{{\ttfamily arXiv:1702.00404 [hep-th]}}.
%%CITATION = ARXIV:1702.00404;%%.

\bibitem{Rychkov:2009ij}
V.~S. Rychkov and A.~Vichi, ``{Universal Constraints on Conformal Operator
  Dimensions},'' \href{http://dx.doi.org/10.1103/PhysRevD.80.045006}{{\em Phys.
  Rev.} {\bfseries D80} (2009) 045006},
\href{http://arxiv.org/abs/0905.2211}{{\ttfamily arXiv:0905.2211 [hep-th]}}.
%%CITATION = 0905.2211;%%.

\bibitem{Caracciolo:2009bx}
F.~Caracciolo and V.~S. Rychkov, ``{Rigorous Limits on the Interaction Strength
  in Quantum Field Theory},''
  \href{http://dx.doi.org/10.1103/PhysRevD.81.085037}{{\em Phys. Rev.}
  {\bfseries D81} (2010) 085037},
\href{http://arxiv.org/abs/0912.2726}{{\ttfamily arXiv:0912.2726 [hep-th]}}.
%%CITATION = 0912.2726;%%.

\bibitem{Poland:2010wg}
D.~Poland and D.~Simmons-Duffin, ``{Bounds on 4D Conformal and Superconformal
  Field Theories},'' \href{http://dx.doi.org/10.1007/JHEP05(2011)017}{{\em
  JHEP} {\bfseries 1105} (2011) 017},
\href{http://arxiv.org/abs/1009.2087}{{\ttfamily arXiv:1009.2087 [hep-th]}}.
%%CITATION = ARXIV:1009.2087;%%.

\bibitem{Rattazzi:2010gj}
R.~Rattazzi, S.~Rychkov, and A.~Vichi, ``{Central Charge Bounds in 4D Conformal
  Field Theory},'' \href{http://dx.doi.org/10.1103/PhysRevD.83.046011}{{\em
  Phys. Rev.} {\bfseries D83} (2011) 046011},
\href{http://arxiv.org/abs/1009.2725}{{\ttfamily arXiv:1009.2725 [hep-th]}}.
%%CITATION = 1009.2725;%%.

\bibitem{Vichi:2011ux}
A.~Vichi, ``{Improved bounds for CFT's with global symmetries},''
  \href{http://dx.doi.org/10.1007/JHEP01(2012)162}{{\em JHEP} {\bfseries 1201}
  (2012) 162},
\href{http://arxiv.org/abs/1106.4037}{{\ttfamily arXiv:1106.4037 [hep-th]}}.
%%CITATION = ARXIV:1106.4037;%%.

\bibitem{Poland:2011ey}
D.~Poland, D.~Simmons-Duffin, and A.~Vichi, ``{Carving Out the Space of 4D
  CFTs},'' \href{http://dx.doi.org/10.1007/JHEP05(2012)110}{{\em JHEP}
  {\bfseries 1205} (2012) 110},
\href{http://arxiv.org/abs/1109.5176}{{\ttfamily arXiv:1109.5176 [hep-th]}}.
%%CITATION = ARXIV:1109.5176;%%.

\bibitem{Rychkov:2011et}
S.~Rychkov, ``{Conformal Bootstrap in Three Dimensions?},''
\href{http://arxiv.org/abs/1111.2115}{{\ttfamily arXiv:1111.2115 [hep-th]}}.
%%CITATION = 1111.2115;%%.

\bibitem{Iliesiu:2015akf}
L.~Iliesiu, F.~Kos, D.~Poland, S.~S. Pufu, D.~Simmons-Duffin, and R.~Yacoby,
  ``{Fermion-Scalar Conformal Blocks},''
  \href{http://dx.doi.org/10.1007/JHEP04(2016)074}{{\em JHEP} {\bfseries 04}
  (2016) 074},
\href{http://arxiv.org/abs/1511.01497}{{\ttfamily arXiv:1511.01497 [hep-th]}}.
%%CITATION = ARXIV:1511.01497;%%.

\bibitem{Dymarsky:2017xzb}
A.~Dymarsky, J.~Penedones, E.~Trevisani, and A.~Vichi, ``{Charting the space of
  3D CFTs with a continuous global symmetry},''
\href{http://arxiv.org/abs/1705.04278}{{\ttfamily arXiv:1705.04278 [hep-th]}}.
%%CITATION = ARXIV:1705.04278;%%.

\bibitem{Liendo:2012hy}
P.~Liendo, L.~Rastelli, and B.~C. van Rees, ``{The Bootstrap Program for
  Boundary CFT${}_d$},'' \href{http://dx.doi.org/10.1007/JHEP07(2013)113}{{\em
  JHEP} {\bfseries 1307} (2013) 113},
\href{http://arxiv.org/abs/1210.4258}{{\ttfamily arXiv:1210.4258 [hep-th]}}.
%%CITATION = ARXIV:1210.4258;%%.

\bibitem{Gaiotto:2013nva}
D.~Gaiotto, D.~Mazac, and M.~F. Paulos, ``{Bootstrapping the 3d Ising twist
  defect},'' \href{http://dx.doi.org/10.1007/JHEP03(2014)100}{{\em JHEP}
  {\bfseries 1403} (2014) 100},
\href{http://arxiv.org/abs/1310.5078}{{\ttfamily arXiv:1310.5078 [hep-th]}}.
%%CITATION = ARXIV:1310.5078;%%.

\bibitem{Gliozzi:2015qsa}
F.~Gliozzi, P.~Liendo, M.~Meineri, and A.~Rago, ``{Boundary and Interface CFTs
  from the Conformal Bootstrap},''
  \href{http://dx.doi.org/10.1007/JHEP05(2015)036}{{\em JHEP} {\bfseries 05}
  (2015) 036},
\href{http://arxiv.org/abs/1502.07217}{{\ttfamily arXiv:1502.07217 [hep-th]}}.
%%CITATION = ARXIV:1502.07217;%%.

\bibitem{Paulos:2015jfa}
M.~F. Paulos, S.~Rychkov, B.~C. van Rees, and B.~Zan, ``{Conformal Invariance
  in the Long-Range Ising Model},''
  \href{http://dx.doi.org/10.1016/j.nuclphysb.2015.10.018}{{\em Nucl. Phys.}
  {\bfseries B902} (2016) 246--291},
\href{http://arxiv.org/abs/1509.00008}{{\ttfamily arXiv:1509.00008 [hep-th]}}.
%%CITATION = ARXIV:1509.00008;%%.

\bibitem{Behan:2017dwr}
C.~Behan, L.~Rastelli, S.~Rychkov, and B.~Zan, ``{Long-range critical exponents
  near the short-range crossover},''
  \href{http://dx.doi.org/10.1103/PhysRevLett.118.241601}{{\em Phys. Rev.
  Lett.} {\bfseries 118} no.~24, (2017) 241601},
\href{http://arxiv.org/abs/1703.03430}{{\ttfamily arXiv:1703.03430
  [cond-mat.stat-mech]}}.
%%CITATION = ARXIV:1703.03430;%%.

\bibitem{Behan:2017emf}
C.~Behan, L.~Rastelli, S.~Rychkov, and B.~Zan, ``{A scaling theory for the
  long-range to short-range crossover and an infrared duality},''
  \href{http://dx.doi.org/10.1088/1751-8121/aa8099}{{\em J. Phys.} {\bfseries
  A50} no.~35, (2017) 354002},
\href{http://arxiv.org/abs/1703.05325}{{\ttfamily arXiv:1703.05325 [hep-th]}}.
%%CITATION = ARXIV:1703.05325;%%.

\bibitem{Hofman:2008ar}
D.~M. Hofman and J.~Maldacena, ``{Conformal collider physics: Energy and charge
  correlations},'' \href{http://dx.doi.org/10.1088/1126-6708/2008/05/012}{{\em
  JHEP} {\bfseries 05} (2008) 012},
\href{http://arxiv.org/abs/0803.1467}{{\ttfamily arXiv:0803.1467 [hep-th]}}.
%%CITATION = ARXIV:0803.1467;%%.

\bibitem{Buchel:2009sk}
A.~Buchel, J.~Escobedo, R.~C. Myers, M.~F. Paulos, A.~Sinha, and M.~Smolkin,
  ``{Holographic GB gravity in arbitrary dimensions},''
  \href{http://dx.doi.org/10.1007/JHEP03(2010)111}{{\em JHEP} {\bfseries 03}
  (2010) 111},
\href{http://arxiv.org/abs/0911.4257}{{\ttfamily arXiv:0911.4257 [hep-th]}}.
%%CITATION = ARXIV:0911.4257;%%.

\bibitem{Hofman:2016awc}
D.~M. Hofman, D.~Li, D.~Meltzer, D.~Poland, and F.~Rejon-Barrera, ``{A Proof of
  the Conformal Collider Bounds},''
  \href{http://dx.doi.org/10.1007/JHEP06(2016)111}{{\em JHEP} {\bfseries 06}
  (2016) 111},
\href{http://arxiv.org/abs/1603.03771}{{\ttfamily arXiv:1603.03771 [hep-th]}}.
%%CITATION = ARXIV:1603.03771;%%.

\bibitem{Hartman:2016lgu}
T.~Hartman, S.~Kundu, and A.~Tajdini, ``{Averaged Null Energy Condition from
  Causality},'' \href{http://dx.doi.org/10.1007/JHEP07(2017)066}{{\em JHEP}
  {\bfseries 07} (2017) 066},
\href{http://arxiv.org/abs/1610.05308}{{\ttfamily arXiv:1610.05308 [hep-th]}}.
%%CITATION = ARXIV:1610.05308;%%.

\bibitem{Simmons-Duffin:2015qma}
D.~Simmons-Duffin, ``{A Semidefinite Program Solver for the Conformal
  Bootstrap},'' \href{http://dx.doi.org/10.1007/JHEP06(2015)174}{{\em JHEP}
  {\bfseries 06} (2015) 174},
\href{http://arxiv.org/abs/1502.02033}{{\ttfamily arXiv:1502.02033 [hep-th]}}.
%%CITATION = ARXIV:1502.02033;%%.

\bibitem{Costa:2011mg}
M.~S. Costa, J.~Penedones, D.~Poland, and S.~Rychkov, ``{Spinning Conformal
  Correlators},'' \href{http://dx.doi.org/10.1007/JHEP11(2011)071}{{\em JHEP}
  {\bfseries 1111} (2011) 071},
\href{http://arxiv.org/abs/1107.3554}{{\ttfamily arXiv:1107.3554 [hep-th]}}.
%%CITATION = ARXIV:1107.3554;%%.

\bibitem{Kravchuk:2016qvl}
P.~Kravchuk and D.~Simmons-Duffin, ``{Counting Conformal Correlators},''
\href{http://arxiv.org/abs/1612.08987}{{\ttfamily arXiv:1612.08987 [hep-th]}}.
%%CITATION = ARXIV:1612.08987;%%.

\bibitem{Dymarsky:2013wla}
A.~Dymarsky, ``{On the four-point function of the stress-energy tensors in a
  CFT},'' \href{http://dx.doi.org/10.1007/JHEP10(2015)075}{{\em JHEP}
  {\bfseries 10} (2015) 075},
\href{http://arxiv.org/abs/1311.4546}{{\ttfamily arXiv:1311.4546 [hep-th]}}.
%%CITATION = ARXIV:1311.4546;%%.

\bibitem{Costa:2011dw}
M.~S. Costa, J.~Penedones, D.~Poland, and S.~Rychkov, ``{Spinning Conformal
  Blocks},'' \href{http://dx.doi.org/10.1007/JHEP11(2011)154}{{\em JHEP}
  {\bfseries 1111} (2011) 154},
\href{http://arxiv.org/abs/1109.6321}{{\ttfamily arXiv:1109.6321 [hep-th]}}.
%%CITATION = ARXIV:1109.6321;%%.

\bibitem{Li:2015itl}
D.~Li, D.~Meltzer, and D.~Poland, ``{Conformal Collider Physics from the
  Lightcone Bootstrap},'' \href{http://dx.doi.org/10.1007/JHEP02(2016)143}{{\em
  JHEP} {\bfseries 02} (2016) 143},
\href{http://arxiv.org/abs/1511.08025}{{\ttfamily arXiv:1511.08025 [hep-th]}}.
%%CITATION = ARXIV:1511.08025;%%.

\bibitem{Dobrev:1975ru}
V.~K. Dobrev, V.~B. Petkova, S.~G. Petrova, and I.~T. Todorov, ``{Dynamical
  Derivation of Vacuum Operator Product Expansion in Euclidean Conformal
  Quantum Field Theory},''
\href{http://dx.doi.org/10.1103/PhysRevD.13.887}{{\em Phys. Rev.} {\bfseries
  D13} (1976) 887}.
%%CITATION = PHRVA,D13,887;%%.

\bibitem{Karateev:2017jgd}
D.~Karateev, P.~Kravchuk, and D.~Simmons-Duffin, ``{Weight Shifting Operators
  and Conformal Blocks},''
\href{http://arxiv.org/abs/1706.07813}{{\ttfamily arXiv:1706.07813 [hep-th]}}.
%%CITATION = ARXIV:1706.07813;%%.

\bibitem{Dolan:2011dv}
F.~A. Dolan and H.~Osborn, ``{Conformal Partial Waves: Further Mathematical
  Results},''
\href{http://arxiv.org/abs/1108.6194}{{\ttfamily arXiv:1108.6194 [hep-th]}}.
%%CITATION = ARXIV:1108.6194;%%.

\bibitem{Penedones:2015aga}
J.~Penedones, E.~Trevisani, and M.~Yamazaki, ``{Recursion Relations for
  Conformal Blocks},'' \href{http://dx.doi.org/10.1007/JHEP09(2016)070}{{\em
  JHEP} {\bfseries 09} (2016) 070},
\href{http://arxiv.org/abs/1509.00428}{{\ttfamily arXiv:1509.00428 [hep-th]}}.
%%CITATION = ARXIV:1509.00428;%%.

\bibitem{Zhiboedov:2013opa}
A.~Zhiboedov, ``{On Conformal Field Theories With Extremal a/c Values},''
  \href{http://dx.doi.org/10.1007/JHEP04(2014)038}{{\em JHEP} {\bfseries 04}
  (2014) 038},
\href{http://arxiv.org/abs/1304.6075}{{\ttfamily arXiv:1304.6075 [hep-th]}}.
%%CITATION = ARXIV:1304.6075;%%.

\bibitem{Nachtmann:1973mr}
O.~Nachtmann, ``{Positivity constraints for anomalous dimensions},''
\href{http://dx.doi.org/10.1016/0550-3213(73)90144-2}{{\em Nucl.Phys.}
  {\bfseries B63} (1973) 237--247}.
%%CITATION = NUPHA,B63,237;%%.

\bibitem{Komargodski:2012ek}
Z.~Komargodski and A.~Zhiboedov, ``{Convexity and Liberation at Large Spin},''
  \href{http://dx.doi.org/10.1007/JHEP11(2013)140}{{\em JHEP} {\bfseries 1311}
  (2013) 140},
\href{http://arxiv.org/abs/1212.4103}{{\ttfamily arXiv:1212.4103 [hep-th]}}.
%%CITATION = ARXIV:1212.4103;%%.

\bibitem{Fitzpatrick:2012yx}
A.~L. Fitzpatrick, J.~Kaplan, D.~Poland, and D.~Simmons-Duffin, ``{The Analytic
  Bootstrap and AdS Superhorizon Locality},''
  \href{http://dx.doi.org/10.1007/JHEP12(2013)004}{{\em JHEP} {\bfseries 1312}
  (2013) 004},
\href{http://arxiv.org/abs/1212.3616}{{\ttfamily arXiv:1212.3616 [hep-th]}}.
%%CITATION = ARXIV:1212.3616;%%.

\bibitem{Costa:2017twz}
M.~S. Costa, T.~Hansen, and J.~Penedones, ``{Bounds for OPE coefficients on the
  Regge trajectory},''
\href{http://arxiv.org/abs/1707.07689}{{\ttfamily arXiv:1707.07689 [hep-th]}}.
%%CITATION = ARXIV:1707.07689;%%.

\bibitem{Aharony:2015mjs}
O.~Aharony, ``{Baryons, monopoles and dualities in Chern-Simons-matter
  theories},'' \href{http://dx.doi.org/10.1007/JHEP02(2016)093}{{\em JHEP}
  {\bfseries 02} (2016) 093},
\href{http://arxiv.org/abs/1512.00161}{{\ttfamily arXiv:1512.00161 [hep-th]}}.
%%CITATION = ARXIV:1512.00161;%%.

\bibitem{Hsin:2016blu}
P.-S. Hsin and N.~Seiberg, ``{Level/rank Duality and Chern-Simons-Matter
  Theories},'' \href{http://dx.doi.org/10.1007/JHEP09(2016)095}{{\em JHEP}
  {\bfseries 09} (2016) 095},
\href{http://arxiv.org/abs/1607.07457}{{\ttfamily arXiv:1607.07457 [hep-th]}}.
%%CITATION = ARXIV:1607.07457;%%.

\bibitem{Aharony:2016jvv}
O.~Aharony, F.~Benini, P.-S. Hsin, and N.~Seiberg, ``{Chern-Simons-matter
  dualities with $SO$ and $USp$ gauge groups},''
  \href{http://dx.doi.org/10.1007/JHEP02(2017)072}{{\em JHEP} {\bfseries 02}
  (2017) 072},
\href{http://arxiv.org/abs/1611.07874}{{\ttfamily arXiv:1611.07874
  [cond-mat.str-el]}}.
%%CITATION = ARXIV:1611.07874;%%.

\bibitem{Elkhidir:2014woa}
E.~Elkhidir, D.~Karateev, and M.~Serone, ``{General Three-Point Functions in 4D
  CFT},'' \href{http://dx.doi.org/10.1007/JHEP01(2015)133}{{\em JHEP}
  {\bfseries 01} (2015) 133},
\href{http://arxiv.org/abs/1412.1796}{{\ttfamily arXiv:1412.1796 [hep-th]}}.
%%CITATION = ARXIV:1412.1796;%%.

\bibitem{Echeverri:2015rwa}
A.~Castedo~Echeverri, E.~Elkhidir, D.~Karateev, and M.~Serone,
  ``{Deconstructing Conformal Blocks in 4D CFT},''
  \href{http://dx.doi.org/10.1007/JHEP08(2015)101}{{\em JHEP} {\bfseries 08}
  (2015) 101},
\href{http://arxiv.org/abs/1505.03750}{{\ttfamily arXiv:1505.03750 [hep-th]}}.
%%CITATION = ARXIV:1505.03750;%%.

\bibitem{Echeverri:2016dun}
A.~Castedo~Echeverri, E.~Elkhidir, D.~Karateev, and M.~Serone, ``{Seed
  Conformal Blocks in 4D CFT},''
  \href{http://dx.doi.org/10.1007/JHEP02(2016)183}{{\em JHEP} {\bfseries 02}
  (2016) 183},
\href{http://arxiv.org/abs/1601.05325}{{\ttfamily arXiv:1601.05325 [hep-th]}}.
%%CITATION = ARXIV:1601.05325;%%.

\bibitem{Costa:2016hju}
M.~S. Costa, T.~Hansen, J.~Penedones, and E.~Trevisani, ``{Projectors and seed
  conformal blocks for traceless mixed-symmetry tensors},''
  \href{http://dx.doi.org/10.1007/JHEP07(2016)018}{{\em JHEP} {\bfseries 07}
  (2016) 018},
\href{http://arxiv.org/abs/1603.05551}{{\ttfamily arXiv:1603.05551 [hep-th]}}.
%%CITATION = ARXIV:1603.05551;%%.

\bibitem{Cuomo:2017wme}
G.~F. Cuomo, D.~Karateev, and P.~Kravchuk, ``{General Bootstrap Equations in 4D
  CFTs},''
\href{http://arxiv.org/abs/1705.05401}{{\ttfamily arXiv:1705.05401 [hep-th]}}.
%%CITATION = ARXIV:1705.05401;%%.

\end{thebibliography}\endgroup
\bibliographystyle{utphys}

\end{document}